\documentclass[prb,twocolumn,superscriptaddress,showpacs,preprintnumbers,floatfix,eqsecnum, ]{revtex4-2}

\usepackage{amsmath,amssymb}

\usepackage{graphicx}
\usepackage{xcolor}

\pdfoutput=1

\begin{document}
\title{Three-body Fermi liquid corrections for an infinite-$U$ SU($N$) 
Anderson impurity model}

\author{Kaiji Motoyama}
\affiliation{Department of Physics, Osaka Metropolitan University, 
Sumiyoshi-ku, Osaka 558-8585, Japan}

\author{Yoshimichi Teratani}
\affiliation{Department of Physics, Osaka City University, 
Sumiyoshi-ku, Osaka 558-8585, Japan
}
\affiliation{
NITEP, Osaka Metropolitan University, Sumiyoshi-ku, Osaka 558-8585, Japan}

\author{Kazuhiko Tsutsumi}
\affiliation{Department of Physics, Osaka City University, 
Sumiyoshi-ku, Osaka 558-8585, Japan
}
\affiliation{
NITEP, Osaka Metropolitan University, Sumiyoshi-ku, Osaka 558-8585, Japan}

\author{Kohei Wake}
\affiliation{Department of Physics, Osaka City University, 
Sumiyoshi-ku, Osaka 558-8585, Japan
}

\author{Ryosuke Kobayashi}
\affiliation{Department of Physics, Osaka City University, 
Sumiyoshi-ku, Osaka 558-8585, Japan
}

\author{Rui Sakano}
\affiliation{Department of Physics, Keio University, 3-14-1 Hiyoshi, Kohoku-ku, Yokohama, Kanagawa 223-8522, Japan
}

\author{Akira Oguri}
\affiliation{Department of Physics, Osaka City University, 
Sumiyoshi-ku, Osaka 558-8585, Japan
}
\affiliation{
NITEP, Osaka Metropolitan University, Sumiyoshi-ku, Osaka 558-8585, Japan}

\date{\today}

\begin{abstract}
We study the three-body Fermi liquid effects in the SU($N$) Anderson impurity model 
within the strong interaction limit $U \to \infty$, where the occupation number 
$N_d^{}$ of the impurity levels varies over the range of  $0<N_d^{}<1$.
The three-body correlations arise among the  impurity electrons 
when the electron-hole symmetry,  the time-reversal symmetry, 
or both are broken by external fields or potentials. 
They play a key role in low-energy transport, particularly
 in the next-to-leading-order terms 
 of power expansion with respect to temperature $T$ and 
bias voltage $eV$.  
Using the numerical renormalization group approach, we calculate the differential 
conductance and nonlinear current noise through quantum dots, 
as well as the thermal conductivity of both quantum dots and magnetic alloys. 
 Specifically, we focus on the SU(2) and SU(4) cases, 
 demonstrating how the three-body correlations evolve 
in the strong interaction limit 
across the $1/N$-filling Kondo regime at  $N_d^{}\simeq 1.0$ 
and the valence fluctuation regime where 
the occupation number $N_d^{}$ varies rapidly  
with the position $\epsilon_d^{}$ of the impurity levels. 
We also show that the three-body correlations strongly couple with asymmetries 
in the tunnel couplings between quantum dots and reservoirs.  
For $N=4$, this coupling significantly affects 
the order $(eV)^3$ nonlinear current through the quarter-filling Kondo state, 
where the electron-hole symmetry is broken. 
In particular, in the $U\to \infty$ limit,  
the expansion coefficients exhibit clear 
plateau structures associated with the Kondo effect  at $N_d^{}\simeq 1.0$.
Our formulation and analysis pave the way for deducing the three-body correlations 
from the low-energy experiments on the next-to-leading-order transport.
\end{abstract}

\maketitle

\section{Introduction}

The Kondo effect is an intriguing quantum phenomenon 
\cite{Kondo2012, hewson_1993} that occurs 
 in dilute magnetic alloys (MAs), quantum dots (QDs), 
and various exotic systems, such as ultracold atomic gases 
 \cite{TakahashiColdGasKondo} and quark matter \cite{QCD_Kondo}.
In these Kondo systems, localized impurity spins are screened 
by surrounding conduction electrons at low energies, 
leading to a highly correlated singlet ground state. 
In the 1970s, 
it was confirmed that the ground state and low-lying energy excitations, 
obtained with the high-accuracy numerical renormalization group (NRG)
 \cite{Wilson1975, KWW1, KWW2},   
can be described by an effective zero-dimensional field theory, 
commonly known as local Fermi liquid theory
\cite{Nozieres1974,YamadaYosida2,YamadaYosida4,ShibaKorringa,Yoshimori}.

The low-energy properties 
of Kondo systems realized in quantum dots  have been studied intensively 
through highly sensitive measurements 
\cite{Goldhaber-Gordon1998nature,Goldhaber-Goldon1998PRL,Cronenwett-Oosterkamp-Kouwenhoven1998,vanderWiel2000,KondoCloud2020,GrobisGoldhaber-Gordon,ScottNatelson,Heiblum,Delattre2009,KobayashiKondoShot,Ferrier2016,Hata2021,Costi2022, Svilansexperimentthermopower}. 
Simultaneously, theories to deduce the universal Fermi liquid behavior from experiments  
 have been developed, 
specifically for electrical currents
\cite{Glatzman_RaiKh1988,Ng_Lee1988,Hershfield1,MeirWingreen,Izumida2001,AO2001,Sela-Malecki2009,Aligia2012,Aligia2014,Munoz},
 shot noise \cite{Hershfield2,GogolinKomnikPRL,Sela2006,Golub,OguriSakanoFujii2011}, 
and thermal conductivity \cite{CostiZlatic2010, CostiThermo, Costimagthermopower}.
Furthermore, recent developments 
have revealed 
that three-body correlations between impurity electrons 
play an essential role in Fermi-liquid transport 
at low but finite temperatures $T$ or finite bias voltages $eV$ 
\cite{Mora_etal_2009,Mora2009,MoraMocaVonDelftZarand,FMvDM2018,AO2017_I,AO2017_II,AO2017_III}.

Three-body correlations emerge when 
the electron-hole symmetry, the time-reversal symmetry, 
or both are broken by an internal potential or external field, 
and evolve as the impurity level $\epsilon_d^{}$ 
shifts toward or away from the Fermi level $E_F$ of conduction bands. 
These correlations affect the next-to-leading-order terms 
in both  
the linear and nonlinear responses of 
 electrical and thermal currents 
\cite{KarkiMora2018, KarkiKiselev, MocaMora2018, Teratani2020PRL, Oguri2022, Tsutsumi2021,Tsutsumi2023, teratani2024thermoelectric}.   
The degeneracy $N$ of impurity state also 
introduces interesting variations in three-body effects.  
The SU($N$) Kondo state for $N=4$, 
consisting of the spin and orbital degrees of freedoms, 
 has been realized in carbon nanotubes 
and other types 
of quantum dots 
\cite{RMP-Kouwenhoven,Sasaki2000,Pablo2005,Babic2004,Makarovski2007,Cleuziou2013, Ferrier2016}
 and has investigated theoretically by a numerous authors  
\cite{Izumida1998,Borda2003,Choi2005,Eto2005,Sakano2006,Anders2008,SU4_Kondo_ferro_Weymann,Filippone2014,MantelliMocaZarandGrifoni,Teratani2020PRB}.    
In particular, Mora {\it et al.\ \!\!\!} 
derived several 
 Fermi-liquid relations for the three-body correlations in the SU($N$) Kondo model,
 which can be classified according 
to an integer $N_d^{}$ ($=1$, $2$, $\ldots$, $N-1$) that corresponds to 
the number of electrons at the impurity site 
\cite{Mora_etal_2009, Mora2009}.

Fermi liquid theory for three-body correlations has also been developed for   
the Anderson impurity model, 
in which 
the occupation number $N_d^{}$ 
can vary continuously in the range $0<N_d^{}<N$ 
 \cite{MoraMocaVonDelftZarand,FMvDM2018,AO2017_I,AO2017_II,AO2017_III,Teratani2020PRL,Oguri2022,Tsutsumi2021,Tsutsumi2023,teratani2024thermoelectric}.  
For multilevel impurities with $N \geq 4$,
 it has been shown  
that the contribution of three-body correlations 
is particularly pronounced in the Kondo states  
with the electron numbers $N_d^{} \simeq 1$ and $N-1$, 
and that this contribution increases with the Coulomb interaction $U$ 
\cite{Tsutsumi2023, teratani2024thermoelectric}. 
Therefore, it is expected that three-body correlations are maximized 
in the strong interaction limit $U\to \infty$, 
which could
 provide a benchmark for their contributions. 
 Low-energy properties at $U \to \infty$ have been studied for a long time, 
for instance with the NRG,   
 while also taking 
additional inter-site Coulomb interactions into account    
 \cite{TakayamaSakai, Alascio1986, Zhuravlev2007}. 
However, to our knowledge,  three body correlations have not yet been 
explored in this limit.

The purpose of this paper is to demonstrate the role of 
three-body correlations in the SU($N$) Anderson model  
in the strong interaction limit $U \to \infty$ 
over a wide range of the impurity level position $\epsilon_d^{}$,   
across the $1/N$-filling Kondo regime and  
the valence fluctuation regime $0<N_d^{}<1$. 
To this end, we calculate the Fermi liquid parameters  
such as the phase shift $\delta_\sigma^{}$, the linear susceptibility 
$\chi_{\sigma \sigma'}^{}$
and the three-body correlation function $\chi^{[3]}_{\sigma \sigma'\sigma''}$, 
using the NRG for $N=2$ and $N=4$. 
Here, $\sigma$ labels the $N$ impurity levels, 
with $\sigma=1$, $2$, $\ldots$, $N$.   
From these correlation functions,  
 electrical and thermal currents through quantum dots or magnetic alloys    
 can be deduced up to  next-to-leading-order at finite  $T$ and finite $eV$.

The three-body correlation function  $\chi^{[3]}_{\sigma \sigma'\sigma''}$ has 
 three independent components for the SU($N$) Anderson model of $N \geq 4$, 
which approach 
 a single universal value 
in the $1/N$-filling Kondo regime.  
However, as the occupation number $N_d^{}$ decreases from the $1/N$-filling, 
the three independent components 
contribute distinctly to the transport coefficients,
especially in the valence fluctuation regime.
We also find 
that the order $|eV|^3$ term of the shot noise for SU(4) quantum dots  
becomes positive definite throughout the entire range of $\epsilon_d^{}$
 in the limit of $U\to \infty$, 
in contrast to the case of finite interactions, where the order $|eV|^3$ term 
takes a negative value  
in the valence fluctuation regime for small $U$ \cite{teratani2024thermoelectric}.
Additionally, we investigate the effects of tunneling and bias asymmetries 
on the nonlinear conductance through quantum dots 
at $U\to \infty$ \cite{Tsutsumi2021,Tsutsumi2023},  and demonstrate  
that the three-body correlations couple strongly to these asymmetries.

This paper is organized as follows. 
In Sec.\ \ref{sec:formulation}, 
we introduce 
the Anderson impurity model for QDs and MAs, 
along with the Fermi liquid parameters that include three-body correlations. 
We also introduce 
the nonequilibrium steady-state current and current noise through QDs. 
Section \ref{sec:thermoelectric_formulation} is devoted to 
 the formulation of 
thermoelectiric transport for QDs and MAs. 
Section \ref{sec:SUN_FL_description} 
discusses the three-body correlations in SU($N$) case. 
In Sec.\  \ref{sec:NRG_FL_parameters} 
we discuss NRG results for the Fermi liquid parameters obtained 
in the $U\to \infty$ limit for SU(2) and SU(4) symmetric cases. 
In Sec.\ \ref{sec:NRG_dI/dV_QD}, 
 we discuss the behavior of the next-to-leading-order terms 
of the differential conductance $dI/dV$ through QDs, 
and examine their dependence on the tunneling and bias asymmetries. 
Section \ref{sec:NRG_noise_QD} 
addresses 
  the nonlinear current noise through QDs for $U\to\infty$. 
Sections \ref{sec:NRG_thermal_QD}
and \ref{sec:NRG_thermal_MA} describe thermoelectric transport 
of QDs and MAs, respectively.
A summary is provided in Sec.\ \ref{sec:summary}. 
In the Appendix, we briefly describe 
 Fermi liquid relations including the three-body correlations.

\section{Formulation}
\label{sec:formulation}

\subsection{Multilevel Anderson impurity model}

We begin with an $N$-level Anderson impurity model connected 
to two non-interacting leads on the left ($L$) and right ($R$):
\begin{align}
 &
{H}  \, = \,{H}_d^{} + {H}_c^{} + {H}_T^{}, 
\label{eq:Hamilonian}
 \\[2mm]
 & \!\!
{H}_d^{}
 \,= \sum_{\sigma=1}^N \epsilon_{d\sigma}^{} n_{d\sigma}^{} 
 + \sum_{\sigma\neq\sigma'} \frac{U}{2} n_{d\sigma}^{} {n}_{d\sigma'}^{}\,, 
 \\
 &  \!\!
{H}_c^{}
 \,=   
\sum_{\nu=L,R} \, \sum_{\sigma=1}^N \, 
 \int_{-D}^D \! d\epsilon\,{c}_{\epsilon\nu\sigma}^\dag {c}_{\epsilon\nu\sigma}
\,,
 \\
 & \!\!
{H}_T^{}
 \,= 
- \!
\sum_{\nu=L,R} \, \sum_{\sigma=1}^N \, v_\nu^{}
 \int_{-D}^D \! d\epsilon \, \sqrt{\rho_c}^{}\,
 \left({c}_{\epsilon\nu\sigma}^\dag {d}_\sigma + \mathrm{H.c.}\right) ,
\label{eq:H_T}
\end{align}
Here ${d}_{\sigma}^\dag$ for $\sigma=1$, $2$, $\ldots$, $N$ is the creation operator for an impurity electron 
with energy $\epsilon_{d\sigma}^{}$ and 
$U$ represents the Coulomb interaction between two impurity electrons. 
The number operator of impurity electrons is given 
by $n_{d\sigma}^{} \equiv {d}_\sigma^\dag{d}_\sigma$. 
Additionally, ${c}_{\epsilon\nu\sigma}^\dag$ is the creation operator 
for conduction electrons in the noninteracting lead on $\nu=L,R$, 
normalized such that 
$\bigl\{c_{\epsilon\nu\sigma}\,,\,c_{\epsilon'\nu'\sigma'}^\dagger\bigr\} 
=\delta(\epsilon-\epsilon')\,\delta_{\nu\nu'}\delta_{\sigma\sigma'}$.
Charge transfer occurring between impurity level and conduction bands 
is described by ${H}_T^{}$,  where  $\rho_c^{} \equiv 1/(2D)$
 and $D$ is the half bandwidth of conduction bands.
The impurity-level width 
due to the tunnel couplings $v_\nu^{}$ for $\nu=L$ and $R$  
is given by $\Delta \equiv \Gamma_L + \Gamma_R$, with 
$\Gamma_\nu^{}\equiv \pi\rho_c^{} v_\nu^2$ 
[see also Eq.\ \eqref{eq:Green_Dyson} in Appendix]. 
In this paper, we consider the low-energy region 
where all these parameters except for $U$ are much smaller than 
$D$, i.e., $\max(\Delta, |\epsilon_{d\sigma}^{}|, |\omega|,T, |eV|) \ll D$, 
with $\omega$ representing the frequency introduced later. 


The current flowing through the impurity level $\sigma$ satisfies the equation 
of continuity, given by 
\begin{align}
\frac{\partial  n_{d\sigma}^{}}{\partial t} 
+ \widehat{I}_{R,\sigma}^{} - \widehat{I}_{L,\sigma}^{}\, =\, 0 \,. 
\label{eq:current_conservation}
\end{align}
Here  $\widehat{I}_{L,\sigma}^{}$ represents 
the current flowing from the left lead to the impurity level, 
and $\widehat{I}_{R,\sigma}^{}$ represents the current from the impurity level 
to the right lead:  
\begin{align}
\widehat{I}_{L,\sigma} =& \     -i\,  v_L 
\left(
\psi^{\dagger}_{L\sigma} d^{}_{\sigma} 
-d^{\dagger}_{\sigma} \psi^{}_{L\sigma} \right) , 
\\
\widehat{I}_{R,\sigma}^{} =& \  + i\, v_R 
\left(
\psi^{\dagger}_{R\sigma} d^{}_{\sigma} 
-d^{\dagger}_{\sigma} \psi^{}_{R\sigma}\right) .
\end{align}

\subsection{Three-body correlation functions}

The low-energy properties of the Anderson impurity can be 
described by Fermi liquid parameters, 
which include 
the occupation numbers $\langle n_{d\sigma}^{} \rangle$, 
susceptibilities $\chi_{\sigma_1\sigma_2}^{}$, 
and three-body correlation functions $\chi^{[3]}_{\sigma_1\sigma_2\sigma_3}$:  
\begin{align}
  \langle n_{d\sigma}^{} \rangle 
  &= \frac{\partial\Omega}{\partial\epsilon_{d\sigma}^{}}, 
  \\
  \chi_{\sigma_1\sigma_2}^{} 
  &\equiv 
-\frac{\partial\Omega}{\partial\epsilon_{d\sigma_1}^{}\partial\epsilon_{d\sigma_2}^{}}
  = \int_0^\beta \!\! d\tau\,\langle \delta n_{d\sigma_1}^{}(\tau)\,
\delta{n}_{d\sigma_2}^{}\rangle,
  \\
  \chi^{[3]}_{\sigma_1\sigma_2\sigma_3}
  &\equiv 
- \frac{\partial^3\Omega}{\partial\epsilon_{d\sigma_1}^{} \partial\epsilon_{d\sigma_2}^{} \partial\epsilon_{d\sigma_3}^{}}
  \nonumber \\
  & 
\!\!\!\!\!\!\!\!\! 
\!\!\!\!\!\!\!\!
= 
-\int_0^\beta \!\! d\tau \!\! \int_0^\beta \!\! d\tau' 
\langle T_\tau \delta n_{d\sigma_1}^{}(\tau)\, \delta n_{d\sigma_2}^{}(\tau')\, 
\delta{n}_{d\sigma_3}^{}\rangle.
\end{align}
Here, $\langle  \mathcal{O} \rangle
= \mathrm{Tr} \bigl[\,e^{-\beta H} \mathcal{O}\,\bigr]
/\mathrm{Tr}\, e^{- \beta H}$ denotes the thermal average,  
 $\Omega\equiv-\frac{1}{\beta}\mathrm{ln}[e^{-\beta{H}}]$ 
is the free energy, and  $\beta=1/T$ is the inverse temperature. Additionally, 
$\delta n_{d\sigma}^{}\equiv n_{d\sigma}^{} - \langle n_{d\sigma}^{}\rangle$, 
 $\mathcal{O}(\tau) \equiv e^{\tau H} \mathcal{O}\,e^{-\tau H}$, and  
 $T_\tau$ is the imaginary-time ordering operator. 
Specifically, the ground-state values of these correlation functions
--- namely,  
  $\langle n_{d\sigma}^{}\rangle$, 
  $\chi_{\sigma_1\sigma_2}^{}$, and 
  $\chi^{[3]}_{\sigma_1\sigma_2\sigma_3}$ at $T=0$ ---
completely determine the low-energy transport properties 
in the Fermi liquid regime, up to the next-to-leading-order terms.  
The three-body correlations of impurity electrons emerge when 
the electron-hole symmetry, the time-reversal symmetry, or both are broken, 
and affect the next-to-leading-order terms, 
such as  $T^2$ conductance and $(eV)^3$ nonlinear current. 
Physically, the three-body correlations determine the energy shift 
of order $\omega^2$, $T^2$, and  $(eV)^2$ of low-lying excitations, 
i.e., corrections in the same order as those arising from the lifetime of quasiparticles.

\subsection{Nonequilibrium current through quantum dots}

We consider a steady current $I$ under a finite bias voltage $V$: 
\begin{align}
 &I \, =  \,   \frac{e}{h} \sum_{\sigma=1}^N
\int_{-\infty}^\infty \!\! d\omega\, 
\Bigl[ f_L(\omega) - f_R(\omega) \Bigr] \, 
\mathcal{T}_{\sigma}^{}(\omega) \,, 
\label{eq:LandauerFM}
\\
& \mathcal{T}_{\sigma}^{}(\omega) 
\,\equiv  \  
\frac{4\Gamma_L\Gamma_R}{\Gamma_L+\Gamma_R}
 \, \pi A_\sigma(\omega)\,. 
\rule{0cm}{0.6cm}
\label{eq:transmissionPB}
\end{align}
Here, the average 
 $I \equiv 
\sum_{\sigma} \langle  \widehat{I}_{R,\sigma}^{}\rangle_{V}^{} = 
\sum_{\sigma} \langle \widehat{I}_{L,\sigma}^{}\rangle_{V}^{} $ 
is defined with respect to a nonequilibrium steady state, 
constructed at finite bias voltages $eV$ and temperatures $T$,   
using the Keldysh formalism \cite{Hershfield1,MeirWingreen,Caroli}. 
The Fermi distribution function for the conduction band in lead 
  $\nu=L$ or $R$ is given by 
 $f_\nu(\omega) = \big[e^{\beta(\omega-\mu_\nu)} + 1\big]^{-1}$. 
We consider the situation where 
 the chemical potentials of the left and right leads,   
denoted by $\mu_L$ and  $\mu_R$, respectively,  
are shifted from the Fermi level $E_F$ defined 
at the equilibrium ground state ($eV=T=0$), 
such that  $\mu_L=E_F+\alpha_L\,eV$ and $\mu_R=E_F-\alpha_R\,eV$.     
The bias voltage $\mu_L-\mu_R \equiv eV$ varies continuously 
between positive ($eV \geq 0$) and negative ($eV< 0$) values.  
%
The parameters $\alpha_L$ and $\alpha_R$ are positive  
and satisfy the condition $\alpha_L+\alpha_R=1$. 
For instance, when $\alpha_L=\alpha_R =1/2$, this describes 
a situation where the bias voltage is applied symmetrically, 
with the Fermi level $E_F$ at the center of the bias window. 
In contrast, when $\alpha_L=1$ and $\alpha_R=0$,  
the bias is applied entirely to the left lead ($\mu_L= E_F+eV$), 
while the right lead is grounded at the Fermi level ($\mu_R=E_F$).
Similarly, when $\alpha_L=0$ and $\alpha_R=1$, 
the left lead is  grounded at the Fermi level ($\mu_L= E_F$)  
and the entire bias voltage is applied entirely to the right lead ($\mu_R= E_F-eV$). 
This parametrization naturally interpolates between these cases 
\cite{Mora_etal_2009,Aligia2012,Aligia2014,Sela-Malecki2009},  
and the bias asymmetry can be characterized by a single parameter:    
\begin{align}
\alpha_\mathrm{dif}^{}
\,\equiv & \ 
\frac{\mu_L+\mu_R-2E_F}{\mu_L-\mu_R}
 \ = \ 
\alpha_L-\alpha_R\, ,
\label{eq:alpha_dif}
\end{align}
which takes values in the range $-1\leq \alpha_\mathrm{dif}^{} \leq 1$. 
Note that the  inversion of the chemical potentials 
 $(\mu_L^{},\, \mu_R^{})$ $\Rightarrow$ $(\mu_R^{},\, \mu_L^{})$ 
corresponds to  the transformation $(\alpha_\mathrm{dif}^{},\,eV)$ 
$\Rightarrow$ $(-\alpha_\mathrm{dif}^{},\, -eV)$. 
Hereafter, we set the Fermi level at the center of the conduction band, 
$E_F = 0$.

We also introduce a parameter 
that characterizes the asymmetry in the tunnel couplings: 
\begin{align}
\gamma_\mathrm{dif}^{}
\, \equiv & \  
\frac{\Gamma_L-\Gamma_R}{\Gamma_L+\Gamma_R}\,.  
\label{eq:gamma_dif}
\end{align}
Furthermore, we note that the nonequilibrium steady state, 
constructed using the Keldysh formalism,  
exhibits the following symmetry property \cite{Hershfield1,MeirWingreen,Caroli}:    
When both the chemical potentials and tunnel couplings are inverted, 
i.e.,  $(\mu_L,\mu_R, \Gamma_L, \Gamma_R)$ 
$\Rightarrow$ 
$(\mu_R,\mu_L, \Gamma_R, \Gamma_L)$,  
the current is inverted as well, $I \Rightarrow -I$. 


The spectral function $A_{\sigma}^{}(\omega)$   
on the right-hand side of Eq.\ \eqref{eq:transmissionPB}  
is defined as the imaginary part of retarded Green's function:
\begin{align}
&G_\sigma^r(\omega) \, = \, 
-i\int_{0}^{\infty}\!\! dt\,e^{i(\omega+i0^+) t}\,
\left\langle\Bigl\{d_\sigma^{}(t),\,d_\sigma^\dagger\Bigr\} \right \rangle_{V} . 
\label{eG}
\\[2.5mm] 
&A_{\sigma}^{}(\omega) \, = \,  
-\frac{1}{\pi} \,\mathrm{Im}\,G_{\sigma}^{r}(\omega)\;.
\label{eq:spectral_function} 
\end{align}
The behavior of the Green's function at small $\omega$, $T$, and $eV$ determines 
the low-energy Fermi liquid properties. 
In the zero-temperature limit $T=0$, the linear conductance $dI/dV|_{eV=0}^{}$ 
is determined by the spectral function at $\omega=0$, i.e., 
 $A_{\sigma}^{}(0) \xrightarrow{\,T=eV=0\,} 
\sin^2 \delta_{\sigma}^{}/(\pi \Delta)$, 
with $\delta_{\sigma}^{}$ 
the phase shift defined as the argument of the Green's function 
 $G_\sigma^r(0) \xrightarrow{\,T=eV=0\,} 
-\left| G_\sigma^r(0)\right| e ^{i \delta_{\sigma}^{}}$ 
at the equilibrium ground state. 
At finite temperature $T$ or bias voltage $eV$, 
deviations of the current $I$ from its ground-state value 
are governed by the low-lying excited states.
These contributions can be systematically evaluated through an expansion 
of $A_{\sigma}^{}(\omega)$ in powers of $\omega$, $T$, and $eV$.
Specifically, this expansion has been exactly carried out up to 
terms of order  $\omega^2$, $T^2$, and $(eV)^2$ 
\cite{Mora_etal_2009,Mora2009,MoraMocaVonDelftZarand,FMvDM2018,AO2017_I,AO2017_II,AO2017_III}.   
 The corresponding expansion coefficients can be expressed in terms of  
 the phase shift $\delta_{\sigma}^{}$, 
 susceptibilities $\chi_{\sigma_1\sigma_2}^{}$, 
 and three-body correlation functions $\chi^{[3]}_{\sigma_1\sigma_2\sigma_3}$,  
 all defined at the equilibrium ground state, i.e., at $\omega=T=eV=0$ 
 \cite{Tsutsumi2023,teratani2024thermoelectric}, 
 as summarized in Appendix \ref{sec:Ward_identity}. 


We also consider the current noise $S_\mathrm{noise}^\mathrm{QD}$,  
defined as the correlation function for fluctuations of a symmetrized current 
$\delta{\widehat{I}} = {\widehat{I}} - \langle \widehat{I}\rangle_{V}^{}$ 
\cite{Hershfield2,Mora_etal_2009,MoraMocaVonDelftZarand,Teratani2020PRL,Oguri2022}:
\begin{align}
 S_\mathrm{noise}^\mathrm{QD}
 \,\equiv & \  e^2 \int_{-\infty}^{\infty} \!\! dt \, 
\left\langle 
\delta{\widehat{I}}(t)\, \delta{\widehat{I}}(0)
+  \delta{\widehat{I}}(0)\,\delta{\widehat{I}}(t)
\right\rangle_{V}^{} ,
\label{eq:S_noise}
\\ 
 \widehat{I} \,\equiv & \   
\sum_{\sigma} 
\frac{\Gamma_R \,\widehat{I}_{L,\sigma} 
+ \Gamma_L \,\widehat{I}_{R,\sigma}}{\Gamma_L + \Gamma_R}
\,. 
\rule{0cm}{0.7cm}
\end{align}
Specifically, we calculate the terms up to order $|eV|^3$ 
for nonlinear noise, 
based on expansion formulas obtained in Ref.\ \onlinecite{Oguri2022}, 
taking into account all contributions from 
vertex corrections in the Keldysh formalism.

\section{Thermoelectric transport coefficients}
\label{sec:thermoelectric_formulation}

We also consider the thermoelectric transport through 
multilevel quantum dots and magnetic alloys in the linear-response regime,  
the low-energy behaviors of which can be deduced from 
the asymptotic form of the spectral function $A_{\sigma}(\omega)$.  
Specifically,  we focus on 
the three-body Fermi liquid corrections that emerge 
in the next-to-leading-order terms of the electrical and heat currents,  
as discussed later 
in Secs.\ \ref{sec:NRG_thermal_QD} and \ref{sec:NRG_thermal_MA}.
Here, we briefly describe the basic formulations.

When the temperature difference $\delta T$ 
is applied between the two leads such that $\delta T = T_L - T_R$,  
a heat current $I_Q^{} = \kappa\, \delta T$ flows 
from the high-temperature side toward the low-temperature side,
through quantum dots, or magnetic alloys. 
The thermal conductivity $\kappa$ is defined as  
 the linear-response coefficient of $I_Q^{}$.

\subsection{Thermoelectric coefficients for quantum dots}

The linear conductance $g$, 
thermopower $\mathcal{S}_\mathrm{QD}^{}$
and thermal conductance $\kappa_\mathrm{QD}^{}$ of a quantum dot 
 can be expressed in the form 
\cite{GuttmanBergman,CostiZlatic2010,MocaThermo,KarkiKiselev,Aligia2018thermo}:
\begin{align}
g\, \equiv   & \ 
\left. \frac{dI}{dV} \right|_{eV=0}^{} 
 \ = \ 
\frac{e^2}{h}
\sum_\sigma \mathcal{L}_{0,\sigma}^{\mathrm{QD}} \,, 
\label{eq:linear_conductance_QD} \\
\mathcal{S}_\mathrm{QD}^{}\, =&\, \frac{-1}{|e|T} 
\frac{\sum_{\sigma} \mathcal{L}_{1,\sigma}^{\mathrm{QD}}}
{\sum_{\sigma} \mathcal{L}_{0,\sigma}^{\mathrm{QD}}} \,, 
\label{eq:thermopower_QD}\\ 
\kappa_\mathrm{QD}^{} \,= & \  
\frac{1}{h\, T}
\left[\, 
\sum_{\sigma}
\mathcal{L}_{2,\sigma}^{\mathrm{QD}} 
- 
\frac{ \left(
\sum_{\sigma}
\mathcal{L}_{1,\sigma}^{\mathrm{QD}}\right)^2}{\sum_{\sigma}
\mathcal{L}_{0,\sigma}^{\mathrm{QD}}} \,\right] \,. 
\label{eq:thermal_coefficients_QD}
\end{align}
Here,  $\mathcal{L}_{n,\sigma}^\mathrm{QD}$ 
for $n=0,1$, and $2$ is defined at $eV=0$ 
 with respect to the thermal equilibrium, as   
\begin{align}
\mathcal{L}_{n,\sigma}^\mathrm{QD} = 
\int_{-\infty}^{\infty}  
d\omega\, 
\omega^n\, 
 \mathcal{T}_{\sigma}^{}(\omega) \,
\left( -
\frac{\partial f(\omega)}{\partial \omega}
\right) .
\label{eq:Ln_QD}
\end{align}
Here,  $\mathcal{T}_{\sigma}^{}(\omega)$ is the transmission probability 
defined in Eq.\ \eqref{eq:transmissionPB}. 
In the linear-response regime,   
it can be determined by expanding of $A_{\sigma}^{}(\omega)$ 
with respect to $\omega$ and $T$ at $eV=0$, 
as shown in Appendix \ref{sec:Ward_identity}.

\subsection{Thermoelectric coefficients for magnetic alloys}

Within the linear response theory,  
the electrical resistivity $\varrho_{\mathrm{MA}}^{}
=1/\sigma_{\mathrm{MA}}^{}$, 
 the thermopower $\mathcal{S}_\mathrm{MA}^{}$, 
and the thermal conductivity  $\kappa_{\mathrm{MA}}^{}$ 
for magnetic alloys are given by \cite{CostiThermo}:
\begin{align}
\sigma_{\mathrm{MA}}^{}
\ = & \   
\sigma_\mathrm{MA}^{\mathrm{unit}}\,\frac{1}{N}\,
\sum_{\sigma} \mathcal{L}_{0,\sigma}^\mathrm{MA} \,, 
\label{eq:conductivity_MA_L_0_sigma} \\ 
%
\mathcal{S}_\mathrm{MA}^{}\ =&\ \frac{-1}{|e|T} 
\frac{\sum_{\sigma} \mathcal{L}_{1,\sigma}^\mathrm{MA}}
{\sum_{\sigma} \mathcal{L}_{0,\sigma}^\mathrm{MA}} \,,
\label{eq:thermo_power_MA_L_1_sigma}
\\ 
\kappa_\mathrm{MA}^{} \ = & \   
\frac{\sigma_\mathrm{MA}^{\mathrm{unit}}}{e^{2}T}\,\frac{1}{N}\,
\left[\, 
\sum_{\sigma}
\mathcal{L}_{2,\sigma}^\mathrm{MA} 
- 
\frac{ \left(
\sum_{\sigma}
\mathcal{L}_{1,\sigma}^\mathrm{MA}\right)^2}{\sum_{\sigma}
\mathcal{L}_{0,\sigma}^\mathrm{MA}} \,\right] .
\label{eq:kappa_MA_L0_L1_L2}
\end{align}
For magnetic alloys,  
the response functions 
 $\mathcal{L}_{n,\sigma}^\mathrm{MA}$ for $n=0,1,$ and $2$ are 
determined by the inverse spectral function $1/A_{\sigma}(\omega)$  at $eV=0$:
\begin{align}
\mathcal{L}_{n,\sigma}^\mathrm{MA} = 
\int_{-\infty}^{\infty}  
d\omega\, 
\frac{\omega^n}{\pi \Delta A_{\sigma}(\omega)}
\left( -\frac{\partial f(\omega)}{\partial \omega}\right)  .  
\label{eq:Ln_MA}
\end{align}
 Equation \eqref{eq:conductivity_MA_L_0_sigma} 
defines the electrical conductivity relative to its unitary-limit value 
  $\sigma_\mathrm{MA}^{\mathrm{unit}}$. 
In the limit $T\to 0$,  
the electrical and thermal conductivities of the Anderson impurity 
satisfy the Wiedemann-Franz law 
(see Sec.\ \ref{sec:NRG_thermal_MA}):  
\begin{align}
\lim_{T \to 0}\, \frac{\kappa_\mathrm{MA}^{}}{T \,\sigma_\mathrm{MA}^{}}
\, = \,\frac{\pi^{2}}{3\,e^{2}} \,.
\label{eq:kappa_0_unitary_limit}
\end{align}

\section{Three-body correlations for SU($N$)  Anderson impurity}

\label{sec:SUN_FL_description}

The Hamiltonian $H$, defined in Eqs.\ \eqref{eq:Hamilonian}--\eqref{eq:H_T},
exhibits SU($N$) symmetry when the impurity levels are degenerate, i.e.,  
 $\epsilon_{d\sigma}^{} \equiv \epsilon_{d}^{}$ for all  $\sigma$.   
At  zero temperature $T=0$,
the occupation number of the impurity electron 
is determined by the phase shift $\delta_\sigma^{}$ through the Friedel sum rule 
$\langle n_{d\sigma}^{} \rangle = \delta_\sigma /\pi$, 
 which plays an central role in the ground-state property.  
In particular, in the SU($N$) symmetric case, 
  the total number of impurity electrons is given by
\begin{align}
 N_d^{} 
 \,\equiv\, \sum_{\sigma=1}^N\,\langle n_{d\sigma}^{} \rangle
 \, \xrightarrow{\,\mathrm{SU}(N)\,}\, \frac{N}{\pi}\, \delta\,.
\end{align}
Correspondingly,
the linear susceptibilities have two linearly independent components: 
the diagonal one $\chi_{\sigma\sigma}^{}$ 
and the off-diagonal one  $\chi_{\sigma\sigma'}^{}$  for $\sigma \neq \sigma'.$ 
The diagonal component determines 
 a characteristic energy scale $T^*$ of the SU($N$) Fermi liquid, 
which allows the $T$-linear specific heat 
$\mathcal{C}_\mathrm{imp}^\mathrm{heat}$ of impurity electrons 
to be expressed in the following form 
\cite{YamadaYosida2,ShibaKorringa,Yoshimori}: 
\begin{align}
\mathcal{C}_\mathrm{imp}^\mathrm{heat}
\,=\, \frac{N\pi^2}{12}\, \frac{T}{T^*}\,, 
\qquad  \quad  
T^* \, \equiv \,  \frac{1}{4\chi_{\sigma\sigma}^{}} \,. 
\label{eq:Fermiparaorigin}
\end{align}
 Note that the diagonal susceptibility can be expressed as
 $\chi_{\sigma\sigma}^{} = \rho_{d\sigma}^{}/z$, 
where the density of state of impurity electrons is given by  
 $\rho_{d\sigma}^{} =  \sin^2 \delta/(\pi \Delta)$  
and  $z$ is the wavefunction renormalization factor 
defined in Appendix \ref{sec:Ward_identity}.
The off-diagonal susceptibility determines the Wilson ratio $R$,  
or the rescaled one $\widetilde{K}$,  
which is bounded in the range $0 \leq  \widetilde{K} \leq 1$:  
\begin{align}
R \,\equiv\, 1-\frac{\chi_{\sigma\sigma'}^{}}{\chi_{\sigma\sigma}^{}}\,, 
\qquad \quad 
\widetilde{K}\,\equiv\, (N-1)(R-1)\,.  
\label{eq:K_tilde_def}
\end{align}

We also introduce dimensionless parameters for three-body correlations,  
which have three independent components for $N\geq 3$ in SU($N$) symmetric case:
\begin{align}
  \Theta_\mathrm{I}^{} 
  \, = & \ \frac{\sin2\delta}{2\pi}\, 
\frac{\chi^{[3]}_{\sigma\sigma\sigma}}{\chi_{\sigma\sigma}^2}\,, 
\qquad \ 
  \widetilde{\Theta}_\mathrm{I\!I}^{} 
  \,= \frac{\sin2\delta}{2\pi} \,
\frac{\widetilde{\chi}^{[3]}_{\sigma\sigma'\sigma'}}{\chi_{\sigma\sigma}^2}\,, 
\label{eq:Theta_I_II}
\\[1mm]
  \widetilde{\Theta}_\mathrm{I\!I\!I}^{} 
  \,= & \ \frac{\sin2\delta}{2\pi}\,  
\frac{\widetilde{\chi}^{[3]}_{\sigma\sigma'\sigma''}}{\chi_{\sigma\sigma}^2}\,, 
 \qquad \  \sigma\neq\sigma'\neq\sigma''\neq\sigma\, . 
\label{eq:Theta_III}
\end{align}
For the later two components, 
we have introduced rescaled correlation functions 
 $\widetilde{\chi}_{\sigma\sigma'\sigma'}^{[3]}$ and
 $\widetilde{\chi}_{\sigma\sigma'\sigma''}^{[3]}$ defined as follows:   
\begin{align}
\widetilde{\chi}_{\sigma\sigma'\sigma'}^{[3]} \, \equiv & \ 
 (N-1) \,\chi_{\sigma\sigma'\sigma'}^{[3]},   
 \label{eq:chi_II_tilde_def} 
\\[3mm]
\widetilde{\chi}_{\sigma\sigma'\sigma''}^{[3]} \, \equiv & \ 
\frac{(N-1)(N-2)}{2}\, \chi_{\sigma\sigma'\sigma''}^{[3]} \,. 
 \label{eq:chi_III_tilde_def} 
 \end{align}

In this work, we have calculated these three-body correlations 
as well as the two-body correlations $\chi_{\sigma\sigma'}^{}$
\cite{Hewson2004,Nishikawa2010v1,Nishikawa2010v2,OguriSakanoFujii2011,Oguri2012,Teratani2020PRL,teratani2024thermoelectric}, 
using the NRG approach, employing the following relations 
 (see Appendix \ref{sec:NRG_parameters} for details): 
\begin{align}
\chi^{[3]}_{\sigma\sigma\sigma}
  = & \,  \frac{1}{N}\frac{\partial\chi_{\sigma\sigma}^{}}{\partial\epsilon_d^{}} 
- \frac{N-1}{N}\chi_B^{[3]}\,, 
\label{eq:chi3_SUN_1} \\
  \widetilde{\chi}^{[3]}_{\sigma\sigma'\sigma'}
 = & \, \frac{N-1}{N}\frac{\partial\chi_{\sigma\sigma}^{}}{\partial\epsilon_d^{}} 
+ \frac{N-1}{N}\chi_B^{[3]}\,, 
\label{eq:chi3_SUN_2} \\
  \widetilde{\chi}^{[3]}_{\sigma\sigma'\sigma''}
=& \, 
-\frac{N-1}{N}\frac{\partial\chi_{\sigma\sigma}^{}}{\partial\epsilon_d^{}} 
+ \frac{N-1}{2}\frac{\partial\chi_{\sigma\sigma'}^{}}{\partial\epsilon_d^{}} 
 - \frac{N-1}{N}\chi_B^{[3]}\,.
\label{eq:chi3_SUN_3} 
\end{align}
Here  $\chi_\mathrm{B}^{[3]}$ is defined as the derivative of two-body correlations 
with respect to the magnetic field $b$: 
\begin{align}
  \chi_B^{[3]} \equiv 
\left.
\frac{\partial}{\partial b}
\left(
\frac{\chi_{m\uparrow,m\uparrow}^{} 
- \chi_{m\downarrow,m\downarrow}^{}}{2}\right)\right|_{b=0}^{}
  \! 
=\, -\chi_{\sigma\sigma\sigma}^{[3]} + \chi_{\sigma\sigma'\sigma'}^{[3]}.
\label{eq:chi_B3_def_new}
\end{align}
The magnetic field $b$ is applied 
to the impurity levels  $\sigma = (m,s)$, 
for $m=1,2,\ldots, N/2$ and $s=\uparrow, \downarrow$, 
inducing the spin Zeeman splitting, as  
\begin{align}
\epsilon_{d,m,\uparrow}^{} \,=\, \epsilon_{d}^{}\, -\,  b \,, 
\qquad \ \ 
\epsilon_{d,m,\downarrow}^{}\, =\, \epsilon_{d}^{} \,+\,  b \,.  
\label{eq:def_b_field}
\end{align}
In the following sections, we investigate how three-body correlations 
influence low-energy transport in the strong interaction limit 
across a broad range of impurity levels, $\epsilon_{d}^{}$.

\begin{figure}[t]
\begin{tabular}{cc}
  \begin{minipage}[t]{0.50\hsize}
  \centering
    \includegraphics[keepaspectratio, width=47mm]{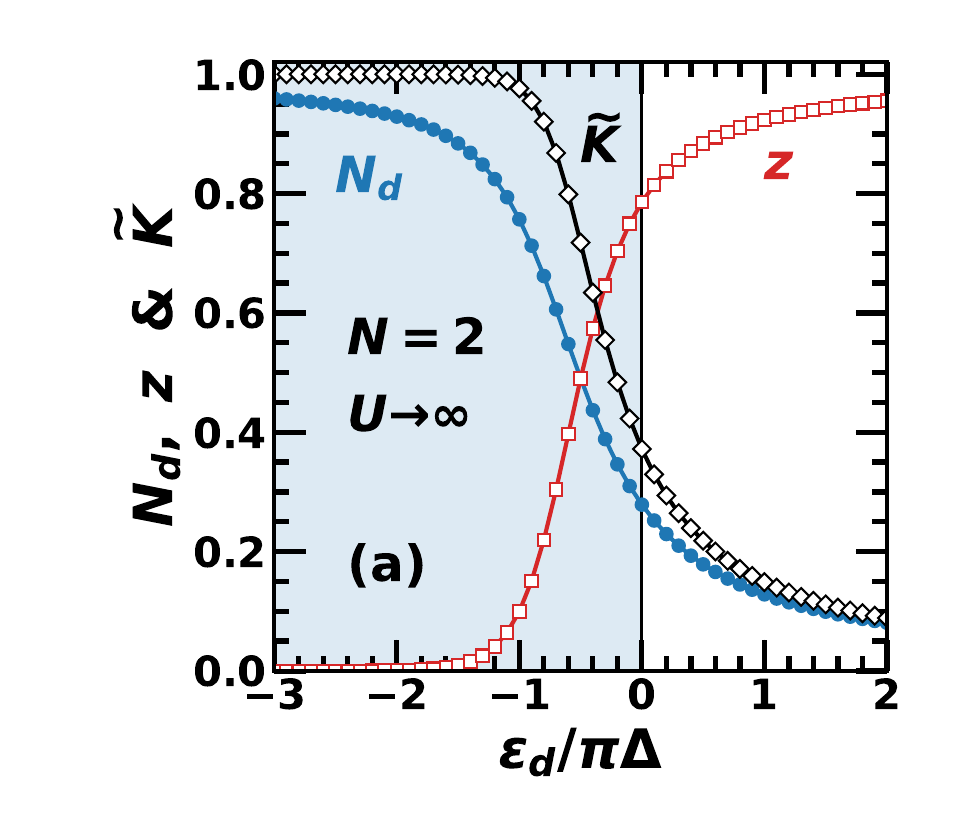}
  \label{su2_nd_z_Ktilde}
  \end{minipage}
  &
  \begin{minipage}[t]{0.50\hsize}
  \centering
    \includegraphics[keepaspectratio, width=47mm]{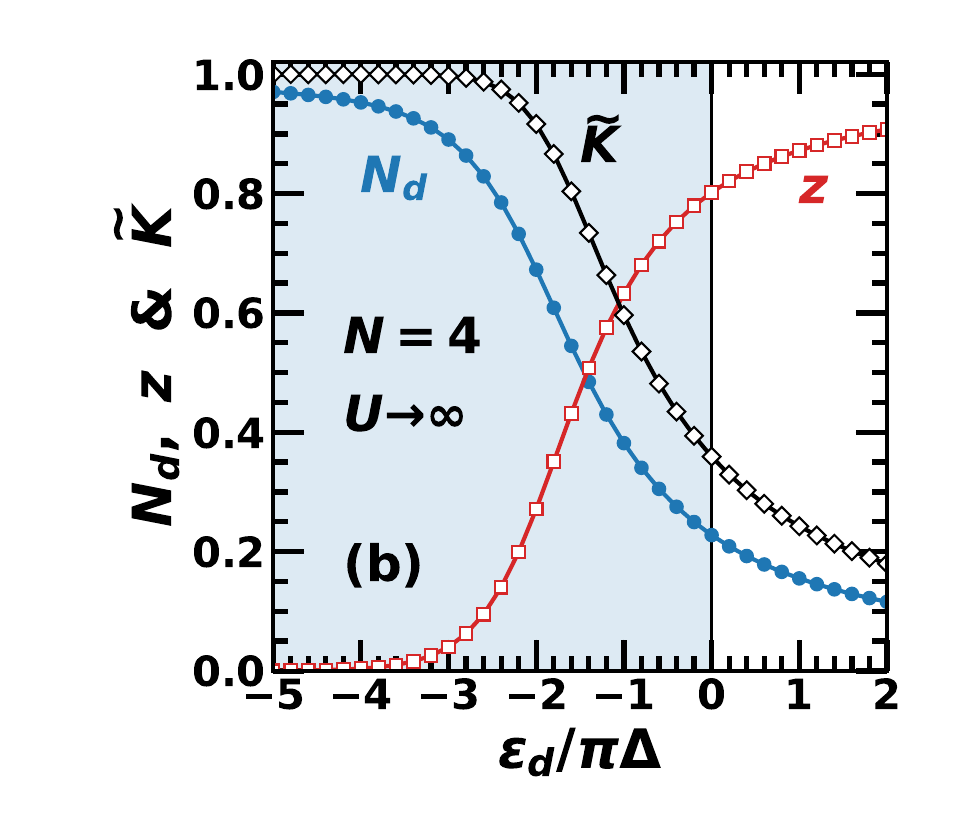}
  \label{su4_nd_z_Ktilde}
  \end{minipage}
\end{tabular}
\vspace{-7mm}
\caption{NRG results for $N_d^{}$, $z,$ and $\widetilde{K}$  
  plotted as functions of $\epsilon_d^{}$ 
in the $U \to \infty$ limit with $\pi \Delta/D = 0.01$: 
 (a) $N=2$  and (b) $N=4$.    
}
\label{fig:nd_z_Ktilde}
\end{figure}

\section{NRG results for SU($N$) Fermi liquid parameters 
for $U \to \infty$ Anderson model}
\label{sec:NRG_FL_parameters}

In this section, we discuss the SU($N$) Fermi-liquid (FL) parameters, 
obtained with the NRG in the $U\to\infty$ limit with $\pi \Delta/D = 0.01$. 
More detailed procedures for the NRG are described in Appendix 
\ref{sec:NRG_parameters}.

\subsection{$\epsilon_d^{}$ dependence 
of $N_d^{}$ and two-body functions}

The NRG results for the number of impurity electrons $N_d^{}$,  
the renormalization factor $z$, and 
the rescaled Wilson ratio $\widetilde{K}$ 
are plotted in Fig.\  \ref{fig:nd_z_Ktilde}, 
as functions of  $\epsilon_d^{}$ for (a) $N=2$ and (b) $N=4$.

The occupation number $N_d^{}$ ($=N \delta /\pi$) increases  
as the impurity level $\epsilon_d^{}$ decreases. 
It approaches the maximum possible value of      
 $N_d^{} \simeq 1$ in the Kondo regime at $\epsilon_d^{}\ll-\Delta$, 
while it becomes nearly empty $N_d^{}\simeq 0$  
in the opposite limit  $\epsilon_d^{}\gg\Delta$.  
In particular,  
$N_d^{}$ increases rapidly in the valence fluctuation regime, 
as $\epsilon_d^{}$ decreases.
Note that in the Kondo regime, 
the phase shift approaches $\delta \to \pi/N $, 
which corresponds to $\delta \to \pi/2$ and $\pi/4$ for 
 $N=2$ and  $4$, respectively.

The rescaled Wilson ratio 
$\widetilde{K}=(N-1)(R-1)$  
approaches its upper bound, $\widetilde{K} \to 1$,  
as $\epsilon_d^{}$ decreases,   
reflecting the strong electron correlations characteristics of the Kondo regime.  
Correspondingly, the renormalization factor $z$   
 becomes significantly small in this regime.
This factor $z$ also determines the width  
of renormalized level width, $\widetilde{\Delta}=z \Delta$, 
and the characteristic energy scale,   
defined in Eq.\ \eqref{eq:Fermiparaorigin}, as 
 $T^*/(\pi \Delta) = z/(4\sin^2 \delta)$.

\begin{figure}[t]
\begin{tabular}{cc}
  \begin{minipage}[t]{0.50\hsize}
  \centering
    \includegraphics[keepaspectratio, width=47mm]{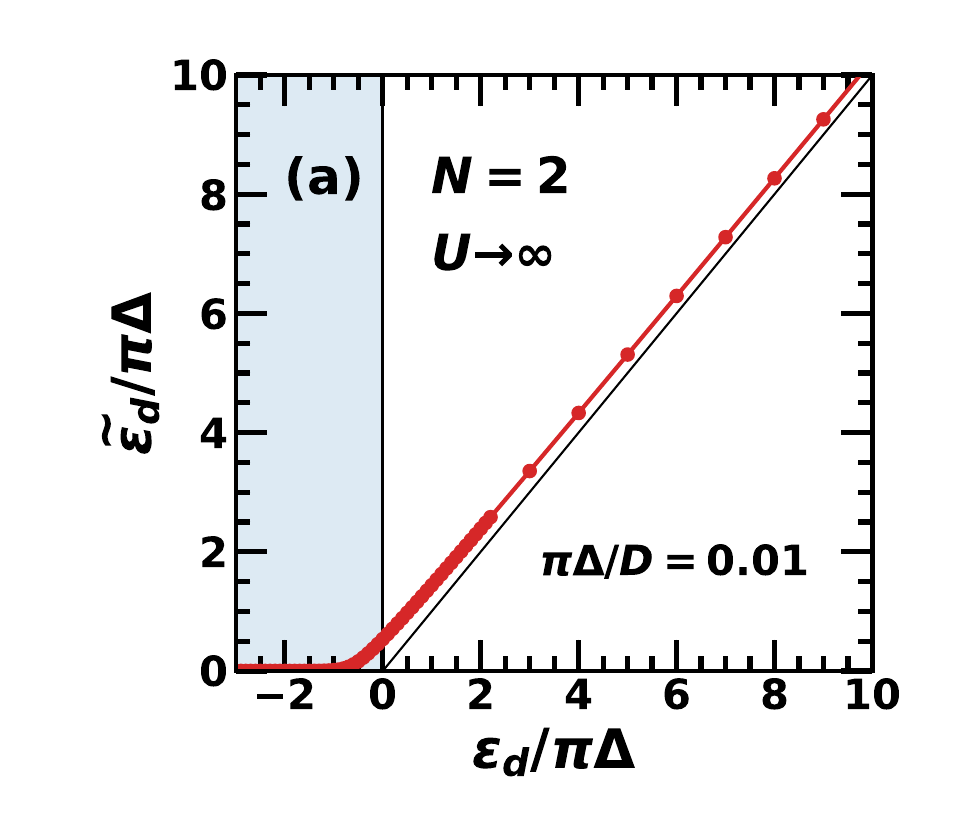}
  \label{su2_Ed_ren}
  \end{minipage}
  &
  \begin{minipage}[t]{0.50\hsize}
  \centering
    \includegraphics[keepaspectratio, width=47mm]{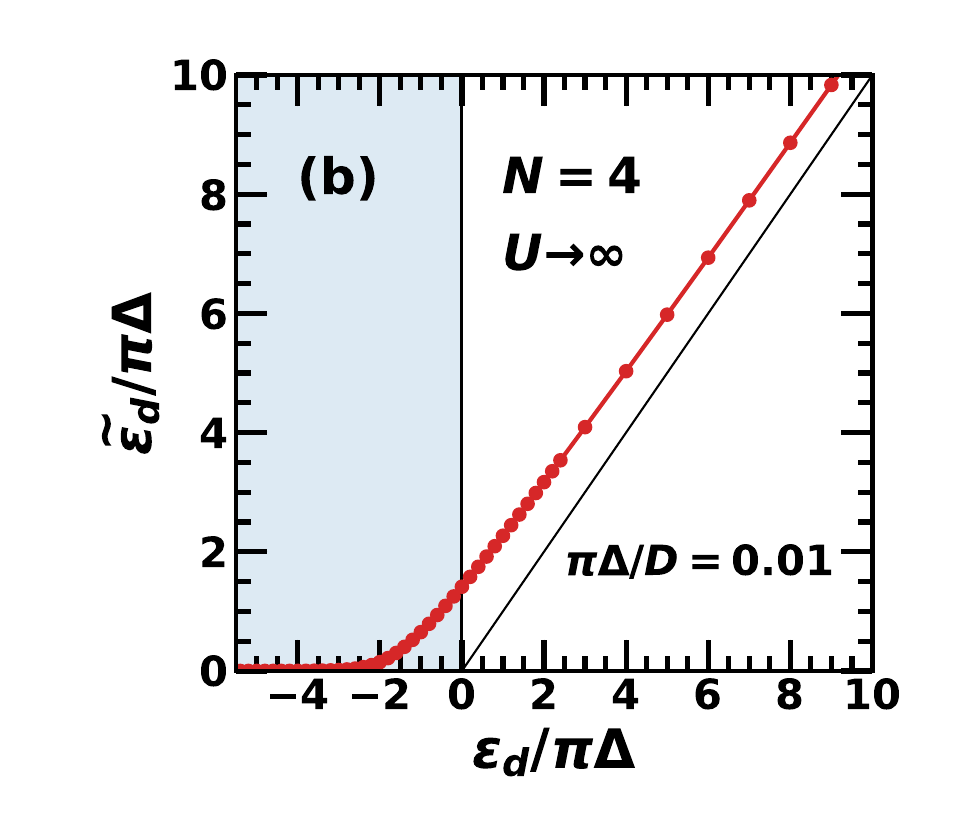}
  \label{su4_Ed_ren}
  \end{minipage}
\end{tabular}
\vspace{-7mm}
\caption{The renormalized impurity level  $\widetilde{\epsilon}_d^{}$,  
defined in Eq.\ \eqref{eq:ed_ren}, is plotted over a wide range 
of  $\epsilon_d^{}$ in the  $U \to \infty$ limit with $\pi \Delta/D = 0.01$: 
 (a) $N=2$ and (b) $N=4$.  
The thin straight line represents the position of bare impurity level 
$\epsilon_d^{}/(\pi\Delta)$.}
\label{fig:EdRen}
\end{figure}

Figure \ref{fig:EdRen} shows the results of renormalized impurity level 
$\widetilde{\epsilon}_{d}^{}=\widetilde{\Delta} \cot \delta$     
defined in Eq.\ \eqref{eq:ed_ren}. 
In the Kondo regime $\epsilon_d^{} \ll  -\Delta$, 
it approaches the Fermi level, $\widetilde{\epsilon}_{d}^{} \to 0 $  
keeping the ratio 
$\widetilde{\epsilon}_{d}^{}/\widetilde{\Delta} \to \cot \pi/N$ 
as a constant (e.g., $1$ for $N=4$ while 0 for $N=2$).  
In the opposite limit, $\epsilon_d^{} \gg  \Delta$, 
 the renormalized level $\widetilde{\epsilon}_{d}^{}$ 
 asymptotically approaches the noninteracting position  
as the occupation number $N_d^{}$ decreases.

\subsection{$\epsilon_d^{}$ dependence of three-body correlations}

\begin{figure}[t]
\begin{tabular}{cc}
  \begin{minipage}[t]{0.50\hsize}
  \centering
    \includegraphics[keepaspectratio, width=47mm]{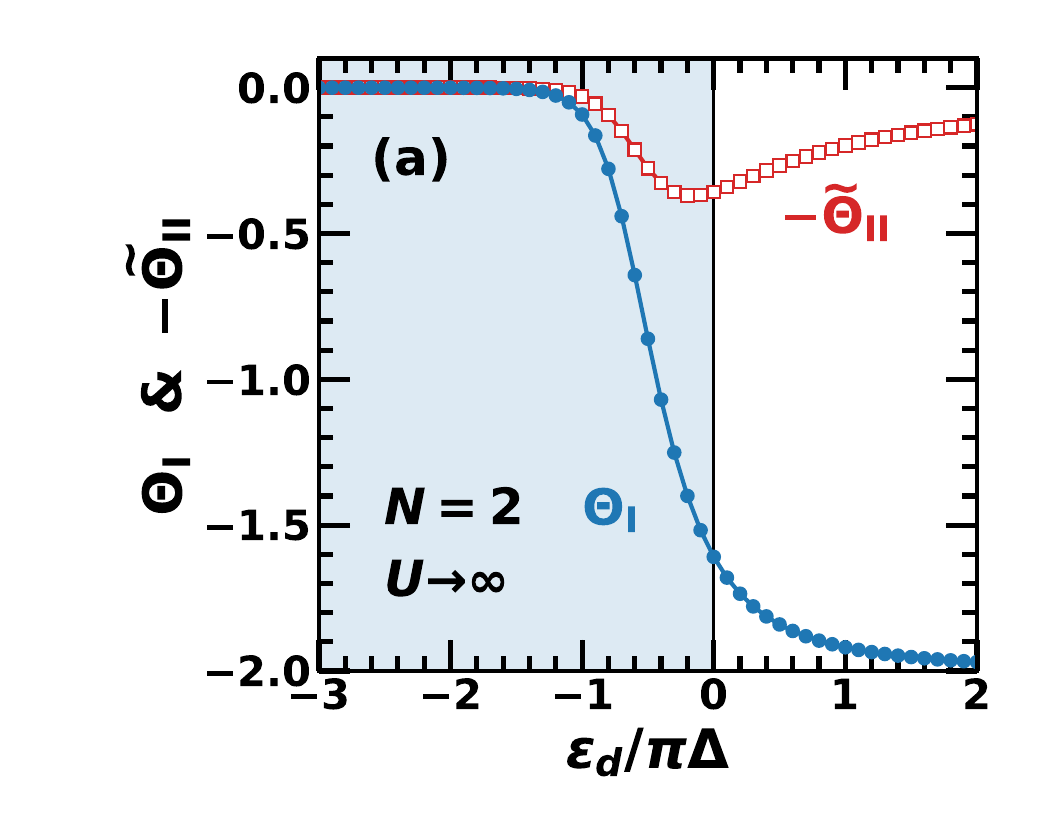}
  \label{su2_3bodycorrelations}
  \end{minipage}
  &
  \begin{minipage}[t]{0.50\hsize}
  \centering
    \includegraphics[keepaspectratio, width=47mm]{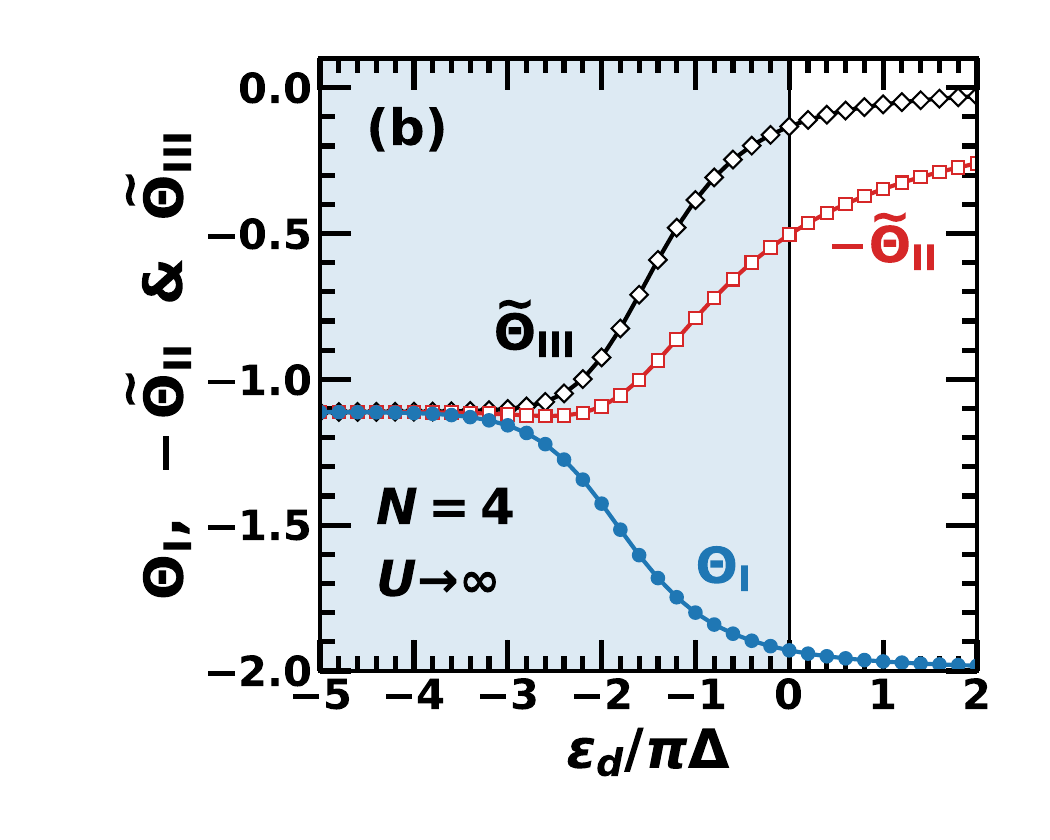}
  \label{su4_3bodycorrelations}
  \end{minipage}
\end{tabular}
\vspace{-7mm}
\caption{Dimensionless three-body correlation functions 
$\Theta_\mathrm{I}^{}$, 
$-\widetilde{\Theta}_\mathrm{I\!I}^{}$, and 
$\widetilde{\Theta}_\mathrm{I\!I\!I}^{}$,
defined in Eqs.\ \eqref{eq:Theta_I_II} and \eqref{eq:Theta_III},
 are  plotted as functions of $\epsilon_d^{}$ 
 in the $U \to \infty$ limit with $\pi \Delta/D = 0.01$: 
 (a) $N=2$ and (b) $N=4$.
}
\label{fig:3bodycorrelations}
\end{figure}

NRG results for the dimensionless 
three-body correlation functions $\Theta_\mathrm{I}^{}$, 
$ -\widetilde{\Theta}_\mathrm{I\!I}^{}$, and 
$\widetilde{\Theta}_\mathrm{I\!I\!I}^{}$ 
are plotted in Fig.\ \ref{fig:3bodycorrelations}.
Among these three components,
the last one, $\widetilde{\Theta}_\mathrm{I\!I\!I}^{}$, 
which is absent for $N=2$ by definition,  
contributes to the next-to-leading-order transport coefficients  
of multilevel impurities with $N \geq 3$.    

In the limit $\epsilon_d^{}\to \infty$,  
the intra-level component $\Theta_\mathrm{I}^{}$ 
approaches the noninteracting value, while the other two components vanish    
as the impurity level becomes nearly empty, $N_d^{} \to 0$:   
\begin{align}
\!\! 
\Theta_{\mathrm{I}}^{} 
\xrightarrow{\epsilon_d^{}\to \infty\,}\, -2 ,  
\quad
\widetilde{\Theta}_{\mathrm{II}}^{} 
\xrightarrow{\epsilon_d^{}\to \infty\,}\,  0,
\quad
\widetilde{\Theta}_{\mathrm{III}}^{}  
\xrightarrow{\epsilon_d^{}\to \infty\,}\,  0.
\label{eq:Theta_I_empty_limit}
\end{align}

Figure \ref{fig:3bodycorrelations} clearly shows that, 
in the Kondo regime $\epsilon_d^{}\to -\infty$,  
all three components of the three-body correlation function 
converge to the same value for $N \geq 3$: 
 \begin{align}
\!\!\!\!
\lim_{\epsilon_d^{} \to - \infty}  \!  \Theta_\mathrm{I}^{}
  = \, - \! \lim_{\epsilon_d^{} \to - \infty} \! \widetilde{\Theta}_\mathrm{I\!I}^{}
  =  \lim_{\epsilon_d^{} \to - \infty} \! \widetilde{\Theta}_\mathrm{I\!I\!I}^{}
\ \ \,  \equiv   \   
\Theta_\mathrm{Kond}^{1/N} \,.
\label{eq:TC_universal}
\end{align}
This occurs because 
the contributions of $\chi_B^{[3]}$,   
which appear in Eqs.\ \eqref{eq:chi3_SUN_1}--\eqref{eq:chi3_SUN_3},    
dominate three-body correlations, while the other two terms,  
${\partial\chi_{\sigma\sigma}^{}}/{\partial\epsilon_d^{}}$ and 
${\partial\chi_{\sigma\sigma'}^{}}/{\partial\epsilon_d^{}}$, 
are suppressed in the Kondo regime \cite{teratani2024thermoelectric}.
The value $\Theta_\mathrm{Kond}^{1/N}$ of 
the three-body correlation in this limit can be compared 
with the one derived 
from the Bethe ansatz solution for the SU($N$) Kondo model 
 by Mora $et\ al.$ \cite{Mora_etal_2009,Mora2009} 
(see Appendix \ref{sec:Mora_formula}).
For $N=4$, our NRG result closely agrees  
with the value $\Theta_\mathrm{Kond}^{1/4} = -1.11$.  
Compared to the previous result obtained 
at finite $U$ \cite{teratani2024thermoelectric}, 
the plateau structure with this height becomes significantly clearer 
in the strong interaction limit.
Note that, for $N=2$,   
the three-body correlations vanish in SU(2) Kondo regime 
due to electron-hole symmetry.

\section{Nonlinear current through $U\to \infty$ quantum dots}

\label{sec:NRG_dI/dV_QD}

In this section, 
we examine the low-energy behavior of the  
differential conductance $dI/dV$ through SU($N$) quantum dots. 
The exact low-energy asymptotic form of the nonlinear current $I$  
has recently been derived up to terms of order $(eV)^3$ at $T=0$, 
taking into account the tunneling and bias asymmetries \cite{Tsutsumi2023},  
\begin{align}
 \!\!\!\!   \frac{dI}{dV} \, =& \  \frac{Ne^2}{h}(1-\gamma_\mathrm{dif}^2) 
    \nonumber\\
    &  \!\!\!   
\times \left[\,\sin^2\delta \,+ C_V^{(2)}\, \frac{eV}{T^*} 
\,- C_V^{(3)} \left(\frac{eV}{T^*}\right)^2 + \cdots \, \right].
\label{eq:CV_SUN}
\end{align}
The coefficient $C_V^{(2)}$ for the order $eV$ term of $dI/dV$, 
is determined by the static susceptibilities: 
\begin{align}
    C_V^{(2)} &= 
\frac{\pi}{4}
\left[\, \alpha_\mathrm{dif}^{}\, (1 - \widetilde{K}) 
\,-\, \gamma_\mathrm{dif}^{} \,\widetilde{K} \,\right] \sin 2\delta\,. 
\label{eq:CV(2)_def}
\end{align}
This coefficient $C_V^{(2)}$ linearly depends on the tunneling and bias asymmetries, 
i.e., $\gamma_\mathrm{dif}^{}$ and $\alpha_\mathrm{dif}^{}$  
Therefore, it vanishes, $C_V^{(2)} =  0$,  
when both of these asymmetries are absent, specifically for 
 $\alpha_\mathrm{dif}^{}=\gamma_\mathrm{dif}^{}=0$. 
The magnitude of $C_V^{(2)}$ also depends on 
the rescaled Wilson ratio $\widetilde{K}$ and $\sin 2 \delta$,   
which arises through the derivative of the spectral function 
 $\rho_{d\sigma}'=\sin2\delta/(4 T^*\Delta)$
defined in Eq.\ \eqref{eq:rho_d_omega_2}.   
Equation \eqref{eq:CV(2)_def} reveals that 
effects of bias asymmetry $\alpha_\mathrm{dif}^{}$ vanish
in the limit of  $\widetilde{K} \to 1$,  
where the charge fluctuations are suppressed 
due to strong electron correlations.

The order $(eV)^2$ term of $dI/dV$ depends on 
the three-body correlation functions:  
\begin{align}
    C_V^{(3)} &=\,\frac{\pi^2}{64} \,\bigl(\,W_V^{} + \Theta_V^{}\,\bigr), 
\label{eq:CV(3)_def}
\\[3mm]
    W_V^{} &\equiv\, -\cos2\delta \Biggl[\,1 + 3\,\alpha_\mathrm{dif}^2 
- 6\big(\alpha_\mathrm{dif}^2 
+ \alpha_\mathrm{dif}^{} \gamma_\mathrm{dif}^{}\big) \widetilde{K} 
\nonumber \\
    & 
+ \bigg\{ \frac{5}{N-1} + 3 \alpha_\mathrm{dif}^2 + 6 \alpha_\mathrm{dif}^{} \gamma_\mathrm{dif}^{} + \frac{3(N-2)}{N-1} \gamma_\mathrm{dif}^2 \bigg\} \widetilde{K}^2 \Biggr], 
\label{eq:WV_def}
\\
    \Theta_V^{} &\equiv\, \Bigl(1 + 3\alpha_\mathrm{dif}^2\Bigr) 
\,\Theta_\mathrm{I}^{} 
+ 3 \Bigl(1 + 3\alpha_\mathrm{dif}^2 
+ 4\alpha_\mathrm{dif}^{} \gamma_\mathrm{dif}^{} \Bigr) 
\,\widetilde{\Theta}_\mathrm{I\!I}^{} 
    \nonumber\\
    & \ \   + 6 \Bigl( \alpha_\mathrm{dif}^2 
+ 2\alpha_\mathrm{dif}^{} \gamma_\mathrm{dif}^{} 
+ \gamma_\mathrm{dif}^2\Bigr) 
\,\widetilde{\Theta}_\mathrm{I\!I\!I}^{}\, . 
\label{eq:ThetaV_def}
\end{align}
The coefficient $C_V^{(3)}$ reflects the properties of 
the low-lying excited states near the Fermi level, 
whose contributions arise through the low-energy expansion 
of $A_{\sigma}^{}(\omega)$ with respect to $\omega$ and $eV$ at $T=0$. 
This coefficient depends on the tunneling and bias asymmetries 
via the quadratic terms,     
 $\alpha_\mathrm{dif}^{2}$,  
$\alpha_\mathrm{dif}^{}\gamma_\mathrm{dif}^{}$,
 and $\gamma_\mathrm{dif}^{2}$ \cite{Tsutsumi2023}.

When both the bias and tunnel asymmetries are inverted, 
 such that 
  $(\alpha_\mathrm{dif}^{}$, $\gamma_\mathrm{dif}^{}) 
  \Rightarrow  
  (-\alpha_\mathrm{dif}^{}$, $-\gamma_\mathrm{dif}^{})$, 
 the coefficients  $C_V^{(2)}$ and $C_V^{(3)}$ 
exhibit odd and even properties, respectively, that is 
$C_V^{(2)} \Rightarrow - C_V^{(2)}$ and $C_V^{(3)} \Rightarrow C_V^{(3)}$. 

\begin{figure}[t]
\begin{tabular}{cc}
  \begin{minipage}[t]{0.50\hsize}
  \centering
    \includegraphics[keepaspectratio, width=47mm]{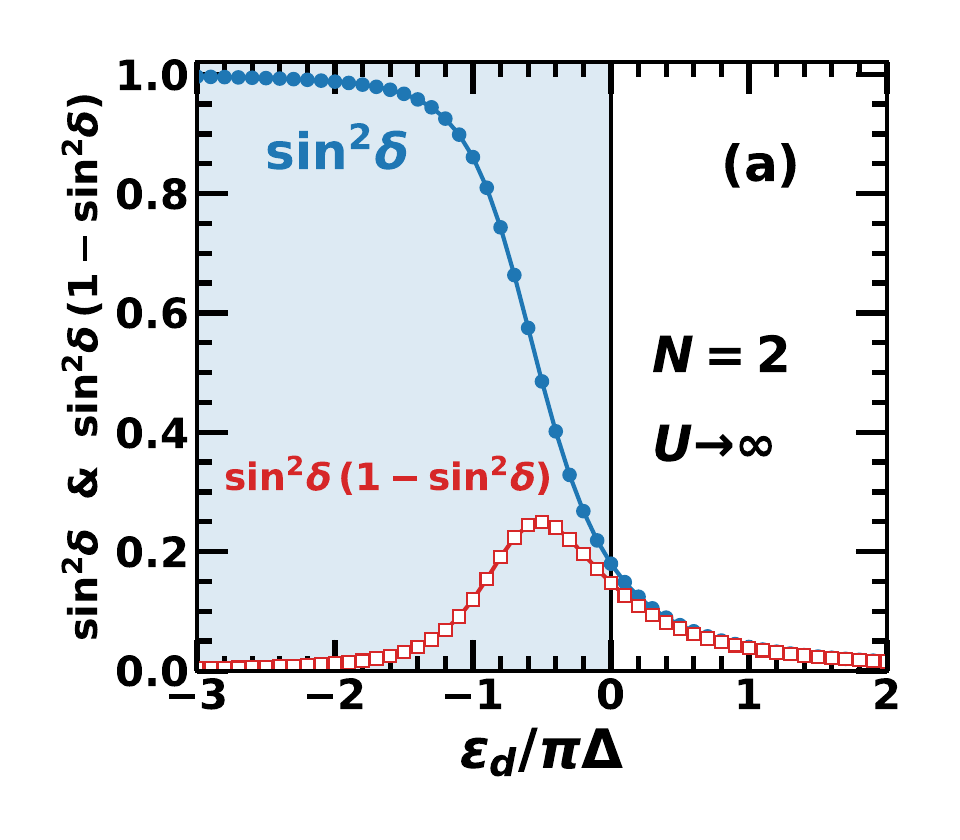}
  \label{su2_sin2delta}
  \end{minipage}
  &
  \begin{minipage}[t]{0.50\hsize}
  \centering
    \includegraphics[keepaspectratio, width=47mm]{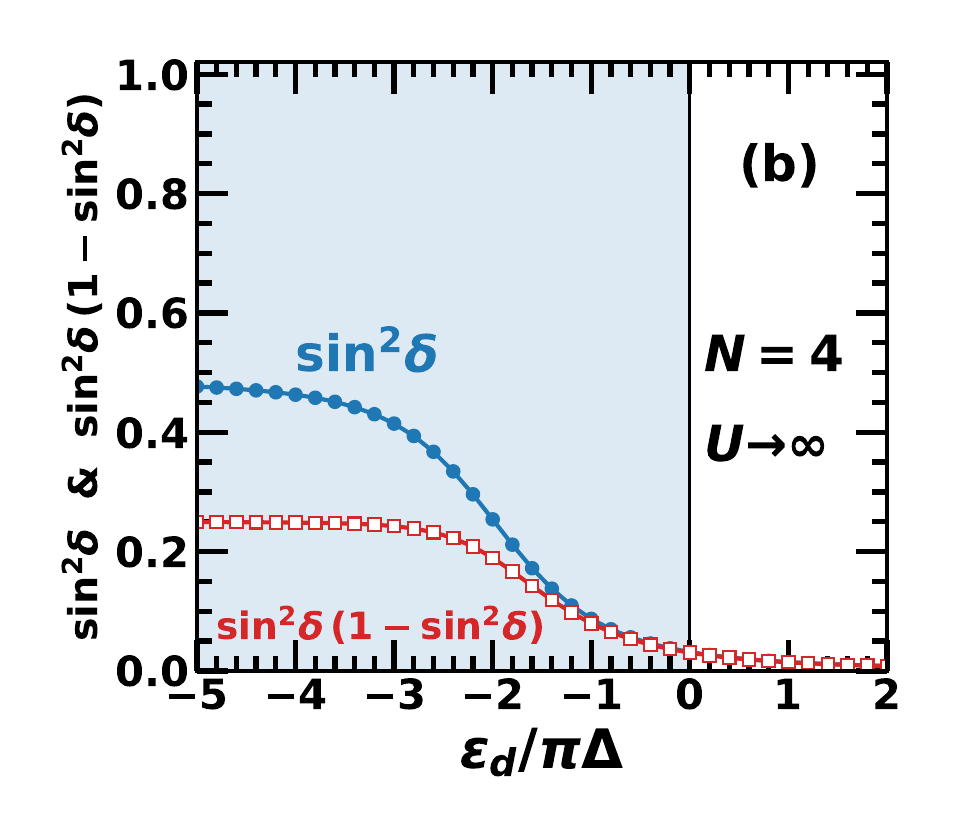}
  \label{su4_sin2delta}
  \end{minipage}
\end{tabular}
\vspace{-7mm}
\caption{Linear conductance $\sin^2\delta$ 
and noise $\sin^2\delta\,(1-\sin^2\delta)$ are 
plotted as functions of  $\epsilon_d^{}$  
in the $U \to \infty$ limit with $\pi \Delta/D = 0.01$: 
 (a) $N=2$ and (b) $N=4$. 
}
\label{fig:sin2delta}
\end{figure}

\begin{figure}[t]
\begin{tabular}{cc}
  \begin{minipage}[t]{0.50\hsize}
  \centering
    \includegraphics[keepaspectratio, width=47mm]{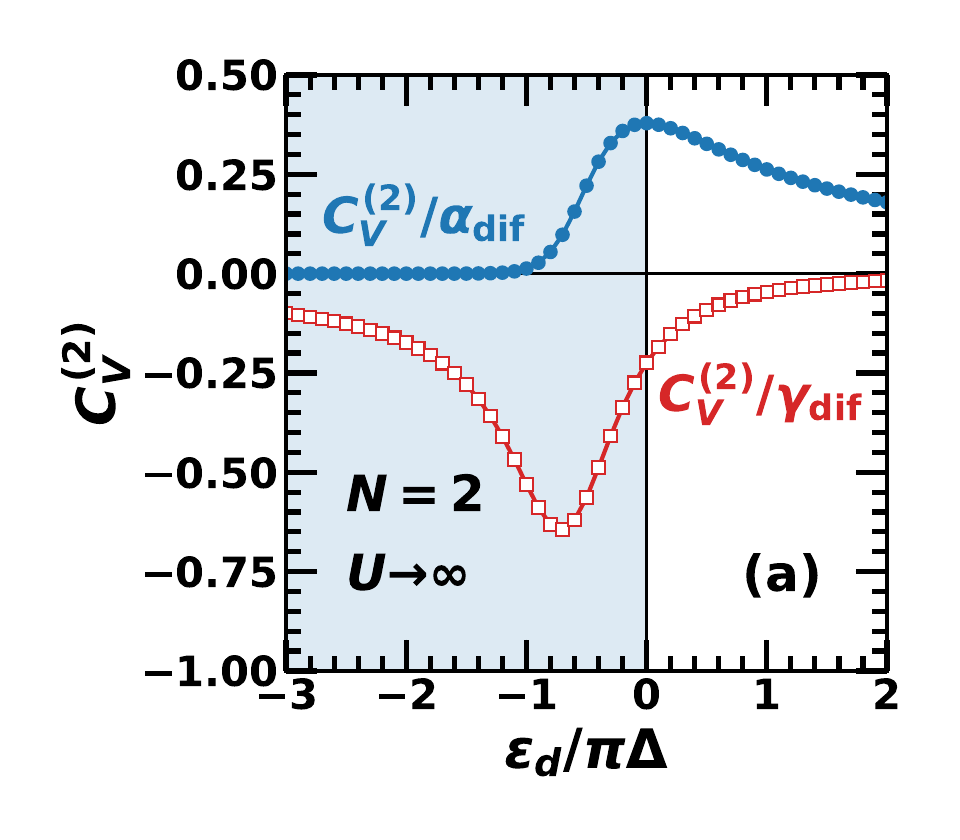}
  \label{su2_cv2}
  \end{minipage}
  &
  \begin{minipage}[t]{0.50\hsize}
  \centering
    \includegraphics[keepaspectratio, width=47mm]{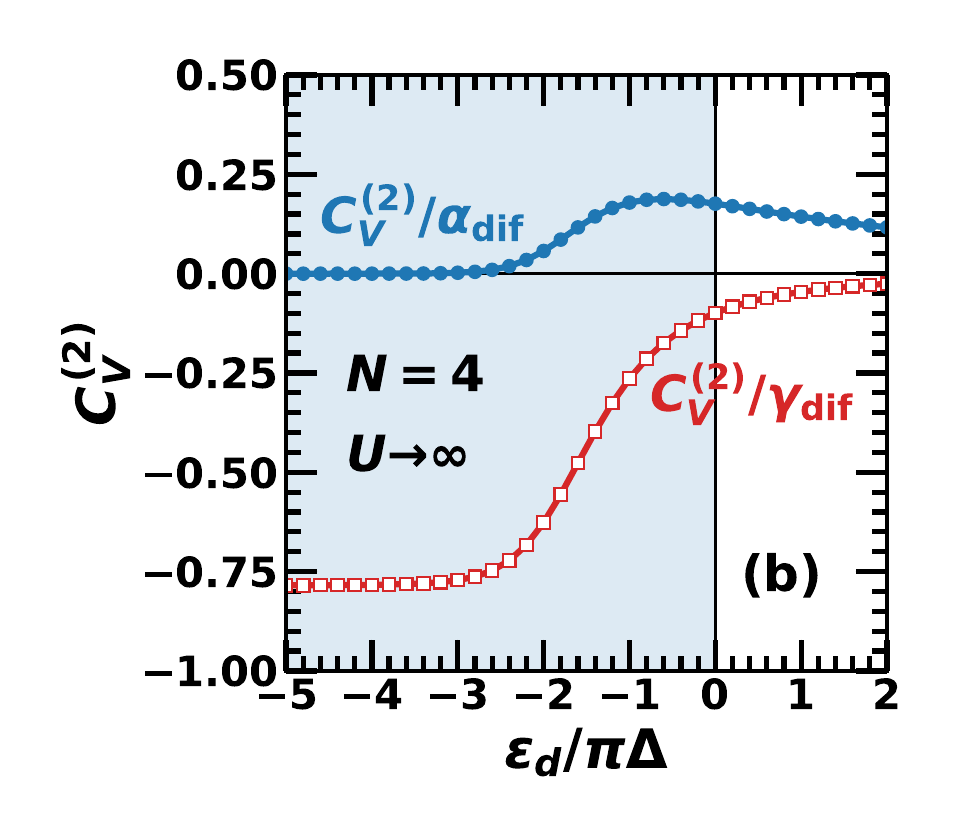}
  \label{su4_cv2}
  \end{minipage}
\end{tabular}
\vspace{-7mm}
\caption{Coefficient for the $V^2$ nonlinear current, 
defined in Eq.\ \eqref{eq:CV(2)_def},
 plotted as a function of $\epsilon_d^{}$,
 for (a) $N=2$ and (b) $N=4$. The results are scaled such that 
$C_V^{(2)}/\gamma_\mathrm{dif}^{}$ corresponds to 
the case with symmetric bias voltages ($\alpha_\mathrm{dif}^{}=0$), 
and $C_V^{(2)}/\alpha_\mathrm{dif}^{}$ 
corresponds to the case with 
symmetric tunnel couplings ($\gamma_\mathrm{dif}^{}=0$).   
}
\label{fig:cv2}
\end{figure}

\subsection{NRG results for $\sin^2 \delta$ and $C_V^{(2)}$ for $U\to \infty$}

 NRG results for 
$\sin^2 \delta$, which determines the linear conductance
 $\left. dI/dV\right|_{V=0}^{}$   
through the $U \to \infty$ Anderson impurity,   
are plotted in Fig.\ \ref{fig:sin2delta}  
for (a) $N=2$ and (b) $N=4$. 
In this strong interaction limit, 
the phase shift varies within the range $0 < \delta < \pi/N$ 
as the occupation number varies in the range $0 < N_d^{} < 1$.
Therefore,  
in the Kondo limit $\epsilon_d^{}\to -\infty$,   
the linear conductance reaches the value $\sin^2 (\pi/N)$, 
which equals $1$ for SU(2) and $1/2$ for SU(4).

The order $eV$ term of $dI/dV$,   
described in Eq.\ \eqref{eq:CV_SUN},
emerges when tunneling asymmetry  $\gamma_\mathrm{dif}^{}$, 
 bias asymmetry $\alpha_\mathrm{dif}^{}$, or both are present.  
The dependence of the coefficient $C_V^{(2)}$ on these asymmetries 
arises through the bias window, i.e., $f_L-f_R$ in  Eq.\ \eqref{eq:LandauerFM},   
and is determined by  
the spectral function $A^{(1)}(\omega)$, 
which is exact up to linear order terms in $\omega$ and $eV$,   
\begin{align}
\pi\Delta A^{(1)}(\omega)\,\equiv & \  \, 
\frac{\widetilde{\Delta}^2}
{\Bigl(\omega-\widetilde{\epsilon}_d^{(1)}\Bigr)^2+\widetilde{\Delta}^2}\,, 
\label{eq:Spectral_1}
\\[2mm] 
\widetilde{\epsilon}_d^{(1)}\,\equiv & \ \,  
\widetilde{\epsilon}_d^{}\,+\, \frac{\alpha_{\mathrm{dif}}^{}
+ \gamma_{\mathrm{dif}}^{}}{2}
\,\widetilde{K}\, eV\,.
\label{eq:EdRen_1}
\end{align}
Here,  $\widetilde{\epsilon}_d^{(1)}$ is 
the renormalized impurity level calculated up to 
terms of order $eV$, 
and $\widetilde{\epsilon}_d^{}=\widetilde{\Delta}\cot\delta$, 
with being $\widetilde{\Delta}$ the renormalized level width
 defined in Appendix \ref{sec:Ward_identity}.

The coefficient $C_V^{(2)}$ vanishes in the limit $\epsilon_d^{}\to +\infty$,   
where $\delta \to 0$: 
\begin{align}
    C_V^{(2)} \xrightarrow{\epsilon_d\to +\infty} 0\,.
\end{align}
In the opposite limit $\epsilon_d\to -\infty$, 
the FL parameters approach 
 $\delta \to \pi/N$ and $\widetilde{K} \to 1$. 
Thus, $C_V^{(2)}$ approaches   
\begin{align}
    C_V^{(2)} \,
 &\xrightarrow{\epsilon_d\to -\infty}  
-\,\frac{\pi}{4}\, \gamma_\mathrm{dif}^{}   
\, \sin \frac{2\pi}{N}
\,, 
\end{align}
which results in $0$ for SU(2) 
and $-(\pi/4)\,\gamma_\mathrm{dif}^{}$  for SU(4).  
In this limit,  $C_V^{(2)}$ becomes independent 
of the bias asymmetry $\alpha_\mathrm{dif}^{}$ 
as charge fluctuations are suppressed 
by the strong electron correlation.

Figure \ref{fig:cv2} presents the NRG results for $C_V^{(2)}$ in two different cases. 
The first case involves the rescaled coefficient 
 $C_V^{(2)}/\gamma_\mathrm{dif}^{}$, 
obtained under symmetric bias voltages ($\alpha_\mathrm{dif}^{}=0$),  
with finite tunneling asymmetries  ($\gamma_\mathrm{dif}^{} \neq 0$).
The second case involves $C_V^{(2)}/\alpha_\mathrm{dif}^{}$, obtained 
under symmetric tunnel couplings ($\gamma_\mathrm{dif}^{}=0$)    
with finite bias asymmetries ($\alpha_\mathrm{dif}^{} \neq 0$).
Note that the coefficient $C_V^{(2)}$ is proportional 
to the factor $\sin 2\delta$ that exhibits a maximum at $\delta=\pi/4$. 
This maximum occurs in the valence fluctuation region for $N=2$, 
while it occurs in the Kondo regime for $N=4$.

For SU(2) quantum dots, 
the peak of $C_V^{(2)}/\alpha_\mathrm{dif}^{}$
and the dip of $C_V^{(2)}/\gamma_\mathrm{dif}^{}$ appearing 
in Fig.\ \ref{fig:cv2} (a)  are mainly caused by 
the maximum of $\sin 2\delta$ that occurs in the valence fluctuation region. 
In contrast, for SU(4) quantum dots, both $\widetilde{K}$
and  $\sin 2\delta$ increase as $\epsilon_d^{}$ decreases. 
Consequently, the magnitude $|C_V^{(2)}/\gamma_\mathrm{dif}^{}|$ 
for the bias symmetric case increases as  $\epsilon_d^{}$ decreases  
from the valence fluctuation region toward the $1/4$-filling Kondo regime.
Physically, since  $\sin 2\delta$ reflects the behavior of  the derivative 
of the spectral function $\rho_{d\sigma}'$, 
as mentioned in the context of Eq.\ \eqref{eq:CV(2)_def}, 
the magnitude $|C_V^{(2)}/\gamma_\mathrm{dif}^{}|$ 
is enhanced in the SU(4) Kondo regime  \cite{Tsutsumi2023},
whereas it is suppressed in the SU(2) Kondo regime. 
The peak that emerges for 
$C_V^{(2)}/\alpha_\mathrm{dif}^{}$  
in the valence fluctuation region for $N=4$ is due to the behavior  
of the factor $1-\widetilde{K}$, 
which decreases as $\epsilon_d^{}$ decreases.

\begin{figure}[t]
\begin{tabular}{cc}
  \begin{minipage}[t]{0.50\hsize}
  \centering
    \includegraphics[keepaspectratio, width=47mm]{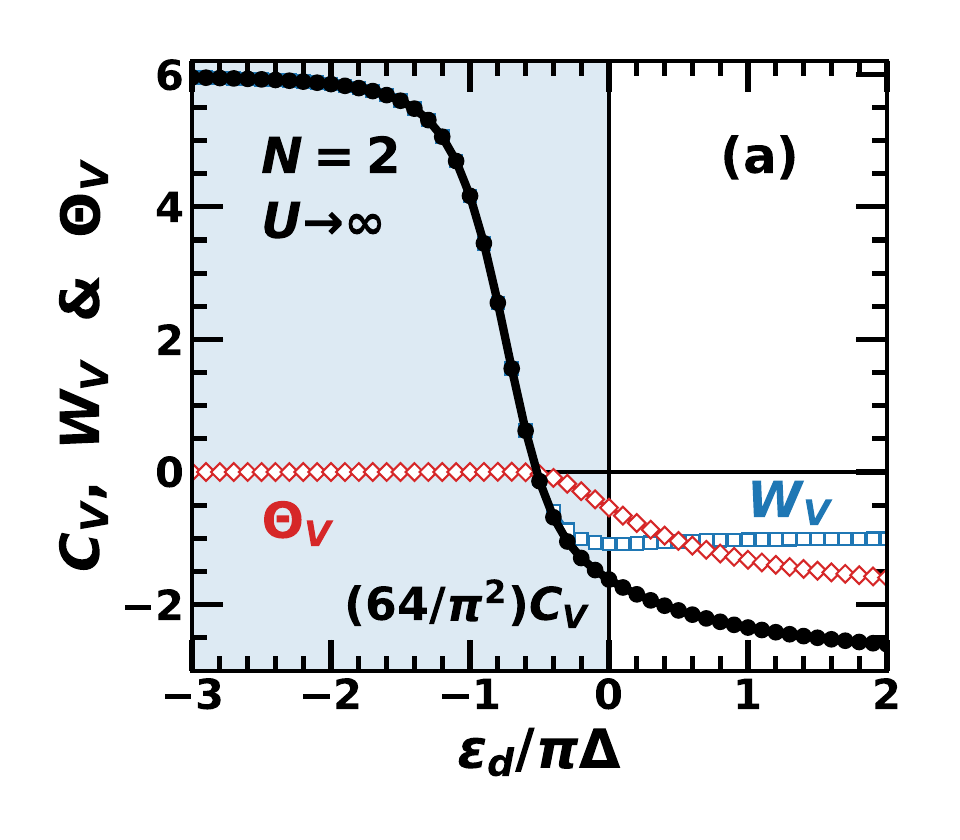}
  \label{su2_cv_wv_thetav}
  \end{minipage}
  &
  \begin{minipage}[t]{0.50\hsize}
  \centering
    \includegraphics[keepaspectratio, width=47mm]{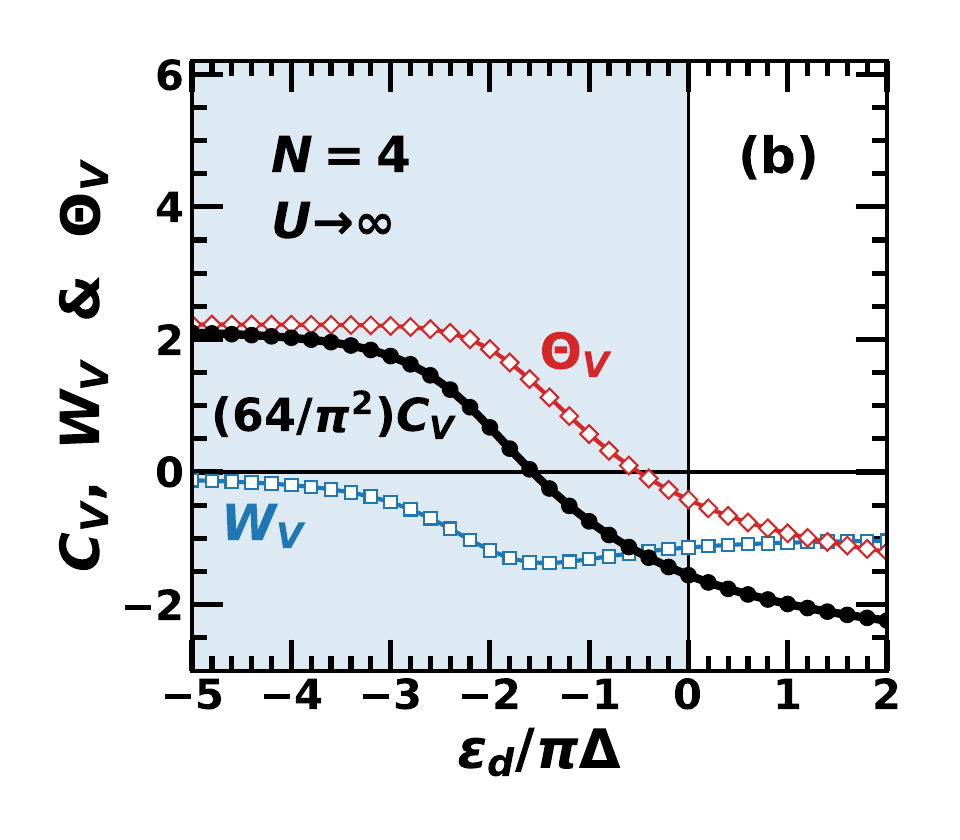}
  \label{su4_cv_wv_thetav}
  \end{minipage}
\end{tabular}
\vspace{-7mm}
\caption{Coefficients $C_V^{(3)}$,  $W_V^{}$, and $\Theta_V^{}$ 
for the $V^3$ nonlinear current, defined in Eq.\ \eqref{eq:CV(3)_def},
plotted as functions of $\epsilon_d^{}$ 
for a symmetric junction ($\gamma_\mathrm{dif}^{}
=\alpha_\mathrm{dif}^{}=0$): 
  (a) $N=2$ and (b) $N=4$.}
\label{fig:cv_wv_thetav}
\end{figure}

\subsection{$C_V^{(3)}$: Order $(eV)^2$ term of $dI/dV$}

In this subsection, we discuss the properties of   
 $C_V^{(3)}$ 
defined in Eq.\ \eqref{eq:CV(3)_def},  
as the coefficient for the order $(eV)^2$ term of  $dI/dV$.   

In the limit $\epsilon_d^{}\to +\infty$, the two-body part $W_V^{}$ 
and the three-body part $\Theta_V^{}$ 
approach their noninteracting values, which take the following forms:  
\begin{align}
    W_V^{} &\xrightarrow{\epsilon_d^{}\to +\infty} 
-\Bigl(1+3\alpha_\mathrm{dif}^{2}\Bigr)\,, \\
    \Theta_V^{} & \xrightarrow{\epsilon_d^{} \to +\infty}
 -2 \Bigl(1+3\alpha_\mathrm{dif}^{2}\Bigr)\,.
\end{align}

In the opposite limit $\epsilon_d^{}\to -\infty$,
 the phase shift and the  rescaled Wilson ratio approach $\delta \to \pi/N$ 
and  $\widetilde{K} \to 1$,  respectively, 
 and the three-body correlation functions 
converge to the same value 
$\Theta_{\mathrm{I}}^{}
= -\widetilde{\Theta}_{\mathrm{II}}^{}
= \widetilde{\Theta}_{\mathrm{III}}^{}\to \Theta_\mathrm{Kond}^{1/N}$, 
as described in Eq.\ \eqref{eq:TC_universal}.
Consequently,   $C_V^{(3)}$ 
becomes independent of the bias asymmetry $\alpha_\mathrm{dif}^{}$ 
in the Kondo regime, and in the $1/N$-filling case it takes the following form:
\begin{align}
& \!\!\!\!\!\!\!\! 
W_V^{} \xrightarrow{\epsilon_d\to -\infty}
\, - \!  \left[1+\frac{5}{N-1}
+\frac{3(N-2)}{N-1}\gamma_\mathrm{dif}^{2}\right]
\cos \frac{2\pi}{N}, 
 \label{eq:WV_Kondo}
\\
& \!\!\!\!\!\!\!
\Theta_V^{} \xrightarrow{\epsilon_d\to -\infty}
\,-2 \Bigl(1-3\gamma_\mathrm{dif}^{2}\Bigr)\,
\Theta_\mathrm{Kond}^{1/N} \,.
 \label{eq:ThetaV_Kondo}
\end{align}
Here,  $\Theta_\mathrm{Kond}^{1/2}=0$ for $N=2$ 
due to the electron-hole symmetry,  
and $\Theta_\mathrm{Kond}^{1/4}=-1.11$ for $N=4$.
The two-body part 
$W_V^{}$  depends on the factor $-\cos (2\pi/N)$,  
which reaches a maximum for $N=2$ but vanishes for $N=4$. 
Consequently, the three-body correlation $\Theta_V^{}$ dominates 
in the $1/4$-filling Kondo regime of SU(4) quantum dots, 
whereas in the SU(2) Kondo limit,
 $C_V^{(3)}$ is determined solely by 
the two-body correlation $W_V^{}$.

The NRG results for $C_V^{(3)}$, $W_V^{}$, and $\Theta_V^{}$ 
for symmetric junctions,  
 $\alpha_\mathrm{dif}^{}=0$ and $\gamma_\mathrm{dif}^{}=0$,  
are plotted vs $\epsilon_d^{}$ in Fig.\ \ref{fig:cv_wv_thetav}.
For SU(2) quantum dots, 
 $W_V^{}$ becomes the dominant term 
and approaches $W_V^{} \to 6$ in the $1/2$-filling Kondo regime
 $\epsilon_d^{}\to - \infty$,  
where $\Theta_V^{}$ vanishes due to electron-hole symmetry.  
Outside the Kondo regime, however, 
 $W_V^{}$ and $\Theta_V^{}$ become comparable 
in the valence fluctuation region, $\epsilon_d^{}/\pi\Delta \gtrsim -0.5$. 
In contrast, for SU(4) quantum dots, 
the three-body contribution dominates $C_V^{}$ 
 in the $1/4$-filling Kondo regime, where  
 it reaches the value $C_V^{} \to
-2\,\Theta_\mathrm{Kond}^{1/4} = 2.22\cdots$. 
In the valence fluctuation region, $\epsilon_d^{}/\pi\Delta \gtrsim -3$, 
both the two-body $W_V^{}$ and  three-body  $\Theta_V^{}$ parts give 
contribute comparably to $C_V^{}$.

\begin{figure}[t]
  \begin{minipage}[t]{0.70\hsize}
  \centering
  \includegraphics[keepaspectratio, width=60mm]{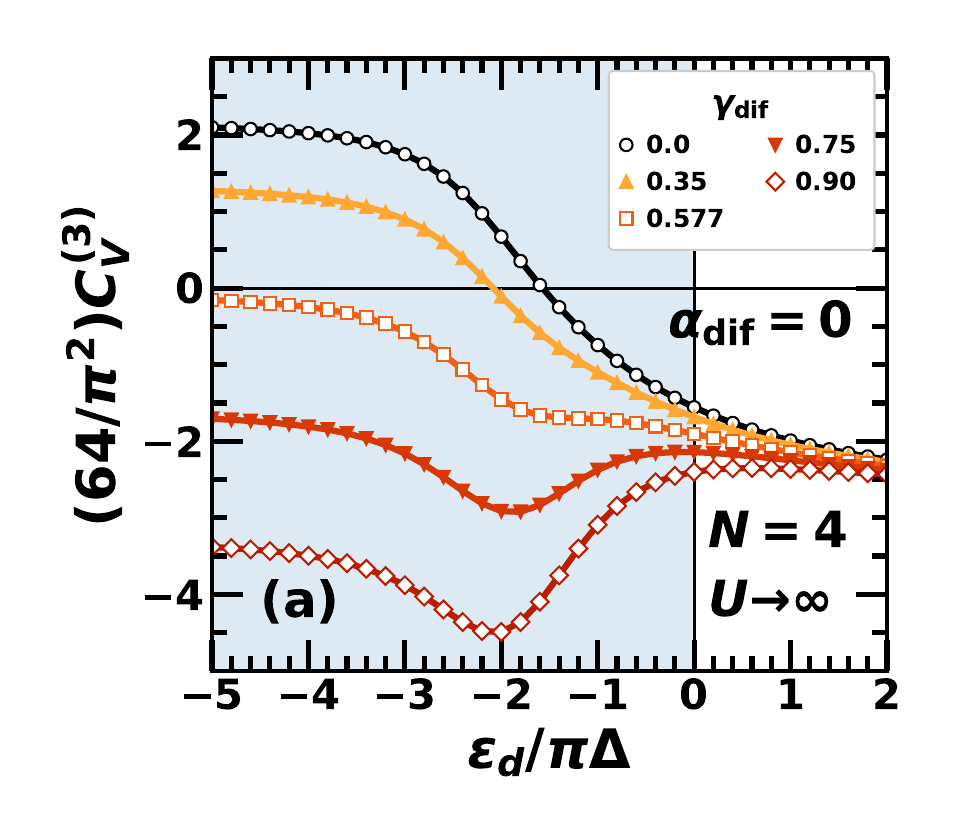}
  \label{su4_cv3_a0}
  \end{minipage} 
\begin{tabular}{cc}
\begin{minipage}[t]{0.50\hsize}
  \centering
    \includegraphics[keepaspectratio, width=47mm]{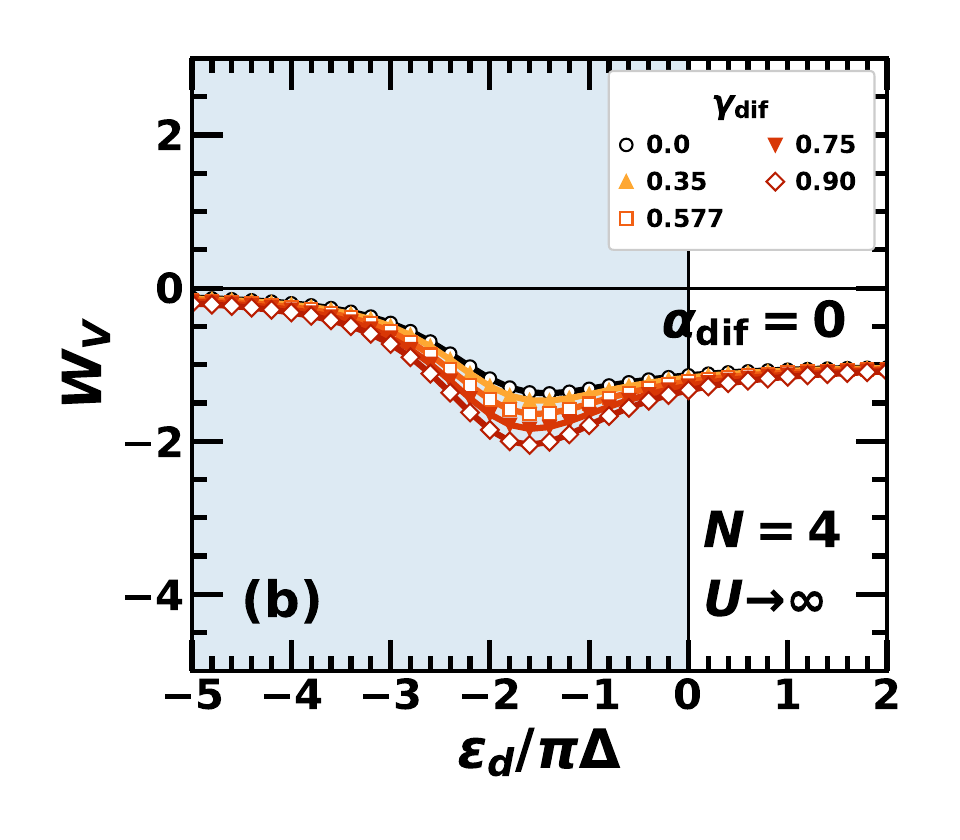}
  \label{su4_Wv_a0}
  \end{minipage} 
  &
  \begin{minipage}[t]{0.50\hsize}
  \centering
    \includegraphics[keepaspectratio, width=47mm]{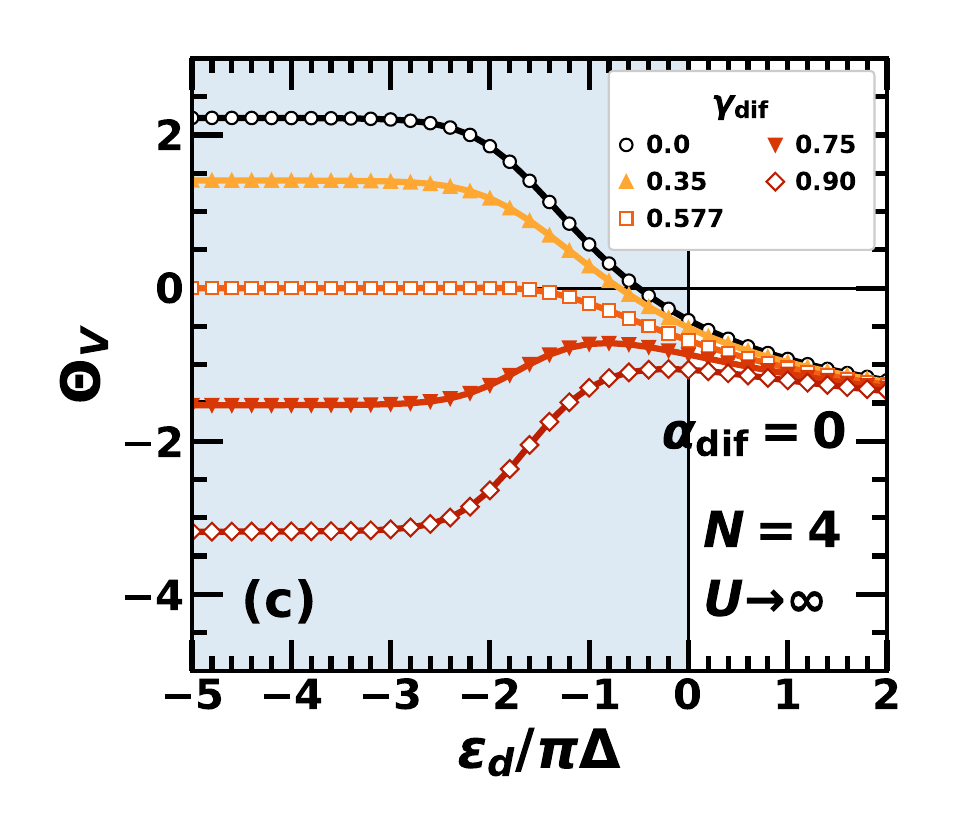}
  \label{su4_Thv_a0}
  \end{minipage} 
\end{tabular}
\vspace{-7mm}
\caption{$C_V^{(3)}$,  $W_V$, and $\Theta_V$ 
 for SU(4) QDs are plotted vs $\epsilon_d^{}$,  
with varying tunneling asymmetry $\gamma_\mathrm{dif}^{}=0 (\circ)$, 
$0.35(\blacktriangle)$, $1/\sqrt{3}(\square)$, $0.75(\blacktriangledown)$, 
and $0.9(\diamondsuit)$, while keeping the bias voltage symmetric 
($\alpha_\mathrm{dif}=0$). 
}
\label{fig:cv3_a0}
\end{figure}

\subsubsection{$C_V^{(3)}$ in bias symmetric case: $\alpha_\mathrm{dif}^{}=0$}

We now consider the behavior of  $C_V^{(3)}$ 
in the bias-symmetric case $\alpha_\mathrm{dif}^{}=0$, 
where the tunneling asymmetry affects $W_V^{}$ and $\Theta_V^{}$ 
through the order $\gamma_\mathrm{dif}^{2}$ terms  
that appears in Eqs.\ \eqref{eq:WV_def} and \eqref{eq:ThetaV_def}.
The NRG results for $C_V^{(3)}$ of SU(4) quantum dots 
at $\alpha_\mathrm{dif}^{}=0$ are 
plotted against $\epsilon_d^{}$  in Fig.\ \ref{fig:cv3_a0}, 
for several values of $\gamma_\mathrm{dif}^{}$ values. 
The coefficient  $C_V^{(3)}$ depends significantly 
on $\gamma_\mathrm{dif}^2$  
in both the valence fluctuation and Kondo regimes, where  
$\epsilon_d^{}/(\pi\Delta) \lesssim 0$. 
In particular, the two-body part $W_V^{}$ vanishes in the 
1/4-filling SU(4) Kondo regime, where $\delta=\pi/4$ 
[see Fig.\ \ref{fig:cv3_a0}(b)].   
Thus, the plateau structure 
of $C_V^{(3)}$ in the Kondo regime 
is determined solely by the three-body part  $\Theta_V^{}$, 
as shown in Fig.\ \ref{fig:cv3_a0}(c).   
The plateau height of  $\Theta_V^{}$ 
decreases as the tunneling asymmetry $\gamma_\mathrm{dif}^{}$  
increases from 0 and changes sign at $\gamma_\mathrm{dif}^{2}=1/3$. 
In the valence fluctuation regime,  
$C_V^{(3)}$ exhibits a local minimum for large tunneling asymmetries, 
 primarily due to the two-body part $W_V^{}$.    
Note that  $C_V^{(3)}$ for SU(2) quantum dots 
is unaffected by tunneling asymmetry in the bias symmetric 
case $\alpha_\mathrm{dif}^{}=0$, as 
the  $\gamma_\mathrm{dif}^2$ term  vanishes  for $N=2$.  
 Therefore, the curves shown in Fig.\ \ref{fig:cv_wv_thetav} (a) for $N=2$ 
remain unchanged 
in this case, even at finite $\gamma_\mathrm{dif}^{}$.

\begin{figure}[t]
\begin{tabular}{cc}
  \begin{minipage}[t]{0.50\hsize}
  \centering
    \includegraphics[keepaspectratio, width=47mm]{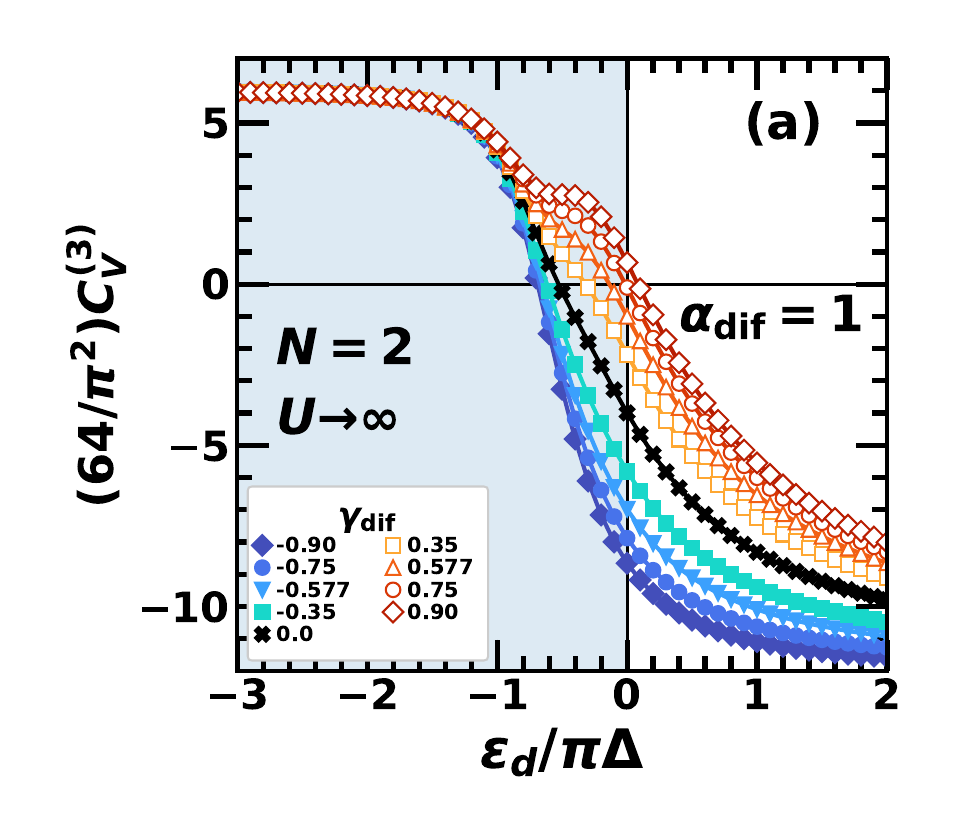}
  \label{su2_cv3_a1}
  \end{minipage}
  &
  \begin{minipage}[t]{0.50\hsize}
  \centering
    \includegraphics[keepaspectratio, width=47mm]{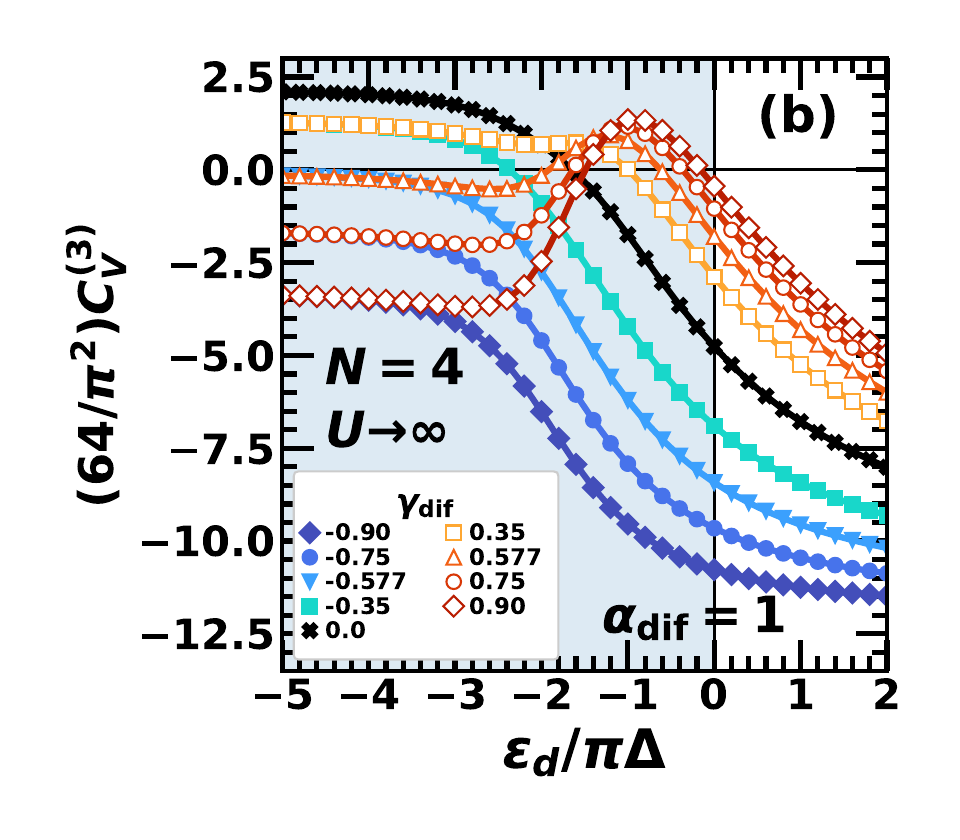}
  \label{su4_cv3_a1}
  \end{minipage}
  \\
  \begin{minipage}[t]{0.50\hsize}
  \centering
    \includegraphics[keepaspectratio, width=47mm]{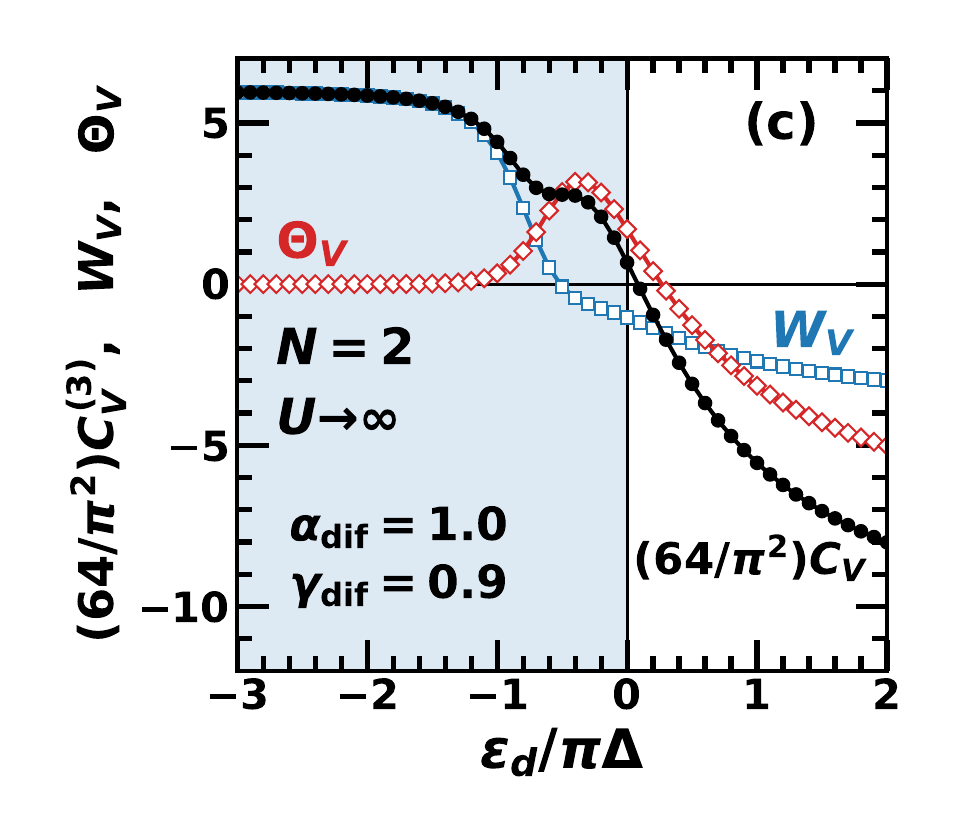}
  \label{su2_Cv3_Wv_Thv_a1_gp09}
  \end{minipage}
  &
  \begin{minipage}[t]{0.50\hsize}
  \centering
    \includegraphics[keepaspectratio, width=47mm]{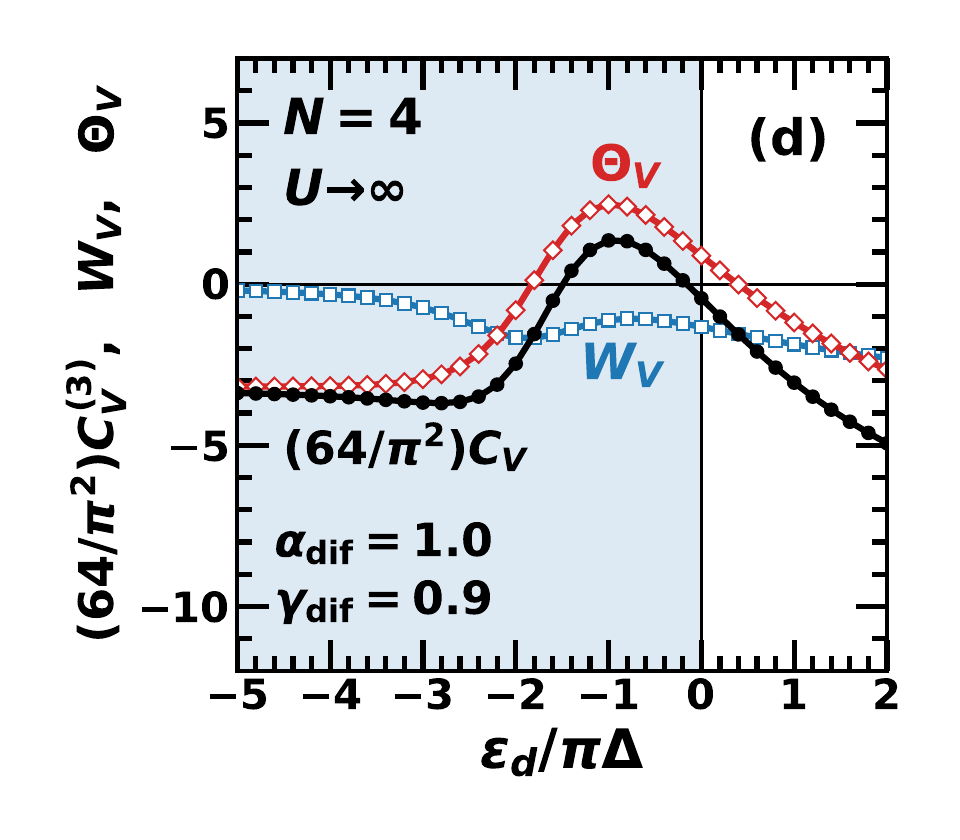}
  \label{su4_Cv3_Wv_Thv_a1_gp09}
  \end{minipage}
  \\
  \begin{minipage}[t]{0.50\hsize}
  \centering
    \includegraphics[keepaspectratio, width=47mm]{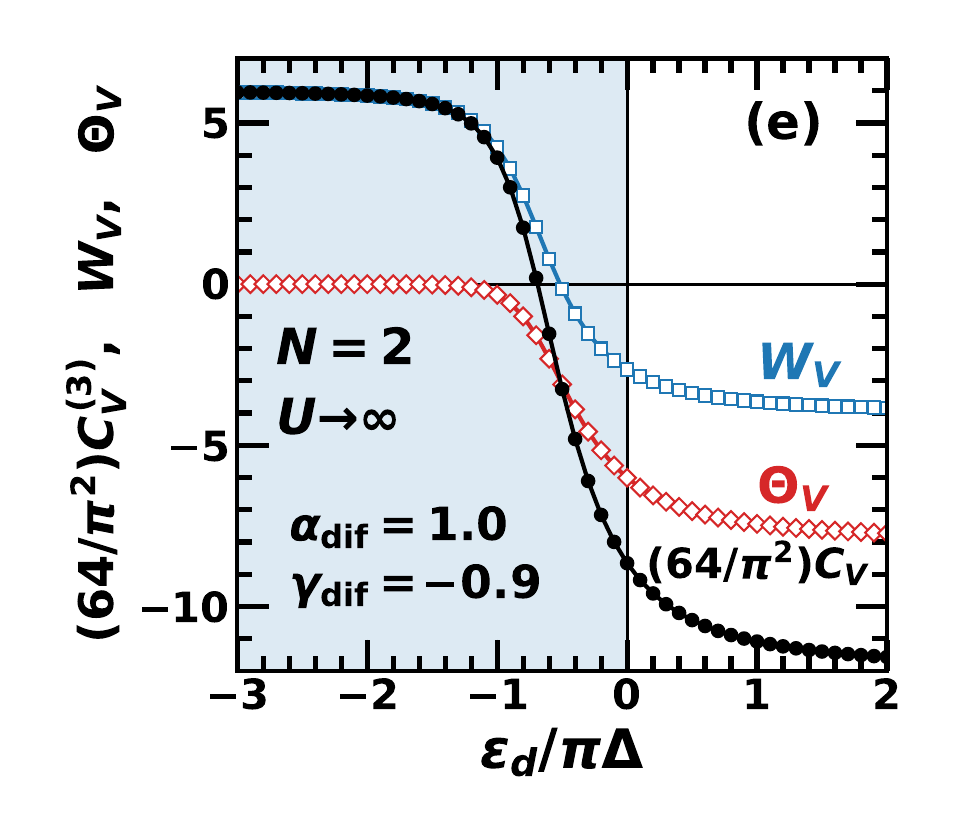}
  \label{su2_Cv3_Wv_Thv_a1_gm09}
  \end{minipage}
  &
  \begin{minipage}[t]{0.50\hsize}
  \centering
    \includegraphics[keepaspectratio, width=47mm]{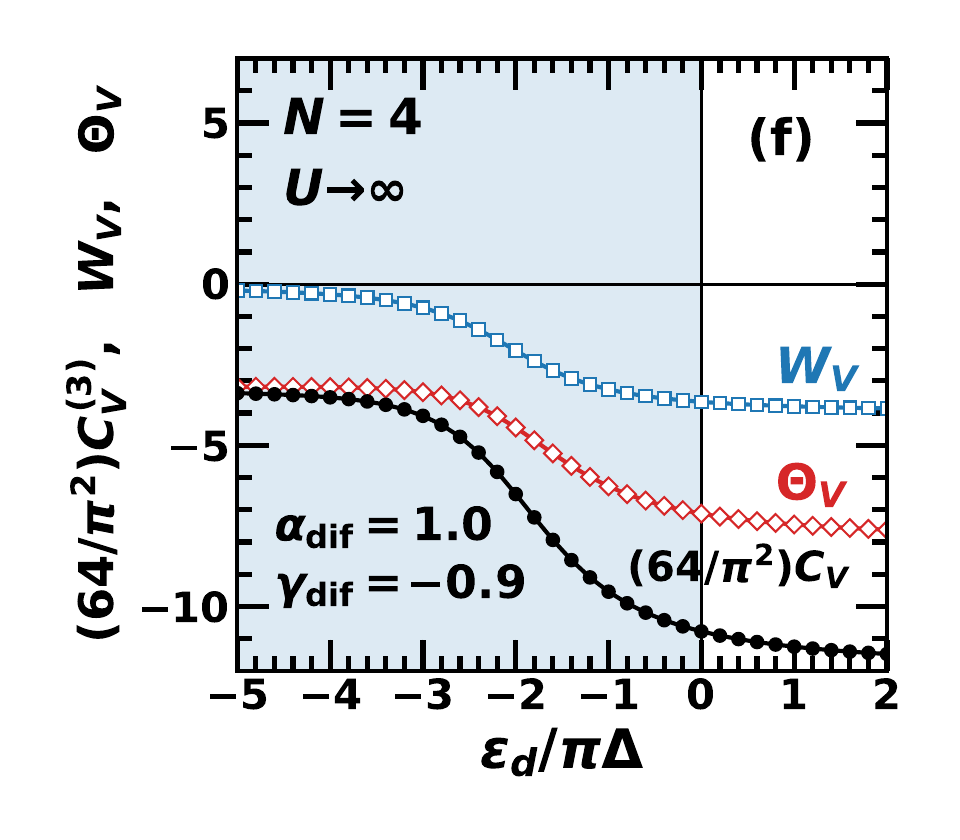}
  \label{su4_Cv3_Wv_Thv_a1_gm09}
  \end{minipage}
\end{tabular}
\vspace{-7mm}
\caption{Behavior of $C_V^{(3)}$ at large bias asymmetry 
$\alpha_\mathrm{dif}^{}=1$ 
is described as a function of  $\epsilon_d^{}$ 
for (left panel) $N=2$ and (right panel) $N=4$.  
Top panels: Tunneling asymmetry is varied as 
$\gamma_\mathrm{dif}^{}=-0.9(\blacklozenge)$, 
$-0.75(\bullet)$, 
$-1/\sqrt{3}(\blacktriangledown)$, 
$-0.35(\blacksquare)$, $0(\times)$, $0.35(\square)$, $1/\sqrt{3}(\triangle)$, 
$0.75(\circ)$, and $0.9(\diamondsuit)$. 
In addition, 
two-body part $W_V^{}$ and 
three-body part $\Theta_V^{}$ are plotted together with $C_V^{(3)}$ 
for two large opposite tunneling asymmetries: 
(middle panels)  $\gamma_\mathrm{dif}^{}=0.9$, and 
(bottom panels)  $\gamma_\mathrm{dif}^{}=-0.9$. 
}
\label{fig:cv3_a1}
\end{figure}

\subsubsection{$C_V^{(3)}$ 
under large bias asymmetry: $\alpha_\mathrm{dif}^{}=1$}

Bias and tunneling asymmetries affect the 
coefficient $C_V^{(3)}$ through the quadratic terms 
 $\alpha_\mathrm{dif}^{2}$, 
$\alpha_\mathrm{dif}^{}\gamma_\mathrm{dif}^{}$, 
 and $\gamma_\mathrm{dif}^{2}$, which appear  
in Eqs.\ \eqref{eq:WV_def} and \eqref{eq:ThetaV_def}. 
In order to clarify the effects due to bias asymmetry, 
we set the bias parameter to be $\alpha_\mathrm{dif}^{}=1$, 
representing the situation 
where  one of the two leads  is grounded, i.e., 
$\mu_R^{}=0$ and  $\mu_L^{}=eV$. 
In this case, the cross term 
$\alpha_\mathrm{dif}^{}\gamma_\mathrm{dif}^{}$ 
changes sign depending on whether   
 $\gamma_\mathrm{dif}^{}$ is positive or negative.

NRG results for $C_V^{(3)}$ 
are plotted versus $\epsilon_d^{}$ 
for several values of $\gamma_\mathrm{dif}^{}$ 
in Figs.\ \ref{fig:cv3_a1}(a) and \ref{fig:cv3_a1}(b) 
for $N=2$ and $N=4$, respectively.   
The plateau structure of $C_V^{(3)}$  
in the Kondo regime, $\epsilon_d^{} \to -\infty$, 
is determined by 
Eqs.\ \eqref{eq:WV_Kondo} and \eqref{eq:ThetaV_Kondo}; 
therefore the plateau height does not depend on 
 the bias asymmetry $\alpha_\mathrm{dif}^{}$. 
For $N=4$,  
the plateau value of $C_V^{(3)}$  decreases 
 as the tunneling asymmetry 
$\gamma_\mathrm{dif}^{2}$ increases, 
as shown in Fig.\ \ref{fig:cv3_a1}(b).  
In the valence fluctuation regime,  
contributions from the cross term $\mathrm{dif}^{}\gamma_\mathrm{dif}^{}$   
become significant. 
In particular, $C_V^{(3)}$ 
exhibits a peak near $\epsilon_d^{}/(\pi\Delta) \simeq -1$ 
for large positive $\gamma_\mathrm{dif}^{}$, 
where $\alpha_\mathrm{dif}^{}\gamma_\mathrm{dif}^{}>0$, 
and the bias and tunneling asymmetries 
cooperatively enhance charge transfer from one of the electrodes 
(the left lead in this case). 
The peak emerges more prominently for SU(4) quantum dots 
than for SU(2).
The two-body $W_V^{}$ part and three-body part $\Theta_V^{}$ 
for a large positive cross term  ($\alpha_\mathrm{dif}^{}=1.0$ 
and  $\gamma_\mathrm{dif}^{}=0.9$) are plotted
in Figs.\ \ref{fig:cv3_a1}(c) and \ref{fig:cv3_a1}(d).  
The results clearly show that 
the peak structure of $C_V^{(3)}$ is caused by 
three-body contributions  $\Theta_V^{}$.  
In contrast, for a negative cross term 
$\alpha_\mathrm{dif}^{}\gamma_\mathrm{dif}^{}<0$, 
neither  $W_V^{}$ nor  $\Theta_V^{}$, exhibits a peak
 in the valence fluctuation regime, 
and thus $C_V^{(3)}$ exhibits monotonous $\epsilon_d^{}$ dependence,  
as shown in Figs.\ \ref{fig:cv3_a1}(e) and \ref{fig:cv3_a1}(f)  
for  $\alpha_\mathrm{dif}^{}= 0$ and
  $\gamma_\mathrm{dif}^{}=-0.9$.

\section{Nonlinear noise of current through $U\to \infty$ quantum dots}
\label{sec:NRG_noise_QD}

We next consider the low-bias behavior of the current noise 
 $S^\mathrm{QD}_\mathrm{noise}$ 
for symmetric junctions, 
 $\gamma_\mathrm{dif}^{}=\alpha_\mathrm{dif}^{}=0$,  
in the SU($N$) case.  
The current-current correlation function 
defined in Eq.\ \eqref{eq:S_noise} can be expanded  
up to terms of order $|eV|^3$ at $T=0$, as 
\begin{align}
 S_\mathrm{noise}^\mathrm{QD}
 = \frac{2Ne^2|eV|}{h} \left[\sin^2\delta\,(1-\sin^2\delta) 
 + C_S^{}\bigg(\frac{eV}{T^*}\bigg)^2  \! + \cdots\right] \!. 
\label{eq:noise_SUN}
\end{align}

The first term in the bracket 
corresponds to the order $|eV|$ shot noise, which   
can also be expressed in the form  
$\sin^2\delta\,(1-\sin^2\delta)=(1-\cos 4\delta)/8$.  
Note that this term is maximized at the 1/4-filling point, where $\delta=\pi/4$,  
and the phase shift $\delta$
varies in the range $0<\delta<\pi/N$ 
 in the strong interaction limit $U\to \infty$. 
The $\epsilon_d^{}$ dependence of
 this linear noise in this case 
is shown in Fig.\ \ref{fig:sin2delta} for (a) $N=2$ and (b) $N=4$. 
For SU(2) quantum dots, 
 the linear noise exhibits a sharp peak at $1/4$ filling, 
which occurs in the valence fluctuation regime 
where the electron correlation is not significant. 
 In contrast, for  SU(4),  the maximum of linear noise  emerges 
in the $1/4$-filling Kondo regime as a wide plateau.

The coefficient $C_S^{}$ for the order $|eV|^3$ term 
can be divided into the two-body part  $W_S^{}$ 
and the three-body part $\Theta_S^{}$ 
 \cite{Oguri2022,teratani2024thermoelectric}:
\begin{align}
& \!\!\!
C_S^{} = \, \frac{\pi^2}{192}\,\bigl(W_S^{} + \Theta_S^{}\bigr), 
\\[2.8mm]
&\!\!\!\!
W_S^{} \equiv \,   \cos4\delta 
\nonumber\\
 & \ \ 
+ \left[ 4+5\cos4\delta 
+\frac{3}{2}(1-\cos4\delta)(N-2)\right]\frac{\widetilde{K}^2}{N-1}, 
\\[2.8mm]
&\!\!\! 
\Theta_S^{} \equiv \,   
-  \,\Theta_V^{}
\cos2\delta 
\ = \,   - \,\frac{\sin4\delta}{4\pi\chi_{\sigma\sigma}^2}
\left(
\chi^{[3]}_{\sigma\sigma\sigma}
+3\widetilde{\chi}^{[3]}_{\sigma\sigma'\sigma'}
\right) \!.
\label{eq:ThetaS_SUN}
\end{align}
Note that  $\Theta_S^{}$ exhibits a $\sin 4 \delta$ 
dependence because  
$\Theta_V^{}$, 
defined by  Eqs.\ \eqref{eq:Theta_I_II} and \eqref{eq:ThetaV_def}, 
has  an extra factor of $\sin2 \delta$.

These two parts, $W_S^{}$ and $\Theta_S^{}$, 
approach their noninteracting values 
in the limit of  $\epsilon_d^{}\to +\infty$:  
\begin{align}
    W_S^{} \xrightarrow{\epsilon_d^{}\to +\infty} 1\,,
    \qquad \quad 
\Theta_S^{} \xrightarrow{\epsilon_d^{}\to +\infty} 2\,.
\end{align}
In contrast, in the $1/N$-filling Kondo limit $\epsilon_d^{}\to -\infty$,
the Wilson ratio and three-body correlation functions take the strong coupling values:
  $\widetilde{K} \to 1$ 
and  $\Theta_\mathrm{I}^{} + \widetilde{\Theta}_\mathrm{I\!I}^{}\to 0$,  
and thus  
\begin{align}
    W_S^{} &\xrightarrow{\epsilon_d^{}\to -\infty} 
\frac{1}{2} \left( 3+\frac{5}{N-1}\right)
+
\frac{1}{2} \left(\frac{13}{N-1} -1\right) \cos \frac{4\pi}{N} ,
\label{eq:WS_Kondo}
    \\
    \Theta_S^{} &\xrightarrow{\epsilon_d^{}\to -\infty} \, 
2\,\Theta_\mathrm{Kond}^{1/N} \, \cos \frac{2\pi}{N}
\,.
\label{eq:ThetaS_Kondo}
\end{align}
Here, $\Theta_\mathrm{Kond}^{1/2}=0$ for $N=2$
and $\Theta_\mathrm{Kond}^{1/4}=-1.11$ for $N=4$ 
(see Appendix \ref{sec:Mora_formula}) .

\begin{figure}[t]
\begin{tabular}{cc}
  \begin{minipage}[t]{0.50\hsize}
  \centering
    \includegraphics[keepaspectratio, width=47mm]{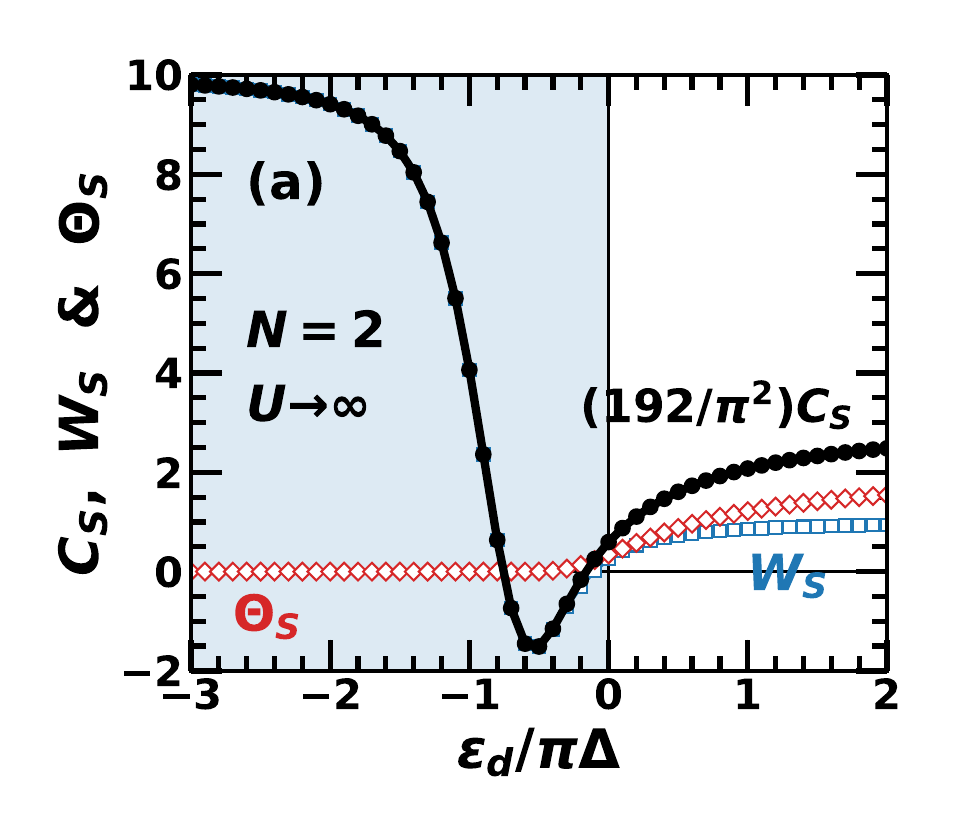}
  \label{su2_cs_ws_thetas}
  \end{minipage}
  &
  \begin{minipage}[t]{0.50\hsize}
  \centering
    \includegraphics[keepaspectratio, width=47mm]{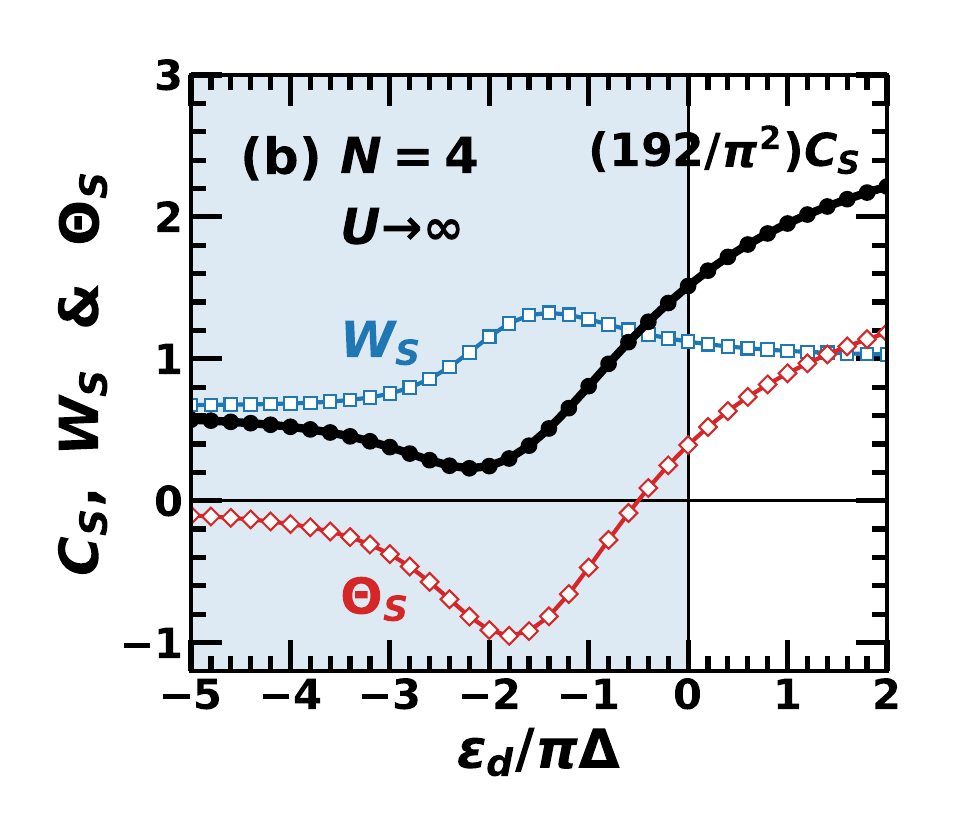}
  \label{su4_cs_ws_thetas}
  \end{minipage}
\end{tabular}
\vspace{-7mm}
\caption{Coefficients 
$C_S^{}$,  $W_S^{}$, and  $\Theta_S^{}$  for the nonlinear noise plotted  
as functions of $\epsilon_d^{}$, for (a) $N=2$, and (b) $N=4$.}
\label{fig:cs_ws_thetas}
\end{figure}

NRG results for 
 $C_S^{}$, $W_S^{}$, and $\Theta_S^{}$
 are plotted as  functions of  $\epsilon_d^{}$ in Fig.\ \ref{fig:cs_ws_thetas}.   
For SU(2) quantum dots,  
$C_S^{}$ exhibits the plateau structure in the Kondo regime, 
which in this case is determined by the two-body contributions $W_S^{}$ 
since  $\Theta_S^{}$ vanishes 
due to the electron-hole symmetry.
The coefficient $C_S^{}$ for $N=2$ also
exhibits a dip with a negative value 
at  $\epsilon_d^{}/\pi \Delta \simeq -0.5$ in the valence fluctuation regime. 
The structure of this dip is determined by the oscillatory $\cos 4\delta$ term in 
the two-body part, 
$ W_{S}^{}\xrightarrow{N=2} 
4\widetilde{K}^2 + \bigl(1+5\widetilde{K}^2\bigr) \cos 4\delta$, 
which reaches a minimum near $\delta \simeq \pi/4$.
As $\epsilon_d^{}$ rises above the Fermi level,  
i.e., for $\epsilon_d^{}/\pi \Delta \gtrsim 0$, 
both  $W_S^{}$ and  $\Theta_S^{}$ 
contribute comparably.

In the $1/4$-filling Kondo regime for SU(4) quantum dots,
 three-body contribution $\Theta_S^{}$ vanishes  
since  $\cos 2 \delta \xrightarrow{\delta \to \pi/4} 0$,     
although the correlation function $\Theta_\mathrm{Kond}^{1/4}$ itself is 
finite [see Eq.\ \eqref{eq:ThetaS_Kondo}], which  
 is one of the characteristics of the $1/4$-filling Kondo state.
Thus, the Kondo plateau for  $C_S^{}$, 
which emerges  in Fig.\ \ref{fig:cs_ws_thetas}(b) 
for $\epsilon_d^{}/(\pi\Delta) \lesssim 3.5$, 
is determined by the two-body contribution $W_S^{}$
through Eq.\ \eqref{eq:WS_Kondo}.
Note that for $U$-finite quantum dots with $N>4$ levels, 
the Kondo effects occur at a number of integer-filling points where  
the phase shift takes the values of $\delta/\pi =1/N$, $2/N$, $\ldots$, $(N-1)/N$. 
Hence, the three-body part $\Theta_S^{}$  contributes 
to the nonlinear current noise for most of the SU($N$) Kondo states, 
except for those at $1/2$- and $1/4$-fillings 
\cite{Mora_etal_2009, teratani2024thermoelectric}.

The results shown in Fig.\ \ref{fig:cs_ws_thetas}(b) 
also reveal that the coefficient $C_S^{}$ for infinite-$U$ SU(4) quantum dots 
is positive throughout the entire range of $\epsilon_d^{}$. 
In particular,  $C_S^{}$ remains positive 
at the local minimum that emerges 
in the valence fluctuation regime. 
Note that the minimum of $C_S^{}$ for finite $U$  
reaches a negative value for small interactions, 
as demonstrated in Ref.\  \onlinecite{teratani2024thermoelectric}. 
The structure of this minimum 
  is determined by comparable contributions from the two-body and three-body parts.  
As a result, the peak that appears in the two-body part near $\delta \simeq \pi/8$, 
due to the balance between the first term and the second $\cos 4\delta$ term in
\begin{align}
W_{S}^{} 
\,\xrightarrow{N=4\,}\,
\frac{1}{3} \, \left[\, 
7\widetilde{K}^2  + \bigl(3+2\widetilde{K}^2\bigr) \cos 4\delta
\, \right ] \,, 
\label{eq:WsSU4}
\end{align}
becomes dominant, 
making the next-to-leading-order term 
of current noise positive $C_S^{}>0$.

\section{Thermoelectric transport through $U\to \infty$ quantum dots}
\label{sec:NRG_thermal_QD}

Thermoelectric transport coefficients $\mathcal{L}_{n,\sigma}^\mathrm{QD}$ 
for quantum dots, defined in Eq.\  \eqref{eq:Ln_QD}, 
can be calculated using  
 the  low-energy asymptotic form of $A_{\sigma}^{}(\omega)$  
that was obtained exactly up to terms of order $\omega^2$ and $T^2$, 
with the self-energy $\Sigma_{\sigma}^{r}(\omega)$ 
described in Appendix \ref{sec:Ward_identity}  
 \cite{Tsutsumi2023, teratani2024thermoelectric}.
Specifically,  $\mathcal{L}_{0,\sigma}^\mathrm{QD}$  and 
 $\mathcal{L}_{2,\sigma}^\mathrm{QD}$  can be  
determined up to the first two terms of the expansion with respect to $T$, 
using  Eqs.\ \eqref{eq:ImSelf} and \eqref{eq:ReSelf}. 
In contrast, solely the lowest-order term can be determined 
for $\mathcal{L}_{1,\sigma}^\mathrm{QD}$, 
 which gives the leading-order term of 
the thermopower $\mathcal{S}_\mathrm{QD}^{}$:
\begin{align}
\mathcal{S}_\mathrm{QD}^{} =    
-\frac{\pi^2}{3|e|} 
\frac{\sum_{\sigma} \rho_{d\sigma}'}{\sum_{\sigma}\rho_{d\sigma}^{}}
\,  T
 +  \cdots 
   \xrightarrow{\mathrm{SU}(N)}  
-\frac{\pi^3 \cot \delta}{6\,|e|}  \frac{T}{T^*} 
 +  \cdots .
\label{eq:ThermoPowerQD_leading}
\end{align}
In this section, we discuss the behavior of the next-to-leading-order terms 
of the linear conductance $g$ and the thermal conductance  
$\kappa_\mathrm{QD}^{}$ through the SU($N$) Anderson impurity 
in the $U\to \infty$ limit.

\subsection{$C_T^{}$: Order $T^2$ term of $g$}

The linear conductance defined in Eq.\ \eqref{eq:linear_conductance_QD} 
can be expanded at low temperatures as follows: 
\begin{align}
g\,=\, 
    \frac{Ne^2}{h} \left[\,\sin^2\delta - C_T^{}\left(\frac{\pi T}{T^*}\right)^2 
 + \cdots \right]. 
\label{eq:g_expansion_SUN}
\end{align}
The coefficient $C_T^{}$ for the  $T^2$ term 
can be divided into two-body part $W_T^{}$  
and three-body part  $\Theta_T^{}$ \cite{teratani2024thermoelectric}:
\begin{align}
 C_T^{} & \,=\, \frac{\pi^2}{48}\,\bigl(W_T^{} + \Theta_T^{}\bigr)\,, 
\\[1.8mm]
  W_T^{} & \,\equiv\, -\left( 1+\frac{2\widetilde{K}^2}{N-1}\right) \cos2\delta\,, 
\\[2mm]
 \Theta_T^{} & \,\equiv\, \Theta_\mathrm{I}^{} 
+ \widetilde{\Theta}_\mathrm{I\!I}^{}
\ = \  
\frac{\sin 2\delta}{2 \pi}\,
(4T^*)^2 \,
\frac{\partial \chi_{\sigma\sigma}^{}}{\partial \epsilon_{d}^{}}
\,.
\label{eq:Th_T}
\end{align}
In the limit of  $\epsilon_d^{}\to +\infty$, the Fermi liquid parameters approach
the values $\delta \to 0$,  $\widetilde{K} \to 0$, 
$\Theta_\mathrm{I}^{} \to -2$, 
and  $\widetilde{\Theta}_\mathrm{I\!I}^{}\to 0$.
Thus,   
\begin{align}
    W_T^{} \xrightarrow{\epsilon_d^{}\to +\infty}  -1,
    \qquad \Theta_T^{} \xrightarrow{\epsilon_d^{}\to +\infty} -2\,.
\end{align}
In contrast, in the $1/N$-filling Kondo regime, $\epsilon_d^{}\to -\infty$, 
the parameters approach the values 
$\delta \to \pi/N$,  $\widetilde{K} \to 1$, 
$\Theta_\mathrm{I}^{} +\widetilde{\Theta}_\mathrm{I\!I}^{}\to 0$, 
so that  
\begin{align}
    W_T^{} &\xrightarrow{\epsilon_d^{}\to -\infty}
\, -  \left( 1+\frac{2}{N-1}\right) \cos \frac{2\pi}{N} \,, 
    \\[1.8mm]
    \Theta_T^{} &\xrightarrow{\epsilon_d^{}\to -\infty} 0 \,.
\end{align}

\begin{figure}[t]
\begin{tabular}{cc}
  \begin{minipage}[t]{0.50\hsize}
  \centering
    \includegraphics[keepaspectratio, width=47mm]{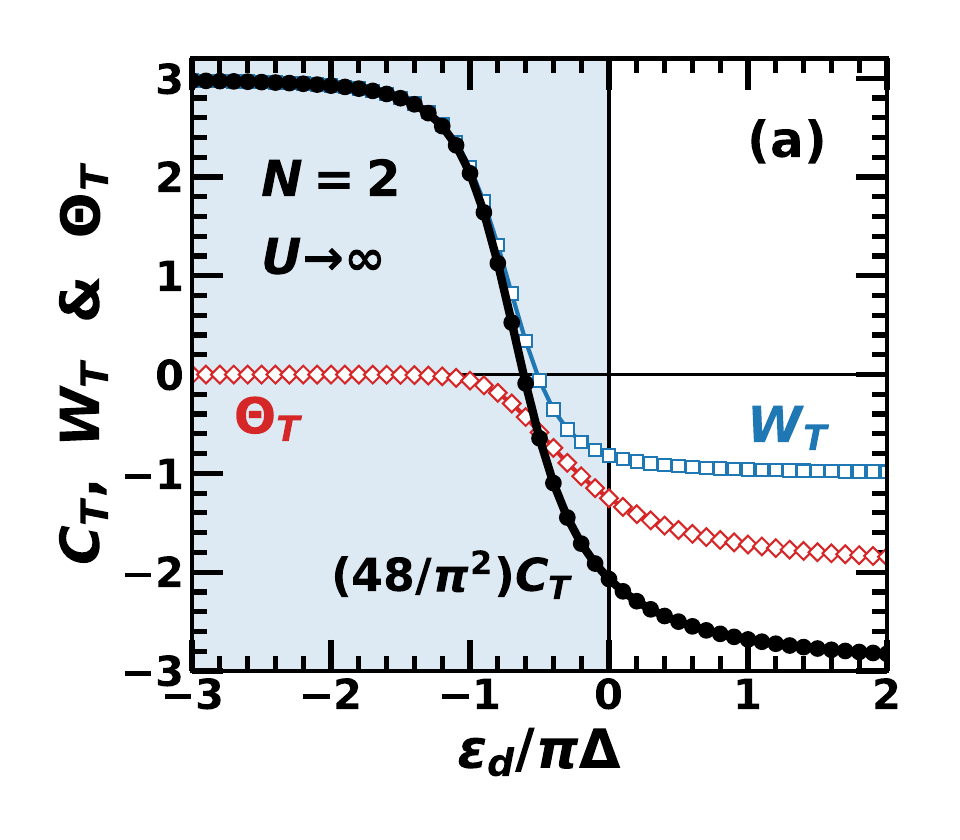}
  \label{su2_ct_wt_thetat}
  \end{minipage}
  &
  \begin{minipage}[t]{0.50\hsize}
  \centering
    \includegraphics[keepaspectratio, width=47mm]{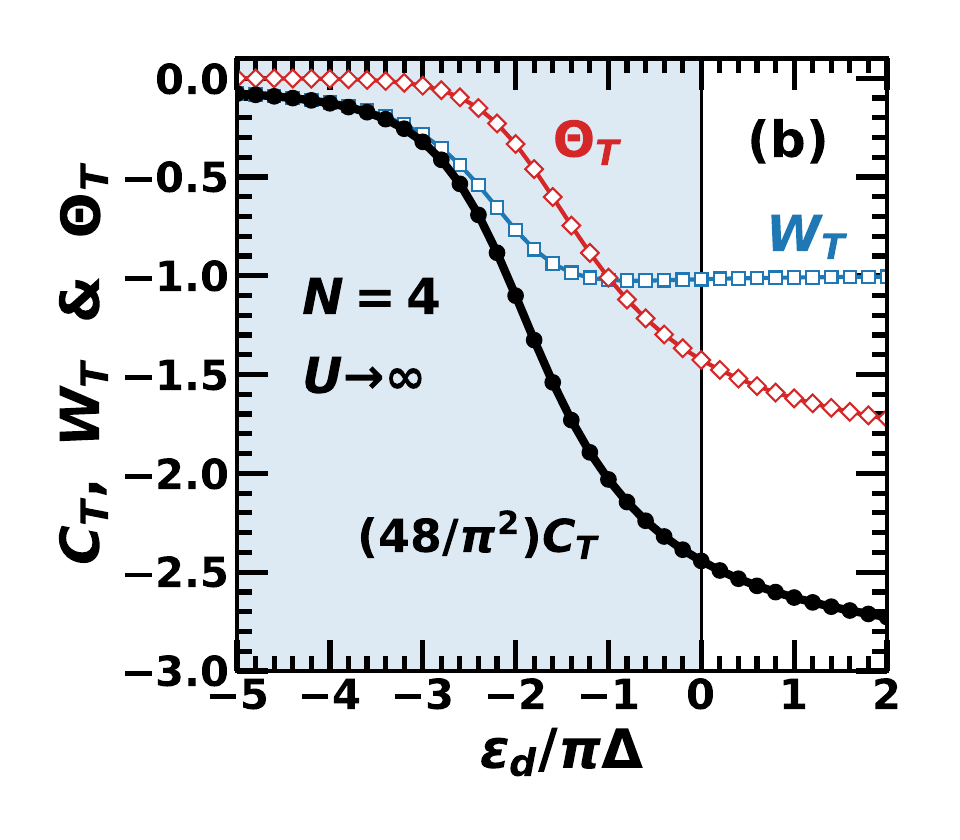}
  \label{su4_ct_wt_thetat}
  \end{minipage}
\end{tabular}
\vspace{-7mm}
\caption{Coefficients $C_T^{}$, $W_T^{}$, and $\Theta_T^{}$ 
for the $T^2$ conductance of QDs plotted as functions of 
 $\epsilon_d^{}$:
 (a) $N=2$, and (b) $N=4$.}
\label{fig:ct_wt_thetat}
\end{figure}

NRG results for 
$C_T^{}$, $W_T^{}$, and $\Theta_T^{}$ are shown in Fig.\ \ref{fig:ct_wt_thetat}.
For  SU(2) quantum dots,  the plateau structure of $C_T^{}$ 
in the half-filled 
Kondo regime is determined by the two-body part $W_T^{}$.  
Outside the plateau region,  the three-body part  $\Theta_T^{}$ 
becomes comparable to the two-body part  $W_T^{}$ as $\epsilon_d^{}$  increases. 
The coefficient  $C_T^{}$ changes sign in the middle of the valence fluctuation region 
and approaches the noninteracting value as $\epsilon_d^{}$ increases further.

The coefficient $C_T^{}$ for SU(4) quantum dots   
takes a negative value 
 in the $U \to \infty$ limit
throughout the entire range of $\epsilon_d^{}$. 
The value of $C_T^{}$ for $\epsilon_d^{}/(\pi \Delta) \lesssim -3$ 
is determined by the two-body part  $W_T^{}$,    
and it vanishes in the Kondo regime at $\epsilon_d^{} \to -\infty$ 
as the phase shift approaches $\delta \to \pi/4$. 
The three-body part $\Theta_T^{}$ for $N=4$ 
decays more rapidly 
than $W_T^{}$ as $\epsilon_d^{} \to -\infty$,  
since the derivative of the diagonal susceptibility 
that appeared in the right-hand of Eq.\ \eqref{eq:Th_T} 
becomes very small, $|\partial \chi_{\sigma\sigma}^{}/\partial \epsilon_{d}^{}| 
\ll  1/(4 T^*)^2$, due to strong electron correlations.   
This result can also be compared to the previous findings obtained 
 at finite $U$ \cite{teratani2024thermoelectric},  
which revealed that $\Theta_T^{}$ is significantly suppressed 
not only in the quarter-filling Kondo regime 
but also over a much broader range of electron filling, 
  $1\lesssim N_d^{} \lesssim N-1$ under strong interactions  
 [e.g., $U/(\pi \Delta) \gtrsim 5$ for $N=4$].

\subsection{ $C_\kappa^\mathrm{QD}$: Order $T^3$ term 
of $\kappa_\mathrm{QD}^{}$}

The low-temperature expansion of 
 thermal conductance  $\kappa_\mathrm{QD}^{}$ 
through quantum dots, 
defined in Eq.\  
 \eqref{eq:thermal_coefficients_QD},  
takes  the  following  form in the SU($N$) symmetric case:   
\begin{align}
    \kappa_\mathrm{QD}^{} \,=\, \frac{N\pi^2 \,T}{3h} 
\left[\,\sin^2\delta \,-\, 
C_\kappa^\mathrm{QD} \left(\frac{\pi T}{T^*}\right)^2 \, 
\cdots \,\right]\,.
\end{align}
The leading-order terms of the thermal conductance 
$\kappa_\mathrm{QD}^{}$ 
 and electrical conductance $g$ satisfy the Wiedemann-Franz law 
in the zero-temperature limit, yielding      
$\kappa_\mathrm{QD}^{}/(T g)
\xrightarrow{T\to 0} \pi^{2}/(3e^{2})$. 
The coefficient $C_\kappa^\mathrm{QD}$
for the next-to-leading-order term of the thermal conductance 
 also consists of the two-body $W_\kappa^\mathrm{QD}$ and 
three-body  $\Theta_\kappa^\mathrm{QD}$ parts:
\begin{align}
    C_\kappa^\mathrm{QD} &
\,= \,\frac{7\pi^2}{80}\,\bigl(\,W_\kappa^\mathrm{QD} 
+ \Theta_\kappa^\mathrm{QD}\,\bigr)\,, 
\\[2mm]
    W_\kappa^\mathrm{QD}  &\,\equiv\, 
\frac{1}{21} \left[ \,10\,-\left(11\,+\,\frac{18\widetilde{K}^2}{N-1} \right) 
\cos2\delta\,\right] , 
\\[2mm]
    \Theta_\kappa^\mathrm{QD} &\,\equiv\, \Theta_\mathrm{I}^{} 
\,+\, \frac{5}{21} \widetilde{\Theta}_\mathrm{I\!I}^{} \,.
\end{align}

In the limit of $\epsilon_d^{}\to +\infty$, where 
$N_d^{} \to 0$,  
the FL parameters approach the values 
$\delta \to 0$,  $\widetilde{K} \to 0$, 
$\Theta_\mathrm{I}^{} \to -2$, 
and  $\widetilde{\Theta}_\mathrm{I\!I}^{}\to 0$. 
Consequently, in this limit, $\Theta_\kappa^\mathrm{QD}$ dominates 
because $W_\kappa^\mathrm{QD}$ becomes very small:  
\begin{align}
    W_\kappa^\mathrm{QD} \xrightarrow{\epsilon_d^{}\to +\infty} -\frac{1}{21},
    \qquad \Theta_\kappa^\mathrm{QD} \xrightarrow{\epsilon_d^{}\to +\infty} -2.
\end{align}
In the opposite limit $\epsilon_d^{}\to -\infty$, 
the $1/N$-filling Kondo effect occurs and the FL parameters take the values 
$\delta \to \pi/N$,  $\widetilde{K} \to 1$, and 
$\Theta_\mathrm{I}^{} + \widetilde{\Theta}_\mathrm{I\!I}^{}\to 0$. 
Thus, we have 
\begin{align}
&   \!\!\!
W_\kappa^\mathrm{QD} \xrightarrow{\epsilon_d^{}\to -\infty}
\frac{1}{21} \left[
10 - \left(11 + \frac{18}{N-1} \right) \cos \frac{2\pi}{N}\, \right], 
    \\[1.8mm]
 &   \!\! 
\Theta_\kappa^\mathrm{QD} \xrightarrow{\epsilon_d^{}\to -\infty} 
\, \frac{16}{21}\,\Theta_\mathrm{Kond}^{1/N} \,. 
\end{align}
Specifically, the dimensionless three-body correlation functions 
 for $N=2$ and $N=4$ are given by 
$\Theta_\mathrm{Kond}^{1/2}=0$ 
and $\Theta_\mathrm{Kond}^{1/4}=-1.11$, respectively, 
as mentioned earlier.

\begin{figure}[t]
\begin{tabular}{cc}
  \begin{minipage}[t]{0.50\hsize}
  \centering
    \includegraphics[keepaspectratio, width=47mm]{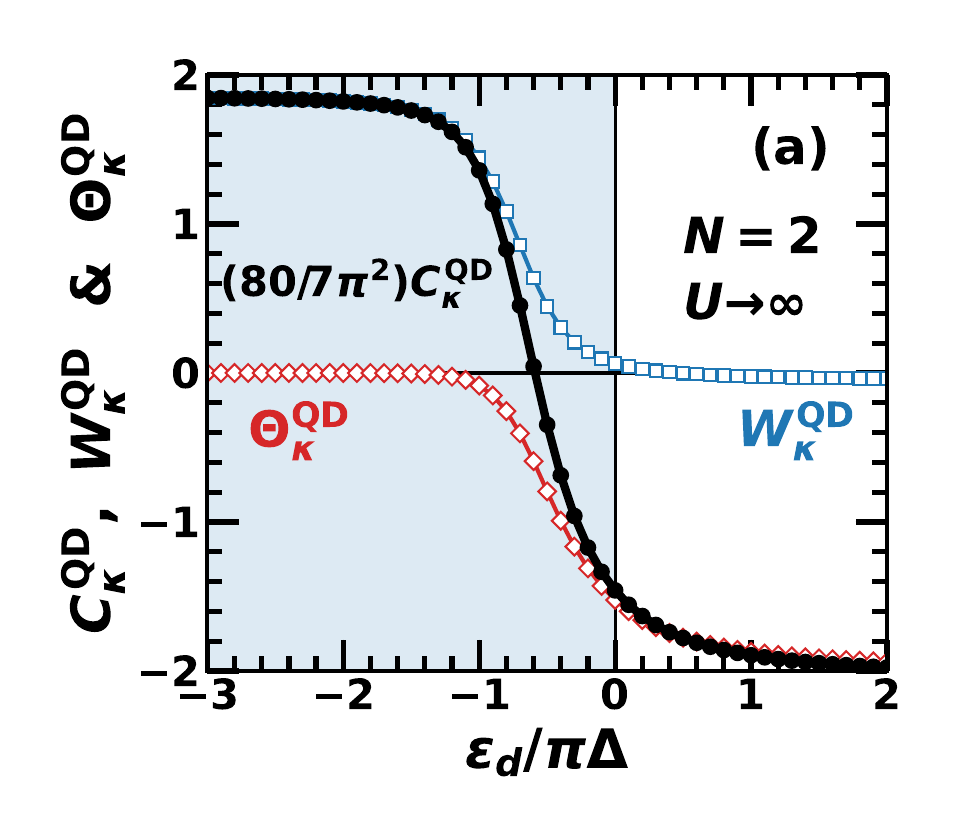}
  \label{su2_ck_QD}
  \end{minipage}
  &
  \begin{minipage}[t]{0.50\hsize}
  \centering
    \includegraphics[keepaspectratio, width=47mm]{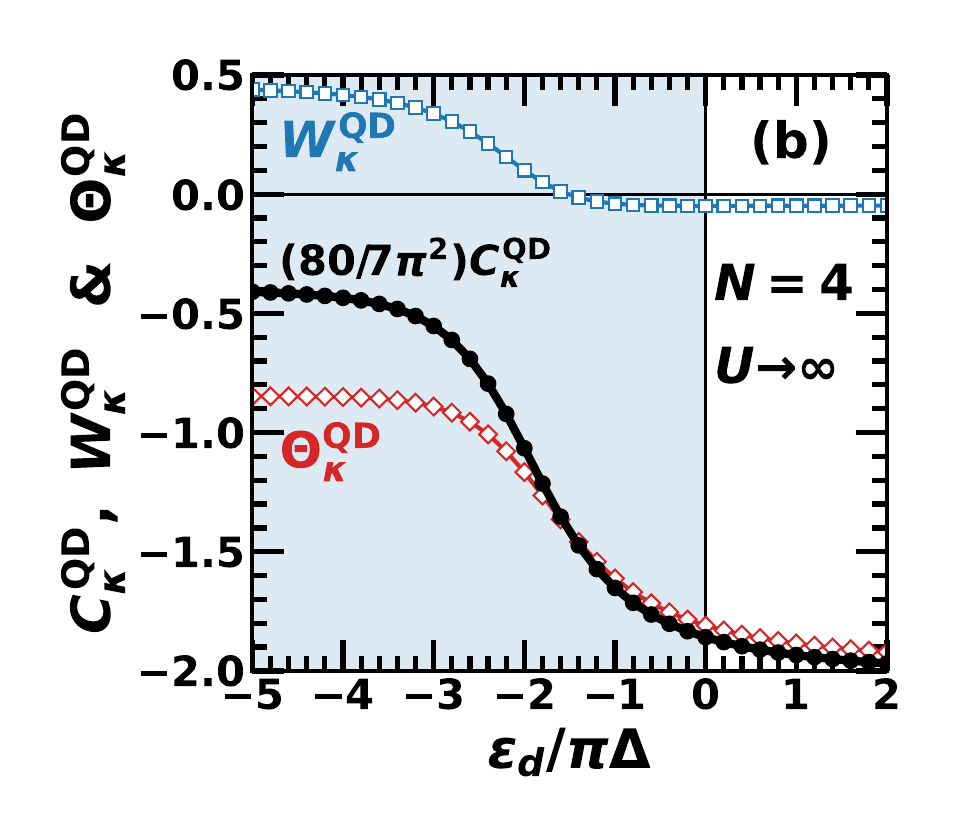}
  \label{su4_ck_QD}
  \end{minipage}
\end{tabular}
\vspace{-7mm}
\caption{Coefficients $C_\kappa^\mathrm{QD}$,  $W_\kappa^\mathrm{QD}$, 
and  $\Theta_\kappa^\mathrm{QD}$ for the $T^3$ thermal conductance of QDs 
plotted as functions 
of $\epsilon_d^{}$: (a) $N=2$,  
and  (b) $N=4$.}
\label{fig:ck_QD}
\end{figure}

NRG results for 
$C_\kappa^\mathrm{QD}$, $W_\kappa^\mathrm{QD}$, and $\Theta_\kappa^\mathrm{QD}$ are plotted as functions of $\epsilon_d^{}$ in Fig.\ \ref{fig:ck_QD}.
For SU(2) quantum dots,  
  $C_\kappa^\mathrm{QD}$ takes a positive value and 
exhibits a plateau structure 
in the half-filling Kondo regime $\epsilon_d^{}/(\pi\Delta) \lesssim -1$,  
which is determined by the two-body part $W_\kappa^\mathrm{QD}$ 
since the three-body part $\Theta_\kappa^\mathrm{QD}$ vanishes 
due to the electron-hole symmetry. 
As $\epsilon_d^{}$ decreases, the coefficient $C_\kappa^\mathrm{QD}$ 
changes sign in the valence fluctuation regime, 
taking a negative value 
that is determined 
by the three-body part $\Theta_\kappa^\mathrm{QD}$. 
In contrast,  for SU(4) quantum dots, 
the three-body part  $\Theta_\kappa^\mathrm{QD}$ dominates 
 throughout the entire range of $\epsilon_d^{}$ 
in the $U \to \infty$ limit,  
although the two-body part 
$W_\kappa^\mathrm{QD}$ also makes competitive contributions 
in the quarter-filling Kondo regime $\epsilon_d^{}/(\pi \Delta) \lesssim -2$.
As a result, the coefficient $C_\kappa^\mathrm{QD}$ 
takes a negative value 
over the whole filling range  $0 < N_d^{}<1$ for $N=4$.

\section{Thermoelectric properties of $U\to \infty$ magnetic alloys}

\label{sec:NRG_thermal_MA}

Three-body Fermi-liquid corrections also play an essential role  
in the low-energy properties of MA. 
Thermoelectric transport coefficients $\mathcal{L}_{n,\sigma}^\mathrm{MA}$ 
for magnetic alloys, defined in Eq.\  \eqref{eq:Ln_MA}, 
can be calculated in the low-temperature Fermi liquid regime 
in a manner similar to those for quantum dots 
\cite{teratani2024thermoelectric}.
For instance, the leading-order term of  
the thermopower of magnetic alloys takes the same form 
as that of QDs, given in Eq.\ \eqref{eq:ThermoPowerQD_leading},  
but with the opposite sign: 
 \begin{align}
 \mathcal{S}_\mathrm{MA}^{} 
 \,=\,  
 \frac{\pi^3 \cot \delta}{6|e|} \, \frac{T}{T^*} 
 \, + \, \cdots \,. 
 \label{eq:ThermoPowerMA_leading}
 \end{align}
In this section, 
 we discuss the behavior of  the next-to-leading-order terms 
of other transport coefficients, specifically   
the electrical resistivity $\varrho_\mathrm{MA}^{}$ 
 and  the thermal conductivity $\kappa_\mathrm{MA}^{}$ 
of magnetic alloys.

\subsection{$C_\varrho^\mathrm{MA}$: 
Order $T^2$ term of $\varrho_\mathrm{MA}^{}$}

The electrical resistivity  for magnetic alloys,  
defined as $\varrho_\mathrm{MA}^{}\equiv 
1/\sigma_\mathrm{MA}^{}$  
in Eq.\  \eqref{eq:conductivity_MA_L_0_sigma}, 
takes the following form at low temperatures in the SU($N$) case: 
\begin{align}
\varrho_\mathrm{MA}^{} \,=\, 
\frac{1}{\sigma_\mathrm{MA}^\mathrm{unit}} 
\,\left[\, \sin^2\delta 
\,-\, C_\varrho^\mathrm{MA} \bigg(\frac{\pi T}{T^*}\bigg)^2 
\,+\, \cdots \, \right] . 
\end{align}
Here, $\sigma_\mathrm{MA}^\mathrm{unit}$ 
is the unitary-limit value of electric conductivity.
The coefficient $C_\varrho^\mathrm{MA}$  for the order $T^2$ term 
consists of two-body  $W_\varrho^\mathrm{MA}$ 
and three-body $\Theta_\varrho^\mathrm{MA}$ parts, 
\begin{align}
    C_\varrho^\mathrm{MA} &\,=\, \frac{\pi^2}{48}\,\bigl(W_\varrho^\mathrm{MA} + \Theta_\varrho^\mathrm{MA}\bigr) \,, 
\\
    W_\varrho^\mathrm{MA} &\,\equiv\, 2 
              + \left( 1 \,-\, \frac{2\widetilde{K}^2}{N-1} \right) \cos2\delta\,, 
\label{eq:W_rho_MA}
\\[1.8mm]
    \Theta_\varrho^\mathrm{MA} &\,\equiv\, \Theta_\mathrm{I}^{}
 + \widetilde{\Theta}_\mathrm{I\!I}^{} 
\ = \  (4T^*)^2 \,
\frac{\sin 2\delta}{2 \pi}\,
\frac{\partial \chi_{\sigma\sigma}^{}}{\partial \epsilon_{d}^{}}
\label{eq:Theta_rho_MA}
\,.
\end{align}
Note that 
$\Theta_\varrho^\mathrm{MA} = \Theta_T^{}$, 
where $\Theta_T^{}$ is
the three-body part of $C_T^{}$ for quantum dots 
given in Eq.\ \eqref{eq:Th_T}.

In the limit of  $\epsilon_d^{}\to +\infty$, the FL parameters approach 
the values 
$\delta \to 0$,  $\widetilde{K} \to 0$, 
$\Theta_\mathrm{I}^{} \to -2$, 
and  $\widetilde{\Theta}_\mathrm{I\!I}^{}\to 0$.
Therefore,   
\begin{align}
    W_\varrho^\mathrm{MA} \xrightarrow{\epsilon_d^{}\to +\infty}\, 3\,,
  \qquad \Theta_\varrho^\mathrm{MA} \xrightarrow{\epsilon_d^{}\to +\infty}\,
 -2\,.
\end{align}
 In the opposite  limit $\epsilon_d^{}\to -\infty$, 
the $1/N$-filling Kondo effect occurs, and the FL parameters take values 
$\delta \to \pi/N$,  $\widetilde{K} \to 1$, and 
$\Theta_\mathrm{I}^{} + \widetilde{\Theta}_\mathrm{I\!I}^{}\to 0$. 
Thus,  
\begin{align}
    W_\varrho^\mathrm{MA} &\xrightarrow{\epsilon_d^{}\to -\infty} 
\, 2 +  \left( 1-\frac{2}{N-1}\right) \cos \frac{2\pi}{N} \,,
    \\[2mm]
    \Theta_\varrho^\mathrm{MA} &\xrightarrow{\epsilon_d^{}\to -\infty}\, 0\,. 
\end{align}

\begin{figure}[t]
\begin{tabular}{cc}
  \begin{minipage}[t]{0.50\hsize}
  \centering
    \includegraphics[keepaspectratio, width=47mm]{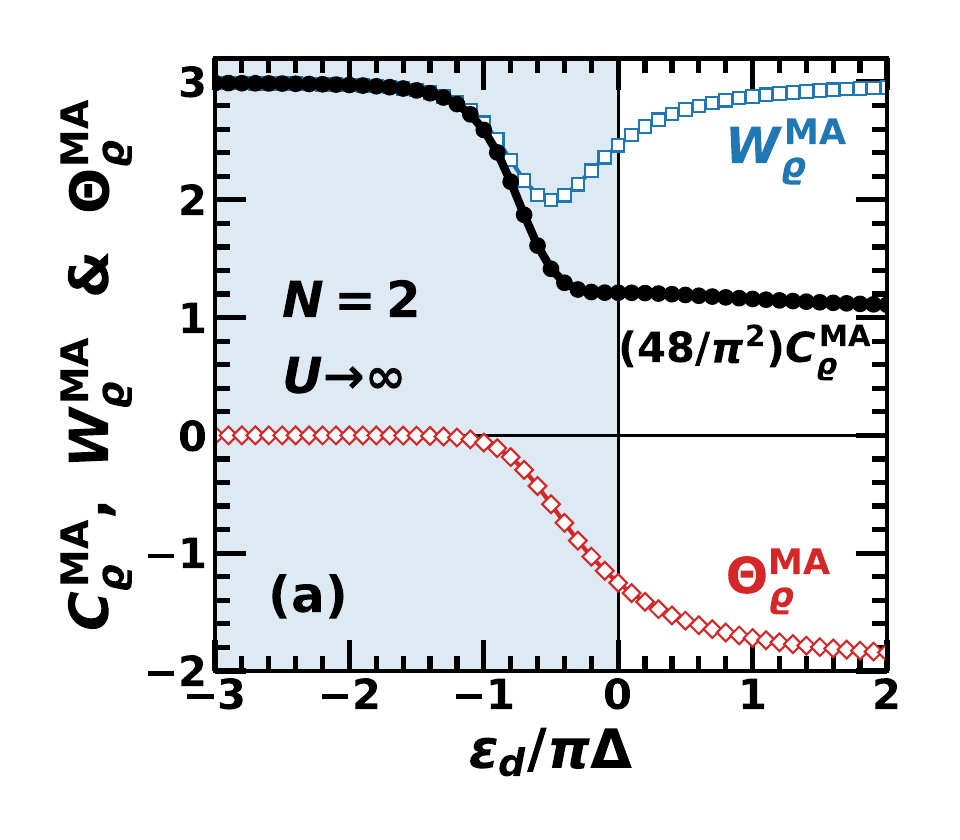}
  \label{su2_cr_MA}
  \end{minipage}
  &
  \begin{minipage}[t]{0.50\hsize}
  \centering
    \includegraphics[keepaspectratio, width=47mm]{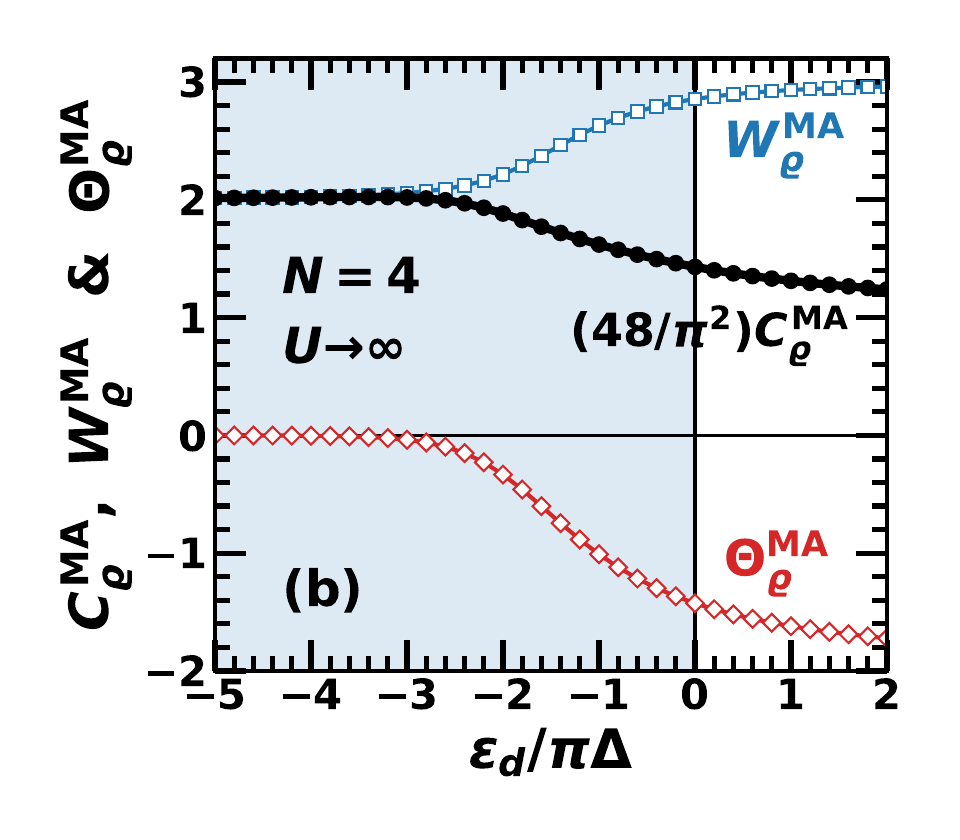}
  \label{su4_cr_MA}
  \end{minipage}
\end{tabular}
\vspace{-7mm}
\caption{Coefficients $C_\varrho^\mathrm{MA}$,  $W_\varrho^\mathrm{MA}$, 
and  $\Theta_\varrho^\mathrm{MA}$ for the $T^2$ resistivity of MAs 
plotted as functions of  $\epsilon_d^{}$:  
(a) $N=2$, and (b) $N=4$.}
\label{fig:cr_MA}
\end{figure}

NRG results for $C_\varrho^\mathrm{MA}$, $W_\varrho^\mathrm{MA}$, 
and $\Theta_\varrho^\mathrm{MA}$ in the $U\to \infty$ limit  
are plotted in Fig.\ \ref{fig:cr_MA}.  
In contrast to the coefficient $C_T^{}$ for the $T^2$ conductance of quantum dots, 
which can change sign, 
$C_\varrho^\mathrm{MA}$ remains positive 
 throughout the entire range of $\epsilon_d^{}$, 
for both $N=2$ and $N=4$. 
In the SU(2) case, it takes the value  
 $(48/\pi^2) C_\varrho^\mathrm{MA} \xrightarrow{\epsilon_d^{}\to -\infty} 3$  
 in the half-filling Kondo regime, at which the two-body part  
 $W_\varrho^\mathrm{MA}$ dominates since  the three-body part 
$\Theta_\varrho^\mathrm{MA}$ vanishes 
due to electron-hole symmetry. 
Note that the prefactor $(1-\frac{2 \widetilde{K}^2}{N-1})$ 
for the  $\cos 2\delta$ term in $W_\varrho^\mathrm{MA}$, 
as described in Eq.\ \eqref{eq:W_rho_MA},  
 changes sign in the case of $N=2$ at the point where 
the Wilson ratio takes the value $\widetilde{K}=1/\sqrt{2}$,
 while the sign remains positive for $N \geq 4$. 
This sign change is due to the wide variation 
of $\widetilde{K}^2/(N-1)$ for $N=2$.
Consequently, in the SU(2) case,  
 $W_\varrho^\mathrm{MA}$ 
exhibits a local minimum at $\epsilon_d^{}/(\pi\Delta ) \simeq -0.5$ 
 in Fig.\ \ref{fig:cr_MA}(a),  
resulting in a steeper variation of $(48/\pi^2) C_\varrho^\mathrm{MA}$ 
in the valence fluctuation region 
compared to $\Theta_\varrho^\mathrm{MA}$.  
In contrast,  in the SU(4) case,  $W_\varrho^\mathrm{MA}$ 
does not exhibit a minimum.
The coefficient for $N=4$ takes the value  
 $(48/\pi^2) C_\varrho^\mathrm{MA} \to 2$  in the quarter-filling Kondo regime, 
which is  determined by the two-body part  $W_\varrho^\mathrm{MA}$. 
The three-body contribution  
 $\Theta_\varrho^\mathrm{MA}$ vanishes in this region 
because the derivative  
$(4T^*)^2 |\partial \chi_{\sigma\sigma}^{}/\partial \epsilon_d^{}|$ is 
significantly suppressed due to strong electron correlations.

\begin{figure}[t]
\begin{tabular}{cc}
  \begin{minipage}[t]{0.50\hsize}
  \centering
    \includegraphics[keepaspectratio, width=47mm]{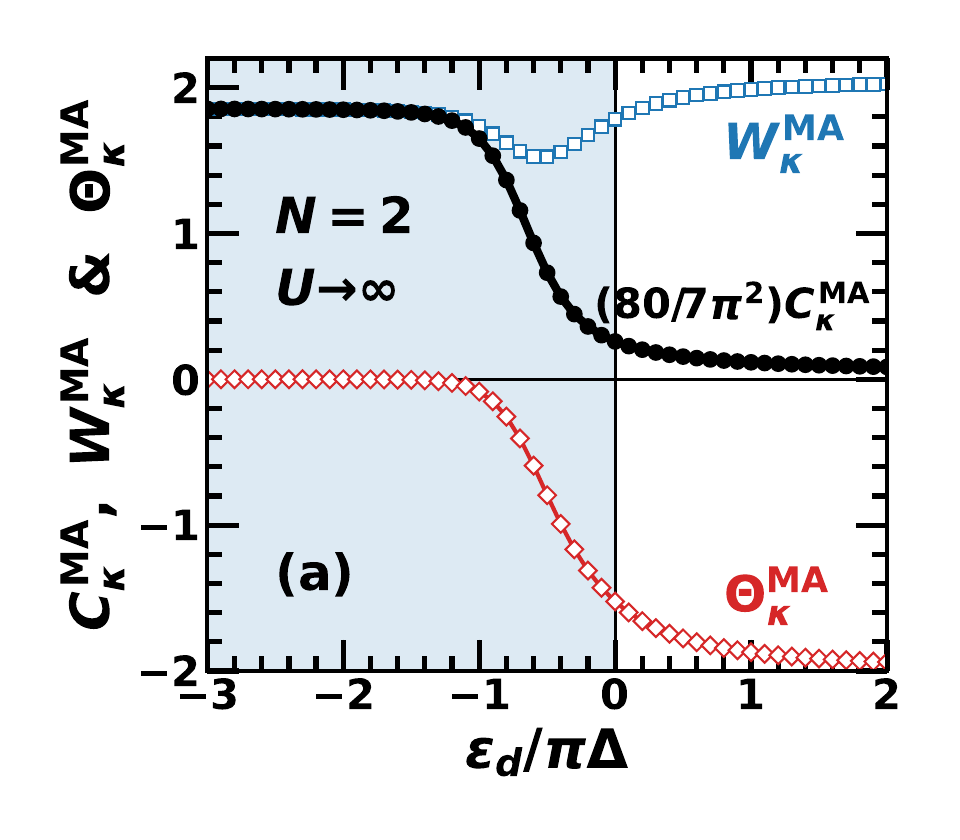}
  \label{su2_ck_wk_thetat}
  \end{minipage}
  &
  \begin{minipage}[t]{0.50\hsize}
  \centering
   \includegraphics[keepaspectratio, width=47mm]{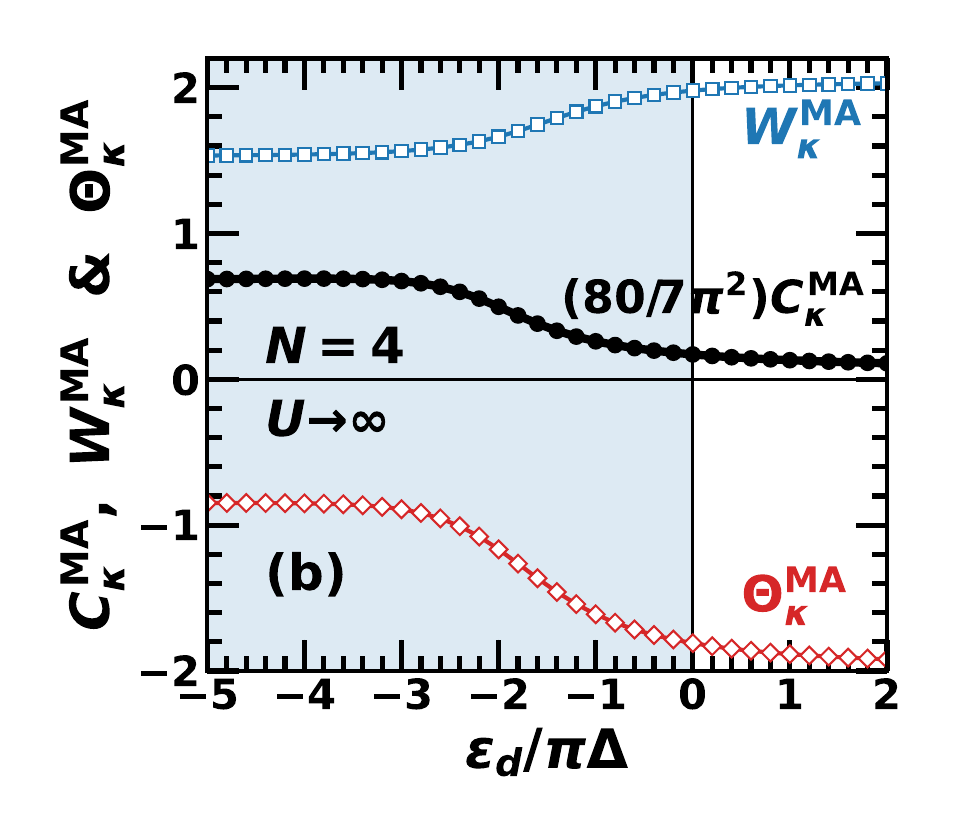}
  \label{su4_ck_MA}
  \end{minipage}
\end{tabular}
\vspace{-7mm}
\caption{Coefficients $C_\kappa^\mathrm{MA}$,  $W_\kappa^\mathrm{MA}$, 
and  $\Theta_\kappa^\mathrm{MA}$ 
for the $T^3$ thermal conductivity of MAs 
 plotted as functions of $\epsilon_d^{}$:  (a) $N=2$, and (b) $N=4$.}
\label{fig:ck_MA}
\end{figure}

\subsection{$C_\kappa^\mathrm{MA}$: Order $T^3$ term of 
$\kappa_\mathrm{MA}^{}$}

The low-temperature expansion of 
the thermal resistivity $1/\kappa_\mathrm{MA}^{}$  
of magnetic alloys, 
derived from Eq.\ \eqref{eq:kappa_MA_L0_L1_L2},  
takes the following form for SU($N$) Anderson impurity: 
\begin{align}
\ \!\!\!\!\! 
    \frac{1}{\kappa_\mathrm{MA}^{}}
   \, =\, \frac{3e^2}{\pi^2\sigma_\mathrm{MA}^\mathrm{unit}} 
\frac{1}{T} \left[\sin^2\delta 
- C_\kappa^\mathrm{MA} \left(\frac{\pi T}{T^*}\right)^2 
+ \cdots \right] .
\end{align}
In the  $T \to 0$ limit, the leading-order terms of 
electrical and thermal conductivities 
satisfy the Wiedemann-Franz law:  
$\kappa_\mathrm{MA}^{}/(T \sigma_\mathrm{MA}^{})
\to  \pi^{2}/(3e^{2})$.
The coefficient $C_\kappa^\mathrm{MA}$ for 
the next-to-leading-order term of the thermal conductivity  
can be divided into two-body $W_\kappa^\mathrm{MA}$ 
and three-body $\Theta_\kappa^\mathrm{MA}$ parts:
\begin{align}
    C_\kappa^\mathrm{MA} &\,=\, 
\frac{7\pi^2}{80}\,\bigl(W_\kappa^\mathrm{MA} 
+ \Theta_\kappa^\mathrm{MA}\bigr)\,, 
\\
    W_\kappa^\mathrm{MA} &\,\equiv \, 
\frac{1}{21}
\left[\,32 \, +  \left( 11 \,-\,  \frac{18\widetilde{K}^2}{N-1} \right) 
 \cos2\delta \,\right] , 
\label{eq:W_kappa_MA}
\\[2mm]
    \Theta_\kappa^\mathrm{MA} &\, \equiv \, \Theta_\mathrm{I}^{}
\, + \,\frac{5}{21} \widetilde{\Theta}_\mathrm{I\!I}^{}\,.
\label{eq:Theta_kappa_MA}
\end{align}

In the limit of  $\epsilon_d^{}\to +\infty$, the FL parameters approach the values 
$\delta \to 0$,  $\widetilde{K} \to 0$, 
$\Theta_\mathrm{I}^{} \to -2$, 
and  $\widetilde{\Theta}_\mathrm{I\!I}^{}\to 0$.
Thus,   
\begin{align}
W_\kappa^\mathrm{MA} \xrightarrow{\epsilon_d^{}\to +\infty}\, \frac{43}{21}\,,
\qquad \Theta_\kappa^\mathrm{MA} \xrightarrow{\epsilon_d^{}\to +\infty} 
\,-2\,.
\end{align}
In the $1/N$-filling Kondo limit $\epsilon_d^{}\to -\infty$, 
the parameters take the following values: 
$\delta \to \pi/N$,  $\widetilde{K} \to 1$, 
$\Theta_\mathrm{I}^{} + \widetilde{\Theta}_\mathrm{I\!I}^{}\to 0$.  
Thus,  
\begin{align}
 & \!\! 
W_\kappa^\mathrm{MA}\xrightarrow{\epsilon_d^{}\to -\infty} 
\frac{1}{21} \left[32 + \left(11 - \frac{18}{N-1} \right) \cos\frac{2\pi}{N}\right], 
 \\[2mm]
 & \!
\Theta_\kappa^\mathrm{MA} \xrightarrow{\epsilon_d^{}\to -\infty}
\, \frac{16}{21}\,\Theta_\mathrm{Kond}^{1/N} \,. 
\end{align}
Here $\Theta_\mathrm{Kond}^{1/2}=0$ for $N=2$ 
and $\Theta_\mathrm{Kond}^{1/4}=-1.11$ for $N=4$, as mentioned.

NRG results for 
$C_\kappa^\mathrm{MA}$, $W_\kappa^\mathrm{MA}$, 
and $\Theta_\kappa^\mathrm{MA}$ are plotted as functions of $\epsilon_d^{}$ 
in Fig.\ \ref{fig:ck_MA}.
In contrast to the coefficient $C_\kappa^\mathrm{QD}$ for quantum dots, 
the coefficient $C_\varrho^\mathrm{MA}$ for magnetic alloys 
does not change sign throughout the entire range of $\epsilon_d^{}$, 
for both $N=2$ and $N=4$. 
In the SU(2) case, the  plateau value of $C_\kappa^\mathrm{MA}$ 
 in the half-filling Kondo regime is determined by 
the two-body part  $W_\kappa^\mathrm{MA}$, since 
the  three-body part 
$\Theta_\kappa^\mathrm{MA}$ vanishes in this case 
due to electron-hole symmetry. 
Note that the prefactor $(11-\frac{18 \widetilde{K}^2}{N-1})$ 
for the $\cos 2\delta$ term in $W_\kappa^\mathrm{MA}$, 
as described in Eq.\ \eqref{eq:W_kappa_MA}, 
changes sign for $N=2$ at the point where the Wilson ratio reaches 
$\widetilde{K}=\sqrt{11/18}$, while this does not occur for $N = 4$. 
The sign change of this prefactor  causes the dip 
that appears in $W_\kappa^\mathrm{MA}$ 
at $\epsilon_d^{}/(\pi\Delta ) \simeq -0.6$ in the SU(2) case,   
and it leads to a steep variation of $C_\kappa^\mathrm{MA}$ 
near $\epsilon_d^{}/(\pi\Delta ) \simeq -0.3$, 
as shown in Fig.\ \ref{fig:ck_MA}(a).
In the SU(4) case, the coefficient $C_\kappa^\mathrm{MA}$  
exhibits a positive plateau structure in the quarter-filling Kondo regime, 
which is determined by  competitive contributions of 
$W_\kappa^\mathrm{MA}$ and $\Theta_\kappa^\mathrm{MA}$. 
Figure \ref{fig:ck_MA}(b) also shows that  
  $C_\kappa^\mathrm{MA}$ for SU(4) magnetic alloys 
 takes the opposite sign compared to 
$C_\kappa^\mathrm{QD}$ for quantum dot, 
described in Fig.\ \ref{fig:ck_QD}(b), over the entire filling range 
$0 < N_d^{} < 1$  in the strong interaction limit  $U \to \infty$.

\section{Summary}
\label{sec:summary}

We have studied low-energy transport 
through the SU($N$) Anderson impurity model 
for quantum dots and magnetic alloys in the strong coupling limit $U\to\infty$  
over a wide range of impurity electron filling $0<N_d^{}<1$,   
across the $1/N$-filling Kondo and valence fluctuation regimes. 
Our analysis is based on the latest version of Fermi liquid theory,  
which reveals the essential role of the three-body correction 
in completely describing 
the next-to-leading leading order terms of the transport coefficients. 
The three-body correlation functions have been calculated 
 for the SU(2) and SU(4) impurities using the NRG approach. 
The results systematically clarify the role of three-body Fermi-liquid corrections  
in the most strongly correlated situations.

In the quarter-filling Kondo regime, 
$\epsilon_d^{} \to -\infty$ for  the SU(4) case, 
the three independent components of the three-body correlation functions, 
 $\Theta_\mathrm{I}^{}$, 
$ -\widetilde{\Theta}_\mathrm{I\!I}^{}$, and 
$\widetilde{\Theta}_\mathrm{I\!I\!I}^{}$, 
converge to 
a single universal value $\Theta_\mathrm{Kond}^{1/4} = -1.11$,  
which agrees with the analytical value obtained by Mora {\it et al.\ } \cite{Mora_etal_2009} for the corresponding function in the SU($4$) Kondo model. 
Compared to the previous result obtained 
at finite $U$ \cite{teratani2024thermoelectric}, 
the plateau structure with this height becomes significantly clearer 
in the strong interaction limit.
However, outside the Kondo regime, such as in the valence fluctuation regime, 
these three components take different values and contribute distinctly 
to the next-to-leading-order terms of the transport coefficients.

It has also been demonstrated 
that these three-body correlations couple strongly to  
the tunneling asymmetry 
$ \gamma_\mathrm{dif}^{} =
(\Gamma_L-\Gamma_R)/(\Gamma_L+\Gamma_R)$ 
 between quantum dots and reservoirs,   
significantly affecting 
the order $(eV)^3$ component 
 of nonlinear current through the quarter-filling Kondo state. 
The order $|eV|^3$ current noise of an SU(4) quantum dot  
exhibits quite different behavior from that of SU(2).  
While the coefficient $C_S^{}$ 
changes sign in the valence fluctuation regime 
in the SU(2) case even in  $U \to \infty$ limit, 
it remains positive in the SU(4) case due to 
the competing contributions of two-body $W_S^{}$ 
and three-body $\Theta_S^{}$ correlations.

We have also further examined the next-to-leading-order terms of thermoelectric 
transport coefficients for both quantum dots and magnetic alloys, 
previously analyzed at finite $U$ \cite{teratani2024thermoelectric}.
In contrast to the coefficients 
$C_T^{}$ and $C_\kappa^\mathrm{QD}$, 
which correspond to the $T^2$ conductance 
and $T^3$ thermal conductance of quantum dots, 
the coefficients $C_\varrho^\mathrm{MA}$ 
and $C_\kappa^\mathrm{MA}$, defined 
with respect to electrical and thermal resistivities of magnetic alloys, 
do not change sign throughout the entire range of impurity electron filling. 
For SU(2) magnetic alloys, 
these coefficients 
exhibit a steep variation in the valence fluctuation region 
as the occupation number increases toward the Kondo regime. 
This behavior is caused by the wide variation range of the Wilson ratio $R$ 
for $N=2$ and does not occur for $N \geq 4$.

The leading and the next-to-leading-order terms of the transport coefficients 
through SU($N$) quantum dots can be fully described by {\it five\/} 
Fermi-liquid parameters: $\delta$, $T^*$, $\widetilde{K}$,  
$\Theta_\mathrm{I}^{}$, and $\widetilde{\Theta}_{\mathrm{II}}$ 
for symmetric tunnel junctions. 
An additional three-body component, $\widetilde{\Theta}_\mathrm{III}^{}$, 
associated with electrons in three different impurity levels, 
also contributes to  $C_V^{(3)}$ for $N \geq 3$, 
when tunneling asymmetry  $\gamma_\mathrm{dif}^{}$,  
bias asymmetry $\alpha_\mathrm{dif}^{}$, or both are present. 
These Fermi-liquid parameters can experimentally be determined 
by measuring the leading and next-to-leading-order terms 
of some transport coefficients. From the observed value, 
it is possible to work backward and deduce the three-body correlations. 
The experimentally determined correlation functions can then be used to 
predict the behavior of other unmeasured transport coefficients.

\begin{acknowledgments}
This work was supported by JSPS KAKENHI
 Grants No.\ JP21K03415 and JP23K03284  
and by JST CREST Grant No.\ JPMJCR1876.
K.M. was supported by JST Establishment of University
Fellowships towards the Creation of Science Technology
Innovation Grant No. JPMJFS2138, 
and by JST SPRING Grant No. JPMJSP2139.
\end{acknowledgments}

\appendix

\section{Higher-order Fermi liquid relations}
\label{sec:Ward_identity}

Here, we provide a brief overview of the microscopic Fermi liquid theory 
for the Anderson impurity model, including recent developments.
The Fermi liquid behavior of quantum impurity systems reflects 
the low-energy asymptotic form of the retarded Green's function 
defined in Eq.\ \eqref{eG},  which can also be expressed in the following form: 
\begin{align}
G_\sigma^r(\omega)\,=& 
\ \frac{1}{\omega-\epsilon_{d\sigma}^{}+i\Delta-\Sigma_\sigma^r(\omega)}\,.
\label{eq:Green_Dyson}
\end{align}
The phase shift $\delta_\sigma^{}$ is given by 
the argument of the Green's function in the complex plane,
 $G_\sigma^r(0) = -\left| G_\sigma^r(0)\right| e ^{i \delta_{\sigma}^{}}$,  
at $\omega=T=eV=0$.
It plays a primary role in the ground-state properties 
through the Friedel sum rule 
 $\langle n_{d\sigma}^{} \rangle \xrightarrow{T \to 0} 
\delta_\sigma /\pi$  \cite{ShibaKorringa}, 
e.g., the spectral weight of impurity levels at the Fermi level is given by 
 \begin{align}
 \rho_{d\sigma}^{} \, \equiv \, \rho_{d\sigma}^{}(0)  
\,= \,  \frac{\sin^2\delta_\sigma^{}}{\pi\Delta}  
\,,
 \label{eq:rho_def} 
\end{align}
where  $\rho_{d\sigma}^{}(\omega) 
\equiv  A_{\sigma}^{}(\omega)\Big|_{T=eV=0}^{}$, with   
 $A_{\sigma}^{}(\omega)$  the nonequilibrium 
spectral function defined in Eq.\ \eqref{eq:spectral_function}.

The contributions of low-energy excitations can be deduced 
from the equilibrium self-energy 
$\Sigma_{\mathrm{eq},\sigma}^r(\omega)
\equiv \left. \Sigma_\sigma^r(\omega) \right|_{T=eV=0}^{}$ 
by expanding it step by step around the Fermi energy $\omega=0$.
The terms up to order $\omega$ determine the structure of 
the renormalized resonance state:     
\begin{align}
G_{\sigma}^{r}(\omega)\,\simeq \,  
\frac{z_\sigma^{}}{\omega-\widetilde{\epsilon}_{d\sigma}^{}
+i\widetilde{\Delta}_{\sigma}^{}} \,. 
\end{align}
The renormalized parameters 
are defined as \cite{YamadaYosida2,HewsonRPT2001}  
\begin{align}
\frac{1}{z_\sigma^{}}\, \equiv & \ 
1-\left.\frac{\partial \Sigma_{\mathrm{eq},\sigma}^r(\omega)}
{\partial \omega}\right|_{\omega=0}^{} 
\,, 
\label{eq:Zdef}
\\
\widetilde{\epsilon}_{d\sigma}^{}
\,\equiv & \ z_\sigma^{}
\left[\,\epsilon_{d\sigma}^{}+\Sigma_{\mathrm{eq},\sigma}^r(0)\,\right] \,,
\qquad \ 
\widetilde{\Delta}_{\sigma}^{} \,\equiv \,  z_\sigma^{}\Delta\,.
\rule{0cm}{0.6cm}
\label{eq:ed_ren}
\end{align}
The Ward identities 
\cite{YamadaYosida2,YamadaYosida4,ShibaKorringa,Yoshimori}, 
which reflect the current conservation described in 
Eq.\ \eqref{eq:current_conservation} \cite{Oguri2022},
can be expressed, at $T=eV=0$, as a relation between the causal self-energy 
$\Sigma_{\mathrm{eq},\sigma}^{--}(\omega)$
and the vertex function $\Gamma_{\sigma\sigma';\sigma'\sigma}^{--;--}
(\omega,\omega';\omega',\omega)$,  
within the standard zero-temperature formalism: 
\begin{align}
\delta_{\sigma\sigma'} \frac{\partial \Sigma_{\mathrm{eq},\sigma}^{--}(\omega) }{\partial \omega} 
+ \frac{\partial \Sigma_{\mathrm{eq},\sigma}^{--}(\omega) }{\partial \epsilon_{d\sigma'}^{}} 
 = \, - \Gamma_{\sigma\sigma';\sigma'\sigma}^{--;--}(\omega,0;0,\omega) 
\,\rho_{d\sigma'}^{}\,.
\label{eq:YYY_T0_causal} 
\end{align}
This self-energy, defined with respect to the causal Green's function 
in the standard $T=0$ formalism, is not analytic by definition, 
as its imaginary part exhibits a discontinuity 
along the real axis of the complex $\omega$ plane \cite{AGD}: 
\begin{align}
\Sigma_{\mathrm{eq},\sigma}^{--}(\omega) \, =\, 
\mathrm{Re}\,\Sigma_{\mathrm{eq},\sigma}^{r}(\omega)\,+\, i\,
\mathrm{Im}\,\Sigma_{\mathrm{eq},\sigma}^{r}(\omega)
\,\mathrm{sgn} \, \omega\,.
\label{eq:T=0selfenergy}
\end{align}

The corresponding causal vertex function has the  property 
that the component with  $\sigma=\sigma'$ vanishes 
 at zero frequencies  $\omega=\omega'=0$ 
\cite{YamadaYosida4,Yoshimori}:  
 \begin{align}
 &\Gamma_{\sigma\sigma;\sigma\sigma}^{--;--}(0,0;0,0) \,=\, 0 \,. 
\label{eq:self_w1_N}
\end{align}
Therefore, the renormalization factor  $z_\sigma^{}$ and the derivative 
$\partial \Sigma_{\mathrm{eq},\sigma}^r(0)/ \partial\epsilon_{d\sigma}^{}$ 
are related to each other through Eq.\ \eqref{eq:YYY_T0_causal}  
\cite{YamadaYosida2}:  
\begin{align}
\frac{1}{z_\sigma^{}}
\, = \, 
\widetilde{\chi}_{\sigma\sigma}^{}
\,, 
\qquad \  
\widetilde{\chi}_{\sigma\sigma'}^{} 
\,\equiv\,
\delta_{\sigma\sigma'}^{}+\frac{\partial \Sigma_{\mathrm{eq},\sigma}^r(0)}
{\partial \epsilon_{d\sigma'}} \,. 
\label{eq:YamadaYsidaRelationAppendix}
\end{align}
The coefficient $\widetilde{\chi}_{\sigma\sigma'}^{}$ 
determines the extent to which the susceptibility $\chi_{\sigma\sigma'}^{}$ 
is enhanced at $T=0$ by the vertex correction, i.e.,   
\begin{align}
\chi_{\sigma\sigma'}^{}\, =\,
- \frac{\partial \bigl\langle n_{d\sigma}^{}\bigr\rangle}
{\partial\epsilon_{d\sigma'}^{}}
\ \xrightarrow{\,T\to 0 \,}\  \rho_{d\sigma}^{} 
\widetilde{\chi}_{\sigma\sigma'}^{}\,.
\label{eq:YamadaYsidaRelationAppendixAdd}
\end{align}
Furthermore, the $\sigma\neq\sigma'$ component of the susceptibility  
is related to the residual interaction between quasiparticles 
through Eq.\ \eqref{eq:YYY_T0_causal} \cite{Yoshimori,HewsonRPT2001}:  
\begin{align}
& \!\!\!\!
\chi_{\sigma\sigma'}^{} =\,
- \Gamma_{\sigma\sigma';\sigma'\sigma}^{--;--}(0,0;0,0) 
\,\rho_{d\sigma}^{}\,\rho_{d\sigma'}^{} , 
\quad  \ \ \sigma\neq \sigma' . 
\end{align}
The derivative of  $\rho_{d\sigma}^{}(\omega)$ also 
contributes to the low-energy transport 
and is related to the susceptibility 
by using Eq.\ \eqref{eq:YamadaYsidaRelationAppendix}, as      
\begin{align}
\rho_{d\sigma}'  \equiv  \left.
\frac{\partial \rho_{d\sigma}^{}(\omega)}{\partial \omega} 
\right|_{\omega=0}^{} 
=\,  -  
\frac{\partial \rho_{d\sigma}^{}}{\partial \epsilon_{d\sigma}^{}} 
\ = \ \frac{\chi_{\sigma\sigma}^{}}{\Delta}\,\sin2\delta_{\sigma}^{}.
\label{eq:rho_d_omega_2}
\end{align}

It has recently been clarified 
that the vertex function for $\sigma=\sigma'$ 
also has the following property 
\cite{AO2017_I,AO2017_II},  
in addition to Eq.\ \eqref{eq:self_w1_N}: 
\begin{align}
& \left. \frac{\partial }{\partial \omega}
\mathrm{Re}\, 
\Gamma_{\sigma\sigma;\sigma\sigma}^{--;--}(\omega,0;0,\omega) 
\right|_{\omega\to 0} =\,0  
\;.
\label{eq:self_w2_N}
\end{align}
This property indicates that the real part of  
$\Gamma_{\sigma\sigma;\sigma\sigma}^{--;--}(\omega,0;0,\omega)$ 
does not contain a linear term in $\omega$.
Based on this and the Ward identity given in Eq.\ \eqref{eq:YYY_T0_causal},  
 the order $\omega^2$ real part of  the self-energy 
has been shown to be expressed in terms of 
the derivative of the susceptibility, 
or the three-body correlation function 
\cite{FMvDM2018,AO2017_I,AO2017_II,AO2017_III}: 
\begin{align}
\!\!\! 
\left.
\frac{\partial^2}{\partial \omega^2}
\mathrm{Re}\,\Sigma_{\mathrm{eq},\sigma}^{r}(\omega)
\right|_{\omega \to 0}^{} 
\,=\  \frac{\partial^2 \Sigma_{\mathrm{eq},\sigma}^{r}(0)}
{\partial \epsilon_{d\sigma}^{2}}
\ = \ 
\frac{\partial \widetilde{\chi}_{\sigma\sigma}^{}}{\partial \epsilon_{d\sigma}^{}} \,.
\label{eq:self_w2}
\end{align}
Furthermore, from Eqs.\ \eqref{eq:YYY_T0_causal}, 
\eqref{eq:self_w1_N}, and \eqref{eq:self_w2_N},  
the vertex function $\Gamma_{\sigma\sigma';\sigma'\sigma}^{--;--}(\omega, \omega'; \omega' ,\omega)$  
can be exactly deduced up to linear-order terms in $\omega$ and $\omega'$  
 at $T=eV=0$ \cite{AO2017_II,AO2017_III,Oguri2022}.   
The result takes the following form, including the imaginary part, 
which is known to exhibit 
nonanalytic 
 $|\omega-\omega'|$ and $|\omega+\omega'|$ dependences 
\cite{YamadaYosida4,ShibaKorringa,Yoshimori,AO2001}: 
The diagonal components ($\sigma = \sigma'$) are given by 
\begin{align}
& 
\Gamma_{\sigma\sigma;\sigma\sigma}^{--;--}
(\omega , \omega'; \omega', \omega) 
\,\rho_{d\sigma}^{2}
\, =\,  
 i \pi 
\sum_{\sigma''(\neq \sigma)}
\chi_{\sigma\sigma''}^2
\,\bigl|\omega-\omega' \bigr| 
+ \cdots ,
\label{eq:GammaUU_general_omega_dash_N}
 \end{align}
and the off-diagonal components ($\sigma \neq \sigma'$) are 
 \begin{align}
&\!\!\!\! 
 \Gamma_{\sigma\sigma';\sigma'\sigma}^{--;--}(\omega, \omega'; \omega' ,\omega) 
\,\rho_{d\sigma}^{}\rho_{d\sigma'}^{}
\nonumber \\
&  \quad \quad  = \     
- \chi_{\sigma\sigma'}^{}
+ 
\rho_{d\sigma}^{}
\frac{\partial \widetilde{\chi}_{\sigma\sigma'}^{}}
{\partial \epsilon_{d\sigma}^{}} \, \omega  
+ 
\rho_{d\sigma'}^{}
\frac{\partial \widetilde{\chi}_{\sigma'\sigma}^{}}
{\partial \epsilon_{d\sigma'}^{}} \, \omega'   
\rule{0cm}{0.6cm}
 \nonumber \\ 
 & \qquad \quad \ 
 + i \pi \,
\chi_{\sigma\sigma'}^2 
\Bigl(
\,\bigl|  \omega - \omega'\bigr| 
-
\,\bigl| \omega + \omega' \bigr| 
\Bigr)
+ \cdots. 
\rule{0cm}{0.5cm}
\label{eq:GammaUD_general_omega_dash_N}
\end{align}
The order $\omega^2$ imaginary part of the self-energy 
has been derived through Eqs.\ \eqref{eq:YYY_T0_causal} and
\eqref{eq:GammaUU_general_omega_dash_N} 
\cite{YamadaYosida4,ShibaKorringa,Yoshimori}: 
\begin{align}
\left.
\frac{\partial^2}{\partial \omega^2}\, 
\mathrm{Im}\,\Sigma_{\mathrm{eq},\sigma}^{r}(\omega)
\right|_{\omega \to 0}^{} 
\,=\ 
 -  \,\frac{\pi}{\rho_{d\sigma}^{}}
\sum_{\sigma''(\neq \sigma)} 
\chi_{\sigma\sigma''}^2  \,.
\end{align}

The order $T^2$ term of the retarded self-energy 
$\Sigma_{\sigma}^r(0)$ can also be deduced from these asymptotic forms   
of the vertex function \cite{AO2017_II},  
by rewriting the proof provided by Yamada in Ref.\ \onlinecite{YamadaYosida4}
 in the following form,  at $eV = 0$:  
\begin{align}
&\Sigma_{\sigma}^r(0) 
-
\left.
\Sigma_{\sigma}^r(0) \right|_{T=0}^{} 
\, = \, \frac{(\pi  T)^2}{6}\, 
\lim_{\omega \to 0^+}
\Psi_{\sigma}^{--}(\omega)   +  \cdots \;,
\label{eq:Psi_result_T2}
\\
&\Psi_{\sigma}^{--}(\omega) 
\,\equiv  \,  
 \lim_{\omega' \to 0}
\frac{\partial}{\partial \omega'} \, 
 \sum_{\sigma'}\,
\Gamma_{\sigma \sigma';\sigma' \sigma}^{--;--}
(\omega , \omega'; \omega' , \omega) 
\rho_{d\sigma'}^{}(\omega') \;.
\label{eq:Psi_T0}
 \end{align}
The right-hand side of Eq.\ \eqref{eq:Psi_T0}
can be calculated by using the low-energy asymptotic forms 
of the vertex function given in  
 Eqs.\ \eqref{eq:GammaUU_general_omega_dash_N} and 
\eqref{eq:GammaUD_general_omega_dash_N} for finite $\omega$, 
and then taking the limit $\omega \to 0$,  
\begin{align}
\lim_{\omega \to 0} \Psi_{\sigma}^{--}(\omega)   =  
 \frac{1}{\rho_{d\sigma}^{}} 
\sum_{\sigma'(\neq \sigma)}
\left[
\frac{\partial \chi_{\sigma\sigma'}^{}}{\partial \epsilon_{d\sigma'}^{}} 
 - i \, \frac{3\pi}{\rho_{d\sigma}^{}} \,\chi_{\sigma\sigma'}^2
\,  \mathrm{sgn}(\omega) 
\right] . 
\label{eq:Psi_result_+_N}
\end{align}
 Here  $\mathrm{sgn}(\omega)$ specifies the sign of the imaginary part, 
which depends on the direction from which the frequency approaches zero, 
 i.e., $\omega\to +0^+$ or  $\omega\to - 0^+$ 
(In our notation $0^+$ denotes a positive infinitesimal,  
and thus $-0^+$ is equivalent to $0^-$). 
It reflects the branch cuts of 
$\Gamma_{\sigma \sigma';\sigma' \sigma}^{--;--}
(\omega , \omega'; \omega' , \omega)$ 
 along the lines  $\omega-\omega'=0$ and $\omega+\omega'=0$
 in the frequency plane.

Similarly, the bias dependence of 
the self-energy can be deduced, up to terms of order $(eV)^2$, 
from the asymptotic form of the vertex function given in 
 Eqs.\ \eqref{eq:GammaUU_general_omega_dash_N} and 
\eqref{eq:GammaUD_general_omega_dash_N}, 
using the Ward identities obtained at $T=0$ 
for the causal self-energy $\Sigma_{\sigma}^{--}(\omega)$ 
in the Keldysh formalism \cite{AO2001,AO2017_III}:     
\begin{align}
&\!\!\!\!\!\!  
 \left.\frac{\partial \Sigma_{\sigma}^{--}(\omega)}{\partial eV}
\right|_{eV= 0}^{}    = \, 
 \alpha   
\sum_{\sigma'}
\Gamma_{\sigma\sigma';\sigma'\sigma}^{--;--}
 (\omega,0; 0,\omega)
\,\rho_{d\sigma'}^{}  \,, 
\label{eq:self_retarded_1st_derivative_in_eV}
\end{align}
\begin{align}
&
\left.
\frac{\partial^2  
\Sigma_{\sigma}^{--}(\omega)}{
\partial (eV)^2}
\right|_{eV= 0}^{}   =    \  \, 
\frac{ \Gamma_L\,\Gamma_R}{ \left( \Gamma_L+ \Gamma_R \right)^2}
   \  \Psi_{\sigma}^{--}(\omega) 
 \nonumber \\
& \qquad \qquad \ 
 -   \alpha^2  
\left(
\frac{\partial}{\partial \omega}
+ \frac{\partial}{\partial \epsilon_{d}^{}}
\right)
\sum_{\sigma'}
\Gamma_{\sigma\sigma';\sigma'\sigma}^{--;--}
 (\omega,0; 0,\omega)
\,\rho_{d\sigma'}^{} .
\rule{0cm}{0.85cm}
\label{eq:derivative_self_2_with_CII_T0}
\end{align}
Here $\partial/\partial \epsilon_{d}^{} \equiv 
  \sum_{\sigma''}  \partial/\partial \epsilon_{d\sigma''}^{}$ 
and  $\Psi_{\sigma}^{--}(\omega)$  
is the correlation function defined in Eq.\ \eqref{eq:Psi_T0}.
The order $\omega\, eV$  term of the self-energy 
follows from Eq.\ \eqref{eq:self_retarded_1st_derivative_in_eV}:
\begin{align}
&
\lim_{\omega \to 0} 
\frac{\partial}{\partial \omega} \left[
\frac{\partial \Sigma^{--}_\sigma(\omega)}{\partial (eV)}\right]_{eV=0} 
 \nonumber  \\
 & 
=  \  
 \alpha 
\sum_{\sigma'(\neq \sigma)} 
 \left[ \,
 \frac{\partial \widetilde{\chi}_{\sigma\sigma'}^{}}{\partial \epsilon_{d\sigma}^{}} 
 +i \, \frac{\pi}{\rho_{d\sigma}^{}} \,\chi_{\sigma\sigma'}^2
\, \mathrm{sgn}(\omega) 
 \,\right] .
\rule{0cm}{0.7cm}
\label{eq:self_w_eV}
\end{align}
Here, the sign of the imaginary part depends on the direction,  
  $\omega\to +0^+$ or $\omega\to - 0^+$,  
as mentioned. 
The second term in the right-hand side of 
Eq.\ \eqref{eq:derivative_self_2_with_CII_T0} 
can also be expressed in the following form, at small frequencies,  
\begin{align}
&\lim_{\omega\to 0}
\left(
\frac{\partial}{\partial \omega}
 + \frac{\partial}{\partial \epsilon_{d}^{}}
\right)
 \sum_{\sigma'}\,
\Gamma_{\sigma \sigma';\sigma' \sigma}^{--;--}(\omega , 0; 0 , \omega) 
\rho_{d\sigma'}^{} 
\nonumber  \\ 
&=
\sum_{\sigma'(\neq \sigma)} 
 \left[ \,
 -
\sum_{\sigma'' (\neq \sigma)} 
\frac{\partial \widetilde{\chi}_{\sigma\sigma'}^{}}{\partial \epsilon_{d\sigma''}^{}} 
 +i  \, \frac{\pi}{\rho_{d\sigma}^{}} \,\chi_{\sigma\sigma'}^2
 \,\mathrm{sgn}(\omega) 
 \,\right] .
\end{align}
Similarly,  $\mathrm{sgn}(\omega)$ 
represents the sign that depends on whether   
  $\omega\to 0^+$ or $\omega\to - 0^+$.  

Note that $\alpha$ is the parameter defined as  
$\alpha\, eV 
\equiv (\Gamma_L \,\mu_L+\Gamma_R\, \mu_R)/ (\Gamma_L+\Gamma_R)$, i.e.,   
\begin{align}
\alpha \, \equiv  \, 
\frac{\alpha_L\Gamma_L-\alpha_R\Gamma_R}{\Gamma_L+\Gamma_R}
\,=\, \frac{1}{2}\Bigl(\alpha_\mathrm{dif}^{}+\gamma_\mathrm{dif}^{}\Bigr)
\,,   
\end{align}
and it affects the nonlinear transport 
when there is tunneling asymmetry, bias asymmetry, or both  
[see  Eqs.\ \eqref{eq:alpha_dif} and \eqref{eq:gamma_dif} 
for the definitions of $\gamma_\mathrm{dif}^{}$ and $\alpha_\mathrm{dif}^{}$].

\begin{widetext}

Using these results for the causal self-energy together with 
Eq.\ \eqref{eq:T=0selfenergy}, the retarded self-energy 
$\Sigma_\sigma^r(\omega)$ has been exactly determined 
up to terms of order $\omega^2$, $T^2$, and $(eV)^2$:
\begin{align}
& 
\mathrm{Im} \, \Sigma_\sigma^r (\omega)\,= \,  
-\frac{\pi}{2} 
\frac{1}{\rho_{d\sigma}^{}}\sum_{\sigma'(\neq \sigma)} 
\chi_{\sigma \sigma'}^2 
 \Biggl[ (\omega-\alpha eV)^2 
+  \frac{3\Gamma_L\Gamma_R}{(\Gamma_L+\Gamma_R)^2} (eV)^2 
 +  (\pi T)^2 \Biggr]+\cdots   ,  
\label{eq:ImSelf}
\end{align}
 \begin{align}
\epsilon_{d\sigma}^{} + 
\mathrm{Re} \, \Sigma_\sigma^r(\omega) \, = &  \    
\Delta \cot \delta_\sigma^{}
-\sum_{\sigma'(\neq \sigma)} \widetilde{\chi}_{\sigma \sigma'}^{}\,\alpha \, eV 
+ (1-\widetilde{\chi}_{\sigma \sigma}^{})\, \omega
+ \frac{1}{6}\frac{1}{\rho_{d\sigma}^{}}\sum_{\sigma'(\neq \sigma)}
\frac{\partial \chi_{\sigma \sigma'}^{}}{\partial \epsilon_{d\sigma'}}
\Biggl[  \frac{3\Gamma_L\Gamma_R}{(\Gamma_L+\Gamma_R)^2} (eV)^2
+(\pi T)^2 \Biggr] 
\nonumber
\\ 
& 
+\frac{1}{2}
\frac{\partial \widetilde{\chi}_{\sigma \sigma}^{}}{\partial \epsilon_{d\sigma}^{}}
\,\omega^2 
+\sum_{\sigma'(\neq \sigma)}
\frac{\partial \widetilde{\chi}_{\sigma \sigma'}^{}}{\partial \epsilon_{d\sigma}^{}} 
\,\alpha \, eV  \omega \,
+\frac{1}{2}\sum_{\sigma'(\neq \sigma)}\sum_{\sigma''(\neq \sigma)}
\frac{\partial \widetilde{\chi}_{\sigma \sigma'}^{}}{\partial \epsilon_{d\sigma''}^{}}
 \,\alpha^2(eV)^2
+\cdots\,.  
\label{eq:ReSelf}
\end{align}
%
%
Note that the imaginary part of the retarded self-energy is related to  
the lesser self-energy $\Sigma_\sigma^{-+}(\omega)$
and the greater self-energy  $\Sigma_\sigma^{+-}(\omega)$
as  
$2i\, \mathrm{Im}\, \Sigma^{r}_{\sigma}(\omega)  
 = \Sigma^{-+}_{\sigma}(\omega) - \Sigma^{+-}_{\sigma}(\omega)$. 
The low-energy asymptotic forms of these two components  
have been derived by Aligia \cite{Aligia2012,Aligia2014}. 
Their results, which are exact up to terms of 
 $\omega^2$, $T^2$, and $(eV)^2$,  
can be expressed in the following form \cite{Oguri2022}: 
\begin{align}
& 
\Sigma^{\mathrm{K}}_{\sigma}(\omega) 
\, \equiv \,  
- \Sigma^{-+}_{\sigma}(\omega) 
- \Sigma^{+-}_{\sigma}(\omega)
\ =  \,   
-i \, \frac{\pi}{\rho_{d\sigma}^{}} \sum_{\sigma'(\neq \sigma)} 
\chi_{\sigma\sigma'}^2 
\  \mathcal{I}_\mathrm{K}^{}(\omega)
\,+\, \cdots  \,, 
\label{eq:self_Keldysh_compoment_table}
\\ 
&
\mathcal{I}_\mathrm{K}^{}(\omega)
\,\equiv \,
2 \! \int_{-\infty}^{\infty}\!\!\!  d\varepsilon_1  
\!\!  \int_{-\infty}^{\infty}
\!\!\! d\varepsilon_2\,
\biggl\{ \Bigl[ 1-f_\mathrm{eff}^{}(\varepsilon_1)\Bigr]
 \Bigl[ 1-f_\mathrm{eff}^{}(\varepsilon_2)\Bigr] \,
 f_\mathrm{eff}^{}(\varepsilon_1+\varepsilon_2-\omega) 
 \,- \,
 f_\mathrm{eff}^{}(\varepsilon_1)  \,f_\mathrm{eff}^{}(\varepsilon_2) \,
\Bigl[1-f_\mathrm{eff}^{}(\varepsilon_1+\varepsilon_2-\omega) \Bigr] \biggr\}, 
\nonumber \\
& \qquad \ \ \,   
 =\,     
 \sum_{j,k,\ell  = L,R} 
\frac{\Gamma_j\Gamma_k\Gamma_\ell}{(\Gamma_L+\Gamma_R)^3}\,
 \Bigl[\, \bigl(\omega-\mu_j-\mu_k + \mu_\ell\bigr)^2 + \bigl(\pi T\bigr)^2 \Bigr]
\, \Bigl[ \,1- 2\,f(\omega-\mu_j-\mu_k + \mu_\ell )\,\Bigr] 
\, .
\label{eq:collision_diff_general_2}
\end{align}
\end{widetext}
Here $f_\mathrm{eff}^{}(\omega) \equiv 
\sum_{j=L,R} \Gamma_j f(\omega-\mu_j)/(\Gamma_L+\Gamma_R)$. 
The counterpart 
$\Sigma^{-+}_{\sigma}(\omega) - \Sigma^{+-}_{\sigma}(\omega)$  
agrees with Eq.\ \eqref{eq:ImSelf}:   
The Fermi distribution functions $f(\omega-\mu_j-\mu_k + \mu_\ell )$, 
which emerge in $\Sigma^{-+}_{\sigma}(\omega)$ and 
$\Sigma^{+-}_{\sigma}(\omega)$ 
 through the collision integral $\mathcal{I}_\mathrm{K}^{}(\omega)$  
defined in Eq.\ \eqref{eq:collision_diff_general_2},
cancel each other out  in $\mathrm{Im}\, \Sigma^{r}_{\sigma}(\omega)$.  
This cancellation reflects a property imposed by causality, i.e.,  
$\Sigma_\sigma^r(\omega)$ is analytic 
in the upper half of the complex $\omega$ plane. 
This requirement places a strong restriction on the form 
of retarded self-energy, and 
the distribution functions such as $f(\omega-\mu_j-\mu_k + \mu_\ell )$ 
cannot remain in $\Sigma_\sigma^r(\omega)$ because  
$f(\omega)$ has a series of poles 
at $\omega=i(2n+1)\pi T$, for integer $n$,  
along the imaginary axis.

In order to investigate the order $|eV|^3$ nonlinear noise,  
it is necessary to calculate the vertex corrections in the Keldysh formalism 
 as well, up to terms of order $|eV|$.   
This has been carried out in Ref.\ \onlinecite{Oguri2022} to derive the formulas 
given in Eqs.\ \eqref{eq:noise_SUN}--\eqref{eq:ThetaS_SUN}.

\section{NRG  procedures}
\label{sec:NRG_parameters}

We have performed 
NRG calculations by dividing the $N$ conduction channels into $N/2$ pairs 
and using the SU(2) spin and U(1) charge symmetries for each pair.
The discretization parameter $\Lambda$ and the number of retained low-lying
excited states $N_\mathrm{trunc}$ are chosen as $(\Lambda, N_\mathrm{trunc})=(2, 4000)$ for $N=2$ and $(\Lambda, N_\mathrm{trunc})=(6, 10\, 000)$ for $N=4$.
Note that the SU(4) symmetry is preserved in our iteration scheme 
because the truncation of higher energy states is performed 
after all new states from these two pairs are added. 

In order to calculate $\chi_{B}^{[3]}$ as defined in Eq.\ \eqref{eq:chi_B3_def_new}, 
we introduced a small external potential $\epsilon_{\mathrm{sp}, k}^{}$ 
that depends on the channel index $k=1, 2$, $\ldots$, $N/2$ 
and shifts the impurity level from  $\epsilon_{d}^{}$.
Specifically, for $N=4$, this potential is applied 
in a way equivalent to a local Zeeman field:  
 $\epsilon_{\mathrm{sp}, 1}^{}=-b$ and 
$\epsilon_{\mathrm{sp}, 2}^{}=b$.    
We then deduced $\chi_{B}^{[3]}$ from the derivatives 
of the susceptibilities with respect to $b$.

\section{Three-body correlations for the $1/N$-filling Kondo state}
\label{sec:Mora_formula}

We briefly describe here 
the relation between the dimensionless 
three-body correlation function 
and the parameters $\alpha_1^{}$ and  $\alpha_2^{}$ 
 introduced by Mora {\it et al.\/} 
for the SU($N$) Kondo model 
in Refs.\ \onlinecite{Mora2009, Mora_etal_2009}.  
In the strong-interaction limit, their notation corresponds to ours as follows: 
$\alpha_{1}^{}/(\pi T_K^{}) \Leftrightarrow    
  \chi_{\sigma\sigma}^{}$ and 
$\alpha_{2}^{}/(\pi T_K)^2 
 \Leftrightarrow   -\chi_{\sigma\sigma\sigma}^{[3]}/(2\pi)$. 
They showed that 
the ratio of their parameters, $\alpha_2^{}/\alpha_1^2$, 
can be determined analytically 
using the Bethe ansatz solution \cite{BazhabovLukyanovTsvelik}. 
Specifically,  
for the SU($N$) Kondo state 
with a single impurity electron, i.e., $\delta  \to \pi /N$,  
which is  realized in the  $\epsilon_d^{} \to -\infty$ limit 
of the infinite-$U$  Anderson model,  
the  three-body correlation 
  $\Theta_\mathrm{I}^{}$  approaches this ratio, as   
\begin{align}
\ \!\!\!\!\!\!
\Theta_\mathrm{Kond}^{1/N} 
\,\equiv & \ 
\lim_{\epsilon_d^{} \to  -\infty} \Theta_\mathrm{I}^{} 
\,= \, -\, \frac{\alpha_2^{}}{\alpha_1^2}\, \sin \frac{2\pi}{N}\,,
\\ \nonumber \\
 \frac{\alpha_2^{}}{\alpha_1^2} \ = & \ \,   
 \frac{N-2}{N-1}\, 
\frac{\Gamma\!\left(\frac{1}{N}\right) }{\sqrt{\pi}\,\Gamma\!\left(\frac{1}{2}+\frac{1}{N}\right)}\,, 
\end{align}
where $\Gamma(x)$ is the Gamma function.
For $N=4$, it takes the value $\Theta_\mathrm{Kond}^{1/4} 
= -1.1128 \cdots$.    


%

\end{document}